\documentclass[12pt,a4paper]{article}

\usepackage{fixltx2e}

\usepackage[font=small]{caption}
\usepackage[text={450pt,650pt},headheight={14.5pt},centering]{geometry}

\usepackage{lmodern}

\usepackage{graphicx}
\usepackage{booktabs}
\usepackage{subfig}
\usepackage{tikz}
\usetikzlibrary{plotmarks,calc}

\usepackage{cite}

\usepackage{hyperref}

\usepackage{amsmath,amssymb}
\usepackage{mathtools}
\usepackage{bbm}
\usepackage{slashed}
\usepackage{braket}

\usepackage{xspace}

\numberwithin{equation}{section}

\makeatletter
 \let\old@startsection=\@startsection
 \let\oldl@section=\l@section
 \renewcommand{\@startsection}[6]{\old@startsection{#1}{#2}{#3}{#4}{#5}{#6\mathversion{bold}}}
 \renewcommand{\l@section}[2]{\oldl@section{\mathversion{bold}#1}{#2}}
\makeatother

\DeclareMathOperator{\tr}{tr}

\def\Xint#1{\mathchoice
  {\XXint\displaystyle\textstyle{#1}}%
  {\XXint\textstyle\scriptstyle{#1}}%
  {\XXint\scriptstyle\scriptscriptstyle{#1}}%
  {\XXint\scriptscriptstyle\scriptscriptstyle{#1}}%
  \!\int}
\def\XXint#1#2#3{{\setbox0=\hbox{$#1{#2#3}{\int}$}
    \vcenter{\hbox{$#2#3$}}\kern-.5\wd0}}
\def\pint{\;\Xint-}

\newcommand{\AdS}{\text{AdS}}
\newcommand{\CFT}{\text{CFT}}
\newcommand{\Sphere}{S}
\newcommand{\CP}{C\kern-.1em P}

\newcommand{\comm}[2]{[#1,#2]}
\newcommand{\acomm}[2]{\{#1,#2\}}

\newcommand{\alg}[1]{#1}
\newcommand{\grp}[1]{\mathrm{#1}}

\newcommand{\algD}[1]{\alg{d}(2,1;#1)}
\newcommand{\grD}[1]{\grp{D}(2,1;#1)}

\newcommand{\algSL}{\alg{sl}}
\newcommand{\grSL}{\grp{SL}}

\newcommand{\grSp}{\grp{Sp}}

\newcommand{\algSU}{\alg{su}}
\newcommand{\grSU}{\grp{SU}}

\newcommand{\grO}{\grp{O}}

\newcommand{\algPSU}{\alg{psu}}

\newcommand{\algOSp}{\alg{osp}}
\newcommand{\grOSp}{\grp{OSp}}

\newcommand{\gen}[1]{#1}

\newcommand{\Integers}{\mathbbm{Z}}
\newcommand{\Reals}{\mathbbm{R}}
\newcommand{\Complex}{\mathbbm{C}}
\newcommand{\Naturals}{\mathbbm{N}}

\newcommand{\order}{\mathcal{O}}

\newcommand{\superN}{\mathcal{N}}

\newcommand{\directsum}{\mathop{\oplus}}
\newcommand{\directprod}{\mathop{\otimes}}

\newcommand{\ie}{\textit{i.e.}\xspace}
\newcommand{\eg}{\textit{e.g.}\xspace}

\newlength\vertdist
\newlength\hordist
\begin{document}

\thispagestyle{empty}
\setlength\textwidth{450pt}

\def\thefootnote{\fnsymbol{footnote}}\begin{flushright}
{\small UUITP-17/11}\\
\end{flushright}\vskip 1cm\begin{center}
\Large{\bfseries\mathversion{bold} Integrability, spin-chains and the $\AdS_3/\CFT_2$ correspondence}
\end{center}\vskip 0.8cm
\begin{center}{\large O. Ohlsson Sax$^{1}$, B. Stefa\'nski, jr.$^{2}$}
\vskip 0.2cm{\itshape $^1$ Department of Physics and Astronomy, Uppsala University \\
SE-751 08 Uppsala, Sweden}
\vskip 0.2cm{\itshape $^2$ Centre for Mathematical Science, City University London,
\\ Northampton Square, London EC1V 0HB, UK}

\end{center}
\vskip 1.0cm
\begin{abstract}\noindent
  Building on arXiv:0912.1723~\cite{Babichenko:2009dk}, in this paper we
  investigate the $\AdS_3/\CFT_2$ correspondence using integrability
  techniques. We present an all-loop Bethe Ansatz (BA) for strings on
  $AdS_3\times S^3\times S^3\times S^1$, with symmetry $\algD{\alpha}^2$, valid
  for all values of $\alpha$. This construction relies on a novel,
  $\alpha$-dependent generalisation of the Zhukovsky map. We investigate the
  weakly-coupled limit of this BA and of the all-loop BA for strings on
  $AdS_3\times S^3\times T^4$. We construct integrable short-range spin-chains
  and Hamiltonians that correspond to these weakly-coupled BAs. The spin-chains
  are alternating and homogenous, respectively. The alternating spin-chain can
  be regarded as giving some of the first hints about the unknown $\CFT_2$ dual
  to string theory on $AdS_3\times S^3\times S^3\times S^1$. We show that, in
  the $\alpha\rightarrow 1$ limit, the integrable structure of the
  $\algD{\alpha}^2$ model is non-singular and keeps track of not just massive
  but also massless modes. This provides a way of incorporating massless modes
  into the integrability machinery of the $\AdS_3/\CFT_2$ correspondence.
\end{abstract}

\vfill

\setcounter{footnote}{0}
\def\thefootnote{\arabic{footnote}}
\newpage

\renewcommand{\theequation}{\thesection.\arabic{equation}}

\section{Introduction}
\label{sec:intro}

The gauge/string
duality~\cite{Maldacena:1997re,Gubser:1998bc,Witten:1998qj,Aharony:1999ti}
offers a fundamentally new approach to understanding strongly-coupled
systems. The strongly coupled system is believed to have a gravitational dual;
in this gravitational dual one is often able to compute quantities which are
physically important but difficult to compute in the original strongly coupled
system. While conceptually striking, it is at present not clear how general this
approach is. Over the last few years some of the most powerful evidence for the
gauge/string correspondence has come from the study of the maximally
supersymmetric example: $3+1$-dimensional $\superN=4$ Super-Yang-Mills (SYM)
superconformal field theory (SCFT) and its dual Type IIB string theory on
$AdS_5\times S^5$.  In this example, using integrability techniques, a proposal
exists which can be used to calculate anomalous dimensions of generic,
unprotected, gauge theory operators and match them with the energies of string
states in the dual gravitational spacetime.\footnote{%
  The results based on integrability techniques are limited to the planar limit
  of the gauge theory, in the spirit of 't Hooft's original suggestion about
  using $1/N^2$ as a small expansion parameter~\cite{'tHooft:1973jz}.%
} %
This proposal has a passed a large number of stringent
tests~\cite{Fiamberti:2007rj,Bajnok:2008bm,Bajnok:2008qj,Fiamberti:2008sh,Bajnok:2009vm}. This
remarkable progress allows for the calculation of quantities at all values
(small, intermediate and large) of the gauge theory coupling (see for
example~\cite{Bajnok:2009vm,Lukowski:2009ce, Arutyunov:2010gb,Balog:2010xa}),
and provides some of the strongest evidence for this particular duality. More
importantly, the integrability approach provides a description of how
gauge/string duality actually works in practice.

Following the success of the integrability approach in the above mentioned
maximally supersymmetric example, other dual pairs have been investigated using
these techniques: these included duals involving $3+1$-dimensional gauge
theories with less supersymmetry, and $2+1$-dimensional super-Chern-Simons with
matter theories. It has been found that in all these examples the anomalous
dimensions of operators in the field theory, and the energies of the
corresponding string states are encoded in a Bethe Ansatz (BA). At small values
of the coupling constant the BA reduces to that of an integrable short-range
interaction spin-chain. At large values of the coupling, in the thermodynamic
limit, it is best described by finite-gap equations which follow from the
classical integrability of the string equations of motion.\footnote{%
  The string equations of motion in these settings turn out to be equivalent to
  a flat Lax connection.%
} %
A recent review of the developments in this field can be found
in~\cite{Beisert:2010jr}. For another recent application of the algebraic
approach to spacetimes with less supersymmetry see~\cite{Sorokin:2011rr}.

In this paper we will investigate string theories on backgrounds with an
$\AdS_3$ factor which preserve 16 supersymmetries and their $\CFT_2$ duals using
the integrability approach.\footnote{%
  The $\AdS_3/\CFT_2$ correspondence has been extensively studied since the
  early days of the gauge/string correspondence; see for
  example~\cite{Giveon:1998ns,Elitzur:1998mm,
    Seiberg:1999xz,Larsen:1999uk,Maldacena:2000hw,
    Maldacena:2000kv,Maldacena:2001km}.%
} %
This programme was initiated in~\cite{Babichenko:2009dk}.\footnote{%
  For earlier work on integrability in this context
  see~\cite{Adam:2007ws,David:2008yk} and more
  recently~\cite{David:2010yg}. Integrability has also recently been
  investigated in the context of the hybrid string formulation on $AdS_3\times
  S^3$ in~\cite{Ashok:2009xx,Benichou:2010rk,Benichou:2010ts} and for classical
  strings in the BTZ black hole background~\cite{David:2011iy}%
} %
One of the main conceptual advantages of this approach is that one can
investigate the dual pairs \emph{without} having to perform an S-duality on the
gravitational side. Most previous investigations of the $\AdS_3/\CFT_2$
correspondence used WZW models for string theory (with NS-NS flux) on $\AdS_3$
which are S-dual to string theory with R-R flux. In the context of the
$\AdS_3/\CFT_2$ correspondence, the strong coupling dual of the $\text{\itshape
  Sym}^N(T^4)$ $\CFT_2$ is IIB string theory on $AdS_3\times S^3\times T^4$ with
R-R flux. As a result, tests of this duality for unprotected quantities should
in the first instance be done for string theory in the R-R background; this is
the analogue of the planar limit in the $\AdS_5/\CFT_4$ duality.

There are in fact two types $\AdS_3$ geometries which preserve 16
supersymmetries: $AdS_3\times S^3\times T^4$ and $AdS_3\times
S^3\times S^3\times S^1$.\footnote{%
  For the purpose of the present paper the $AdS_3\times S^3\times K3$
  background can be treated as an orbifold of $AdS_3\times S^3\times
  T^4$ and shares many of it's features.%
} %
The two backgrounds preserve, respectively, \emph{small} and
\emph{large} $(4,4)$ superconformal symmetry; correspondingly, the
finite-dimensional sub-algebras of these superconformal algebras are
$\algPSU(1,1|2)^2$ and $\algD{\alpha}^2$. The superalgebras
$\algD{\alpha}$ depend on a parameter $\alpha$ which is related to the
relative size of the radii of the
geometry~\cite{Gauntlett:1998kc}. Denoting by $l$ the $\AdS_3$ radius
and by $R_\pm$ the radii of the two $S^3$'s the background solves the
supergravity equations of motion when
\begin{equation}\label{supgraradii}
 \frac{1}{R_+^2}+\frac{1}{R_-^2}=\frac{1}{l^2}\,.
\end{equation}
In terms of these geometric quantities, $\alpha$ is defined as
\begin{equation}\label{radiiphi}
 \alpha =\frac{l^2}{R_+^2}\equiv \sin^2\phi\,,\qquad
 \frac{l^2}{R_-^2}\equiv \cos^2\phi\,.
\end{equation}

A candidate $\CFT_2$ dual to string theory on $AdS_3\times S^3\times T^4$ is a
sigma model on the moduli space of $Q_1\in \Naturals$ instantons in a $U(Q_5)$
gauge theory on $T^4$; this is a natural choice given that the $AdS_3\times
S^3\times T^4$ background arises as the near-horizon limit of $Q_1$ D1-branes
coincident with $Q_5$ D5-branes. On the other hand, very little is known about
the $\CFT_2$ dual of the $AdS_3\times S^3\times S^3\times S^1$
background.\footnote{Early work on this backgorund and its dual description can
  be found in~\cite{Cowdall:1998bu,Boonstra:1998yu,Gauntlett:1998kc,
    Elitzur:1998mm,deBoer:1999rh,Papadopoulos:1999tw,Giveon:2003ku}.} Some
reasons for this ignorance are discussed in~\cite{Gukov:2004ym}; one of the main
obstacles to identifying a suitable $\CFT_2$ is that the supergravity
approximation to the full string theory is not as useful in this case as in
other examples -- for example the BPS states of the finite dimensional
sub-algebra of the full superconformal algebra need not be BPS in the full
\emph{large} super-Virasoro algebra!

In~\cite{Babichenko:2009dk}, it was found that string theory on the two $\AdS_3$
backgrounds could be treated on equal footing: equations of motion for
Green-Schwarz strings on both backgrounds could be written as flatness
conditions for a Lax connection.\footnote{GS actions for strings in curved
  spacetimes were first constructed
  in~\cite{Grisaru:1985fv,Duff:1987bx}. However, these require the knowledge of
  the complete super-geometry rather than just the bosonic spacetime solution,
  making it harder to use them. An algebraic approach to GS actions in
  flat~\cite{Henneaux:1984mh} and curved
  spacetimes~\cite{Metsaev:1998it,Stefanski:2007dp,
    Arutyunov:2008if,Stefanski:2008ik, Babichenko:2009dk,Zarembo:2010sg} leads
  to much simpler, though equivalent~\cite{Gomis:2008jt}, expressions in
  spacetimes with enough (super-)symmetry. This algebraic approach has proven to
  be particularly useful since it leads to Lax-integrable equations of
  motion~\cite{Bena:2003wd} due to the existence of a $\Integers_4$-automorphism
  on the underlying supercurrents~\cite{Berkovits:1999zq}.} From this a set of
finite-gap integral equations was formulated. In the case of the $AdS_3\times
S^3\times T^4$ background as well as the $AdS_3\times S^3\times S^3\times S^1$
background with the radii of the two $S^3$ factors equal it was possible to use
the integral equations to postulate an all-loop discrete BA much as was done for
the case of the $\AdS_4/\CFT_3$ duality in~\cite{Gromov:2008qe}. This BA may be
viewed as giving some of the first concrete prediction for what the elusive
$\CFT_2$ dual to the equal radius $AdS_3\times S^3\times S^3\times S^1$
background should be like. There were however two unresolved puzzles. Firstly,
how to generalise the BA construction when the radii of the two $S^3$ factors
were not equal. Secondly, it was found that the string worldsheet theory on
these backgrounds has a number of massless modes\footnote{The appearance of
  these massless modes is most easily seen from the plane-wave
  analysis~\cite{Berenstein:2002jq,Russo:2002rq,Lu:2002kw,Gomis:2002qi,
    Gava:2002xb,Sommovigo:2003kd,Babichenko:2009dk}.} which were not
incorporated into the finite-gap (and hence also into the BA) equations. This
second problem highlights the fact that at present we do not know how to
incorporate massless worldsheet excitations into the integrability approach to
the gauge/string correspondence.

In this paper we extend the integrability analysis of the $\AdS_3/\CFT_2$
correspondence begun in~\cite{Babichenko:2009dk}. In
section~\ref{sec:BA-d21a-all-loop} we find a discrete all-loop BA, valid for all
values of the radii of the two $S^3$'s. We show that at strong coupling in the
thermodynamic limit this BA reduces to the finite-gap integral equations that
follow from the Lax-pair formuation of the GS string on $AdS_3\times S^3\times
S^3\times S^1$. These results resolve the first of the above puzzles.

In order to gain insight into what a $\CFT_2$ dual to these backgrounds should be
like, in section~\ref{sec:BA-d21a-one-loop} we obtain the weak-coupling limit of
the all-loop BA of section~\ref{sec:BA-d21a-all-loop}. In
section~\ref{sec:alternating-spin-chain} we construct an \emph{alternating}
spin-chain, together with an integrable Hamiltonian. We show that the BA for
this Hamiltonian matches the weak coupling limit of the all-loop BA of
section~\ref{sec:BA-d21a-one-loop}. The construction of the Hamiltonian is first
done in an $\algSL(2|1)$ subsector of the theory and then lifted to the full
$\algD{\alpha}^2$ spin-chain. It relies on the universal R-matrix of
$\algSL(2|1)$ first found in~\cite{Derkachov:2000ne}. A novel feature of the
spin-chain we construct is that it has both left- and right-moving momenta, much
as one would expect for a $\CFT_2$. We expect that these results will be useful
in the eventual identification of the correct $\CFT_2$. At the equal-radius value
$\alpha=1/2$ the superalgebra is in fact $\algOSp(4|2)^2$. Restricting to just
one $\algOSp(4|2)$ factor the spin-chain we construct is closely related to the
ABJM spin-chain constructed in~\cite{Minahan:2008hf}. We discuss the relation of
our BA and spin-chain to that of the ABJM BA and spin-chain in
section~\ref{sec:alternating-spin-chain}.

In section~\ref{sec:t4-limit} we consider the $AdS_3\times S^3\times T^4$ dual
pair. In particular, in section~\ref{sec:weakcouplais1eqs}, we show how the
weakly coupled BA equations for this background~\cite{Babichenko:2009dk} can be
obtained as the $\alpha\rightarrow 1$ limit of the $AdS_3\times S^3\times
S^3\times S^1$ BA equations. In section~\ref{sec:psu112spinchain} we construct a
spin-chain whose energies are described by the weakly coupled $AdS_3\times
S^3\times T^4$ BA equations. The spin-chain in this case is \emph{not}
alternating; instead it is homogenous. It would be very interesting to see how
such a spin-chain emerges from the recent analysis of the weakly-coupled
$\CFT_2$~\cite{Pakman:2009mi}. When restricted to just the left- or right-movers
this spin-chain is closely related to the $\algPSU(1,1|2)$ spin-chain one
encounters in ${\cal N}=4$ SYM~\cite{Beisert:2007sk}.  Finally, in
section~\ref{sec:psu112spinchainasasubchain}, we make a proposal for how to
incorporate massless modes into the integrable description of the gauge/string
correspondence. We argue that the missing massless modes puzzle can be resolved
by keeping track of the integrable structure of the alternating
$\algD{\alpha}^2$ spin-chain in the $\alpha\rightarrow 1$ limit.

In section~\ref{sec:BA-solutions} we explore some of the features of the
weak-coupling BA: we find twist-one solutions in a closed $\algSL(2)$ subsector,
which are important in the identification of the spin-chain as an alternating
chain; we find a degeneracy in the spectrum of states that is similar in nature
the degeneracy of states in the ABJM model~\cite{Zwiebel:2009vb}; we investigate
the behaviour of certain simple solutions to the BA equations in the
$\alpha\rightarrow 1$ limit. We conclude in section~\ref{sec:conclusions}. Some
of the technical details are relegated to the appendices.

\section{All-loop Bethe equations for $\algD{\alpha}^2$}
\label{sec:BA-d21a-all-loop}

In this section we propose an all-loop BA for $\algD{\alpha}^2$. We
show that in the continuum limit this BA reproduces the finite-gap equations
obtained from the GS string constructed in~\cite{Babichenko:2009dk}. We also
obtain the weak-coupling limit of the all-loop BA\@. In this limit, the all-loop
BA reduces to a conventional BA for $\algD{\alpha}^2$ in a particular
representation.\footnote{%
  BA equations for any (super)-algebra in certain classes of representations have been
  proposed in~\cite{Ogievetsky:1986hu}.%
} %
The weights of the representation relevant to our case can then be easily
read-off. This will in turn be useful in the construction of a spin-chain for
the small-coupling BA which we do in the following section.

In~\cite{Babichenko:2009dk} a set of Bethe equations for the coset model on
$\grOSp(4|2)^2$ were proposed:
\begin{subequations}\label{eq:BE-all-loop}
\begin{align}
  \begin{split}
    \left(\frac{x_{1,i}^+}{x_{1,i}^-} \right)^L &= 
    \prod_{\substack{k = 1\\k \neq i}}^{K_1}
    \frac{x_{1,i}^+ - x_{1,k}^-}{x_{1,i}^- - x_{1,k}^+}
    \frac{1- \frac{1}{x_{1,i}^+ x_{1,k}^-}}{1- \frac{1}{x_{1,i}^- x_{1,k}^+}}
    \sigma^2(x_{1,i}, x_{1,k})
    \\ &\phantom{\ = \ }\times
    \prod_{k=1}^{K_2} \frac{x_{1,i}^- - x_{2,k}}{x_{1,i}^+ - x_{2,k}} 
    \prod_{k=1}^{K_{\bar{2}}} \frac{1 - \frac{1}{x_{1,i}^- x_{\bar{2},k}}}{1- \frac{1}{x_{1,i}^+ x_{\bar{2},k}}}
    \prod_{k=1}^{K_{\bar{1}}} \sigma^{-2}(x_{1,i}, x_{\bar{1},k}) \,,
  \end{split} \\
  1 &= 
  \prod_{k=1}^{K_1} \frac{x_{2,i} - x_{1,k}^+}{x_{2,i} - x_{1,k}^-}
  \prod_{k=1}^{K_3} \frac{x_{2,i} - x_{3,k}^+}{x_{2,i} - x_{3,k}^-} 
  \prod_{k=1}^{K_{\bar{1}}} \frac{1 - \frac{1}{x_{2,i} x_{\bar{1},k}^+}}{1- \frac{1}{x_{2,i} x_{\bar{1},k}^-}}
  \prod_{k=1}^{K_{\bar{3}}} \frac{1 - \frac{1}{x_{2,i} x_{\bar{3},k}^+}}{1- \frac{1}{x_{2,i} x_{\bar{3},k}^-}}
  \,, \\
  \begin{split}
    \left(\frac{x_{3,i}^+}{x_{3,i}^-} \right)^L &= 
    \prod_{\substack{k = 1\\k \neq i}}^{K_3}
    \frac{x_{3,i}^+ - x_{3,k}^-}{x_{3,i}^- - x_{3,k}^+}
    \frac{1- \frac{1}{x_{3,i}^+ x_{3,k}^-}}{1- \frac{1}{x_{3,i}^- x_{3,k}^+}}
    \sigma^2(x_{3,i}, x_{3,k})
    \\ &\phantom{\ = \ }\times
    \prod_{k=1}^{K_2} \frac{x_{3,i}^- - x_{2,k}}{x_{3,i}^+ - x_{2,k}} 
    \prod_{k=1}^{K_{\bar{2}}} \frac{1 - \frac{1}{x_{3,i}^- x_{\bar{2},k}}}{1- \frac{1}{x_{3,i}^+ x_{\bar{2},k}^-}}
    \prod_{k=1}^{K_{\bar{3}}} \sigma^{-2}(x_{3,i}, x_{\bar{3},k}) \,,
  \end{split} \\
  \begin{split}
    \left(\frac{x_{\bar{1},i}^-}{x_{\bar{1},i}^+} \right)^L &= 
    \prod_{\substack{k = 1\\k \neq i}}^{K_{\bar{1}}}
    \frac{x_{\bar{1},i}^+ - x_{\bar{1},k}^-}{x_{\bar{1},i}^- - x_{\bar{1},k}^+}
    \frac{1- \frac{1}{x_{\bar{1},i}^+ x_{\bar{1},k}^-}}{1- \frac{1}{x_{\bar{1},i}^- x_{\bar{1},k}^+}}
    \sigma^2(x_{\bar{1},i}, x_{\bar{1},k})
    \\ &\phantom{\ = \ }\times
    \prod_{k=1}^{K_{\bar{2}}} \frac{x_{\bar{1},i}^- - x_{\bar{2},k}}{x_{\bar{1},i}^+ - x_{\bar{2},k}} 
    \prod_{k=1}^{K_2} \frac{1 - \frac{1}{x_{\bar{1},i}^- x_{2,k}}}{1- \frac{1}{x_{\bar{1},i}^+ x_{2,k}^-}}
    \prod_{k=1}^{K_1} \sigma^{-2}(x_{\bar{1},i}, x_{1,k}) \,,
  \end{split} \\
  1 &= 
  \prod_{k=1}^{K_{\bar{1}}} \frac{x_{\bar{2},i} - x_{\bar{1},k}^+}{x_{\bar{2},i} - x_{\bar{1},k}^-}
  \prod_{k=1}^{K_{\bar{3}}} \frac{x_{\bar{2},i} - x_{\bar{3},k}^+}{x_{\bar{2},i} - x_{\bar{3},k}^-} 
  \prod_{k=1}^{K_1} \frac{1 - \frac{1}{x_{\bar{2},i} x_{1,k}^+}}{1- \frac{1}{x_{\bar{2},i} x_{1,k}^-}}
  \prod_{k=1}^{K_3} \frac{1 - \frac{1}{x_{\bar{2},i} x_{3,k}^+}}{1- \frac{1}{x_{\bar{2},i} x_{3,k}^-}}
  \,, \\
  \begin{split}
    \left(\frac{x_{\bar{3},i}^-}{x_{\bar{3},i}^+} \right)^L &= 
    \prod_{\substack{k = 1\\k \neq i}}^{K_3}
    \frac{x_{\bar{3},i}^+ - x_{\bar{3},k}^-}{x_{\bar{3},i}^- - x_{\bar{3},k}^+}
    \frac{1- \frac{1}{x_{\bar{3},i}^+ x_{\bar{3},k}^-}}{1- \frac{1}{x_{\bar{3},i}^- x_{\bar{3},k}^+}}
    \sigma^2(x_{\bar{3},i}, x_{\bar{3},k})
    \\ &\phantom{\ = \ }\times
    \prod_{k=1}^{K_2} \frac{x_{\bar{3},i}^- - x_{\bar{2},k}}{x_{\bar{3},i}^+ - x_{\bar{2},k}} 
    \prod_{k=1}^{K_{\bar{2}}} \frac{1 - \frac{1}{x_{\bar{3},i}^- x_{2,k}}}{1- \frac{1}{x_{\bar{3},i}^+ x_{2,k}^-}}
    \prod_{k=1}^{K_{\bar{3}}} \sigma^{-2}(x_{\bar{3},i}, x_{3,k}) \,.
  \end{split}\end{align}
\end{subequations}
Here $\sigma(x,y)$ is a dressing phase factor\footnote{%
  In general the dressing phase can take a different form than the
  BES/BHL dressing phase~\cite{Beisert:2006ez,Beisert:2006ib} that
  appears in $\superN=4$ SYM and ABJM. As we will see below, the
  classical limit of the BA requires the leading strong coupling
  behavior of $\sigma$ to be given by the AFS
  phase~\cite{Arutyunov:2004vx}.%
}, %
and the variables
$x_{l,i}^\pm$ satisfy
\begin{equation}
  \label{eq:Zhukovsky-OSp}
  x_{l,i}^\pm + \frac{1}{x_{l,i}^\pm} = x_{l,i} + \frac{1}{x_{l,i}} \pm \frac{i}{2h} \,,
\end{equation}
where $h = h(\lambda)$ is a function of the worldsheet coupling constant
$\lambda$. At large coupling it behaves as
\begin{equation}
  h(\lambda) \approx \frac{\sqrt{\lambda}}{2\pi} \,, \quad (\lambda \to \infty) \,.
\end{equation}
We will assume that $h(\lambda) \to 0$ as $\lambda \to 0$, so that there is a
weak coupling limit of the Bethe equations.

Given a solution of \eqref{eq:BE-all-loop}, the corresponding total energy $E$
and momentum $P$ is given by
\begin{gather}
  \label{eq:BE-energy}
  E = ih \sum_{l=1,3,\bar{1},\bar{3}} \sum_j^{K_l} \left( \frac{1}{x_{l,j}^+} - \frac{1}{x_{l,j}^-} \right) \,, \\
  \label{eq:BE-momentum}
  e^{i P} = \frac{\prod_{l=1,3}}{\prod_{l=\bar{1},\bar{3}}} \prod_j \frac{x_{l,j}^+}{x_{l,j}^-} \equiv 1 \,.
\end{gather}

We here propose a generalization of the above equations to the full
symmetry group $\grD{\alpha}^2$ for any
$\alpha$. The Bethe equations, energy and momentum take exactly the
same form as in \eqref{eq:BE-all-loop}, \eqref{eq:BE-energy} and
\eqref{eq:BE-momentum} respectively. However, the Zhukovsky map in
\eqref{eq:Zhukovsky-OSp} is deformed to
\begin{equation}
  \label{eq:Zhukovsky-D21a}
  x_{l,i}^\pm + \frac{1}{x_{l,i}^\pm} = x_{l,i} + \frac{1}{x_{l,i}} \pm \frac{i w_l}{2h} \,,
\end{equation}
where
\begin{equation}
  \begin{aligned}
    w_1 = w_{\bar{1}} &= 2\alpha \,, &
    w_3 = w_{\bar{3}} &= 2(1 - \alpha) \,.
  \end{aligned}
\end{equation}
Since $x_2^\pm$ and $x_{\bar{2}}^\pm$ do not appear in \eqref{eq:BE-all-loop}, we do not need to specify $w_2$ and $w_{\bar{2}}$.
The elementary magnons now have the dispersion relation
\begin{equation}
  \epsilon_l (p) = \sqrt{m_l^2 + 4h^2 \sin^2 \frac{p}{2}} \,,
\end{equation}
where the masses are
\begin{align}
  m_1^2 &= \alpha \,, &
  m_3^2 &= 1 - \alpha \,.
\end{align}

\subsection{Classical Bethe equations for $\algD{\alpha}^2$}

Effectively, the above generalization of the Bethe equations from
$\algOSp(4|2)^2$ to $\algD{\alpha}^2$ takes the form of a rescaling of
the coupling that varies between the different Dynkin nodes. The
classical limit will be almost identical in the two cases. For large
$h$, $x^\pm$ behave as
\begin{equation}
  x_{l,i}^\pm = x_{l,i} \pm \frac{i w_l}{2h} \frac{x_{l,i}^2}{x_{l,i}^2 - 1} + \order(1/h^2) \,.
\end{equation}
Expanding the Bethe equations, the only effect of the extra factor $w_l$ will be
an overall factor for each factor in the equations.

There are two important points that makes this limit work out. Firstly, there is
no direct link between the $x_{2,i}$ and $x_{3,k}$ roots in the Bethe
equations. Secondly, the dressing phase only links nodes in the Dynkin diagram
that are associated with the same value of $w_l$ (\ie, it either involves two
roots $x_{l,i}$ and $x_{l,k}$ at the same node, or one root $x_{l,i}$ from one
copy of $\algD{\alpha}$ and another root $x_{\bar{l},k}$ from the
corresponding node of the second copy of $\algD{\alpha}$.)

The only subtlety is in the coupling dependence of the dressing phase. To get a
nice form in the strong coupling limit we need to rescale the explicit coupling
dependence in the phase coupling $x_{l,i}$ with $x_{l,k}$ or $x_{\bar{l},k}$ by
a factor $1/w_l$. For large $h$, the dressing phase then reduces to the AFS
phase~\cite{Arutyunov:2004vx} 
\begin{equation}
  \sigma(x_l,y_l) \approx \frac{1 - \frac{1}{x_l^- y_l^+}}{1 - \frac{1}{x_l^+ y_l^-}}
  \left[
    \frac{
      \left(1 - \frac{1}{x_l^+ y_l^-}\right) \left(1 - \frac{1}{x_l^- y_l^+}\right)
      }{
      \left(1 - \frac{1}{x_l^+ y_l^+}\right) \left(1 - \frac{1}{x_l^- y_l^-}\right)
      }
  \right]^{\frac{ih}{w_l}\left(x_l + \frac{1}{x_l} - y_l - \frac{1}{y_l}\right)}
\end{equation}

With these modifications, the classical Bethe equations can be derived for
general $\alpha$ in the exact same way as for $\alpha=1/2$. The resulting
equations read
\begin{subequations}\label{eq:BE-d21a-integral}
\begin{align}
  \begin{split}
      \pm 4\pi \alpha \frac{\mathcal{E} x}{x^2 - 1} + 2\pi n_{1,i}^\pm
      &=
        4\alpha \pint dy \frac{\rho_1^\pm}{x-y} 
      - 2\alpha \int dy \frac{\rho_2^\pm}{x-y} \\
      &- 4\alpha \int \frac{dy}{y^2} \frac{\rho_1^\mp}{x-\frac{1}{y}} 
      + 2\alpha \int \frac{dy}{y^2} \frac{\rho_2^\mp}{x-\frac{1}{y}} \,,
  \end{split} \\
  \begin{split}
      2\pi n_{2,i}^\pm
      &=
      - 2\alpha \int dy \frac{\rho_1^\pm}{x-y}
      - 2(1-\alpha) \int dy \frac{\rho_3^\pm}{x-y} \\
      &+ 2\alpha \int \frac{dy}{y^2} \frac{\rho_1^\mp}{x-\frac{1}{y}} 
      + 2(1-\alpha) \int \frac{dy}{y^2} \frac{\rho_3^\mp}{x-\frac{1}{y}} \,,
  \end{split} \\
  \begin{split}
      \pm 4\pi (1-\alpha) \frac{\mathcal{E}  x}{x^2 - 1} + 2\pi n_{3,i}^\pm
      &=
        4(1-\alpha) \pint dy \frac{\rho_3^\pm}{x-y} 
      - 2(1-\alpha) \int dy \frac{\rho_1^\pm}{x-y} \\
      &- 4(1-\alpha) \int \frac{dy}{y^2} \frac{\rho_3^\mp}{x-\frac{1}{y}} 
      + 4(1-\alpha) \int \frac{dy}{y^2} \frac{\rho_1^\mp}{x-\frac{1}{y}} \,.
  \end{split}
\end{align}
\end{subequations}
Setting $\alpha = \sin^2\phi$, these equations exactly agree with the classical
Bethe equations of~\cite{Babichenko:2009dk}.

\subsection{Weak-coupling limit}
\label{sec:BA-d21a-one-loop}

Let us now turn to the weakly-coupled limit of the all-loop BA proposed
above. In this limit, $h \to 0$, the variables $x_i$ will behave as
\begin{equation}
  x_i \approx \frac{u_i}{h} \,,
\end{equation}
where $u_i \sim \order(1)$ as $h \to 0$. Hence
\begin{equation}
  x_i^\pm \approx \frac{u_i \pm iw_i/2}{h}.
\end{equation}
Moreover, we will assume that $\sigma(x, y) \to 1$ as $h \to 0$. This
means that in the weak coupling limit, the Bethe equations in
\eqref{eq:BE-all-loop} decouple into two sets of equations involving
only variables with either un-bared or bared indices. Inserting the
above weak coupling expressions for $x_i$ and $x_i^\pm$ into the first
three equations in \eqref{eq:BE-all-loop} we get
\begin{subequations}
\label{eq:BE-d21a-one-loop}
\begin{align}
  \left( \frac{u_{1,i} + i\alpha}{u_{1,i} - i\alpha} \right)^L &= 
  \prod_{\substack{k = 1\\k \neq i}}^{K_1} \frac{u_{1,i} - u_{1,k} + 2i\alpha}{u_{1,i} - u_{1,k} - 2i\alpha}
  \prod_{k=1}^{K_2} \frac{u_{1,i} - u_{2,k} - i\alpha}{u_{1,i} - u_{2,k} + i\alpha} \,, \\
  1 &= 
  \prod_{k=1}^{K_1} \frac{u_{2,i} - u_{1,k} - i\alpha}{u_{2,i} - u_{1,k} + i\alpha}
  \prod_{k=1}^{K_3} \frac{u_{2,i} - u_{3,k} - i(1-\alpha)}{u_{2,i} - u_{3,k} + i(1-\alpha)} \,, \\
  \left( \frac{u_{3,i} + i(1-\alpha)}{u_{3,i} - i(1-\alpha)} \right)^L &= 
  \prod_{\substack{k = 1\\k \neq i}}^{K_3} \frac{u_{3,i} - u_{3,k} + 2i(1-\alpha)}{u_{3,i} - u_{3,k} - 2i(1-\alpha)}
  \prod_{k=1}^{K_2} \frac{u_{3,i} - u_{2,k} - i(1-\alpha)}{u_{3,i} - u_{2,k} + i(1-\alpha)} \,.
\end{align}
\end{subequations}
We now extract the weights of the $\algD{\alpha}$ representation to which the above BA applies.
Recall, that for a general (super)-Lie algebra with simple
roots $\vec{\alpha}_q$ we can write a set of Bethe equations in the
representation give by weights $\vec{w}$ as~\cite{Minahan:2002ve,Ogievetsky:1986hu}
\begin{equation}
  \label{eq:BE-any-group}
  \left( \frac{u_{l,i} + \frac{i}{2} \vec{\alpha}_l \cdot \vec{w}}{u_{l,i} - \frac{i}{2} \vec{\alpha}_l \cdot \vec{w}} \right)^L
  = \prod_{\substack{k = 1\\k \neq i}}^{K_l} \frac{u_{l,i} - u_{l,k} + \frac{i}{2} \vec{\alpha}_l \cdot \vec{\alpha}_l}{u_{l,i} - u_{l,k} - \frac{i}{2} \vec{\alpha}_l \cdot \vec{\alpha}_l}
  \prod_{l' \neq l} \prod_{k=1}^{K_{l'}} 
  \frac{u_{l,i} - u_{l',k} + \frac{i}{2} \vec{\alpha}_l \cdot \vec{\alpha}_{l'}}{u_{l,i} - u_{l,k} - \frac{i}{2} \vec{\alpha}_l \cdot \vec{\alpha}_{l'}}
\end{equation}
As we will argue shortly the spin-chain we are interested in will be
alternating. In the next section we will compare
equations~\eqref{eq:BE-d21a-one-loop} and~\eqref{eq:BE-any-group} to find that
the spin-chain should have sites which alternate between the representations
$\left(-\tfrac{\alpha}{2};\tfrac{1}{2};0\right)$ and
$\left(-\tfrac{1-\alpha}{2};0;\tfrac{1}{2}\right)$.

\section{An integrable $\algD{\alpha}^2$ spin-chain}
\label{sec:alternating-spin-chain}

In the previous section we have made a proposal for an all-loop Bethe Ansatz for
the energies of massive string states in the $AdS_3\times S^3\times S^3\times
S^1$ background. In equation~\eqref{eq:BE-d21a-one-loop} above, we extracted the
weak-coupling limit of this BA\@. In this section, we will construct an
alternating spin-chain with integrable Hamiltonian whose energies are described
precisely by~\eqref{eq:BE-d21a-one-loop}. This spin-chain Hamiltonian should
provide vital clues in identifying the elusive $\CFT_2$ dual of string theory on
this background.

To see what kind of spin-chain we should consider note the following
observations. In section~\ref{sec:BA-solutions} below, we show that the
weak-coupling BA equations~\ref{eq:BE-d21a-one-loop} have non-trivial solutions
for twist-one operators. More specifically, we identify an $\algSL(2)$ subsector
of the BA~\eqref{eq:BE-d21a-one-loop} and show it has non-trivial solutions of
length $L=1$. This indicates that the spin-chain is not of the conventional
homogenous type, where, by definition, no such solutions exist.\footnote{%
  The fact that the $\algD{\alpha}$ spin-chain is alternating can also be
  anticipated from the fact that it contains two momentum-carrying roots. We
  would like to thanks Kostya Zarembo for a discussion of this point.%
} %
Further, in section~\ref{sec:sec41} we show that at $\alpha=1/2$ the
weak-coupling BA equations match precisely the $\algOSp(4|2)$ subsector of the
weak-coupling spin-chain~\cite{Minahan:2008hf} of the ABJM theory. Both these
observations lead us to conclude that the spin-chain related to the
BA~\eqref{eq:BE-d21a-one-loop} should be alternating.

\subsection{The alternating $\algD{\alpha}$ spin-chain}
\label{sec:alternatingspinchain}

Before constructing the Hamiltonian, we will collect here some facts about
$\algD{\alpha}$ representations. In this discussion we follow
closely~\cite{VanDerJeugt:1985hq}; a very nice review of Lie superalgebras can
be found in~\cite{Frappat:1996pb}. If we pick a suitable real form, the algebra
$\alg{d}(2,1;\alpha)$ has a bosonic subalgebra $\alg{sl}(2,\Reals) \oplus
\alg{su}(2) \oplus \alg{su}(2)$. The corresponding generators are denoted as
$\gen{S}_\mu$ ($\mu = 0,1,2$), $\gen{L}_m$ ($n = 3,4,5$) and $\gen{R}_{\dot{m}}$
($\dot{m} = 6,7,8$). The vector indices of the bosonic generators are raised and
lowered using $\eta_{\mu\nu} = \mathop{\mathrm{diag}}(- + + )$, $\delta_{mn}$
and $\delta_{\dot{m}\dot{n}}$, respectively. There are eight fermionic
generators transforming as a tri-spinor under the bosonic subgroup and denoted
by $\gen{Q}_{a\alpha\dot{\alpha}}$, where each index takes values $+$ or
$-$. The (anti)-commutation relations of $\algD{\alpha}$ are\footnote{%
  The anti-symmetric symbols $\epsilon_{\mu\nu\rho}$, $\epsilon_{mnp}$ and
  $\epsilon_{\dot{m}\dot{n}\dot{p}}$ are normalized so that $\epsilon_{012} =
  \epsilon_{345} = \epsilon_{678} = 1$. The gamma-matrices are given by
  \begin{equation*}
    \left(\gamma^\mu\right)^a{}_b = (-\sigma_3, i\sigma_2, -i\sigma_1) \,, \quad
    \left(\gamma^m\right)^\alpha{}_\beta = (\sigma_1, \sigma_2, \sigma_3) \,, \quad
    \left(\gamma^{\dot{m}}\right)^{\dot{\alpha}}{}_{\dot{\beta}} = (\sigma_1, \sigma_2, \sigma_3) \,.
  \end{equation*}
}%
\begin{equation}
  \label{eq:d21a-comm-rel}
\begin{gathered}
  \begin{aligned}
    \comm{\gen{S}_\mu}{\gen{S}_\nu} &= i \, \epsilon_{\mu\nu\rho} \, \gen{S}^\rho \,, & \qquad
    \comm{\gen{S}_\mu}{\gen{Q}_{a\alpha\dot{\alpha}}} &= \textstyle \frac{1}{2} \gen{Q}_{b\alpha\dot{\alpha}} \gamma^b_{\mu\,a} \,, \\
    \comm{\gen{L}_m}{\gen{L}_n} &= i \, \epsilon_{mnp} \, \gen{L}^p \,, & \qquad
    \comm{\gen{L}_m}{\gen{Q}_{a\alpha\dot{\alpha}}} &= \textstyle \frac{1}{2} \gen{Q}_{a\beta\dot{\alpha}} \gamma^\beta_{m\,\alpha} \,, \\
    \comm{\gen{R}_{\dot{m}}}{\gen{R}_{\dot{n}}} &= i \, \epsilon_{\dot{m}\dot{n}\dot{p}} \, \gen{R}^{\dot{p}} \,, & \qquad
    \comm{\gen{R}_{\dot{m}}}{\gen{Q}_{a\alpha\dot{\alpha}}} &= \textstyle \frac{1}{2} \gen{Q}_{a\alpha\dot{\beta}} \gamma^{\dot{\beta}}_{{\dot{m}}\,\dot{\alpha}} \,,
  \end{aligned} \\
  \begin{aligned}
    \acomm{\gen{Q}_{a\alpha\dot{\alpha}}}{\gen{Q}_{b\beta\dot{\beta}}} &= 
    -
    \Big(
       \gen{S}_\mu \, \gamma^\mu_{ab} \, \epsilon_{\alpha\beta} \, \epsilon_{\dot{\alpha}\dot{\beta}}
       + \alpha \, \gen{L}_m \, \epsilon_{ab} \, \gamma^m_{\alpha\beta} \, \epsilon_{\dot{\alpha}\dot{\beta}}
       + \, (1-\alpha) \, \gen{R}_{\dot{m}} \, \epsilon_{ab} \, \epsilon_{\alpha\beta} \, \gamma^{\dot{m}}_{\dot{\alpha}\dot{\beta}}
     \Big)  \,.
   \end{aligned}
\end{gathered}
\end{equation}
The simple roots for $\algD{\alpha}$ are given by
\begin{equation}
\begin{aligned}
  \vec{\alpha}_1 &= \left( \sqrt{2\alpha} ; 0 ; -\sqrt{2\alpha} \right) \,, \\
  \vec{\alpha}_2 &= \left( -\frac{\sqrt{\alpha} + \sqrt{1 - \alpha}}{\sqrt{2}} ; 1 ; \frac{\sqrt{\alpha} - \sqrt{1 - \alpha}}{\sqrt{2}} \right) \,, \\
  \vec{\alpha}_3 &= \left( \sqrt{2(1 - \alpha)} ; 0 ; \sqrt{2(1 - \alpha)} \right) \,,
\end{aligned}
\end{equation}
where the signature is $(+ \; - \; +)$. The Cartan matrix $A_{ij} =
\vec{\alpha}_i \cdot \vec{\alpha}_j$ is
\begin{equation}
  A =
  \begin{pmatrix}
    \phantom{-}4\alpha & -2\alpha       & \phantom{-}0   \\
    -2\alpha           & \phantom{-}0   & -2(1 - \alpha) \\
    \phantom{-}0       & -2(1 - \alpha) & \phantom{-}4(1 - \alpha)
  \end{pmatrix} \,.
\end{equation}
There are three special values for $\alpha$ where the algebra
simplifies. For $\alpha = 1/2$ it is equivalen to $\alg{d}(2,1) =
\algOSp(4|2)$. When $\alpha = 0$ or $\alpha = 1$ it reduces to
$\algPSU(1,1|2)$.

A highest-weight module of $\algD{\alpha}$ consists of a number of
heighest-weight modules of the bosonic sub-algebra $\algSL(2)\times
\algSU(2)\times \algSU(2)$. The fermionic generators of the algebra map these
into each other while preserving algebra relations; this is the conventional
induced representation way of constructing a superalgebra
module~\cite{Kac:1977em,Kac:1977qb}. Heighest weight representations of
$\algD{\alpha}$ are specified by highest weights of the bosonic
subalgebra~\cite{Kac:1977em,VanDerJeugt:1985hq}. To denote such a representation
we will use the notation
\begin{equation*}
  (p\,;\,q\,;\,r)
\end{equation*}
where $p\in\Complex$ is the spin of the non-compact $\algSL(2)$ and
$q\,,\,r\in\tfrac{1}{2}\Integers$ are the compact spins. The corresponding
Dynkin labels are
\begin{equation*}
  \left[ 2q ; - 2 p - 2\alpha q - 2(1-\alpha) r ; 2r \right] \,.
\end{equation*}
A general representation $(p;q;r)$ decomposes under the bosonic
subalgebra as
\begin{equation}\label{eq:gend21adecomp}
  \begin{split}
    (p ; q ; r) &\to \Big\{ 
      (p,q,r) , 
      (p - \tfrac{1}{2}, q \pm \tfrac{1}{2}, r \pm \tfrac{1}{2}) ,
      (p - \tfrac{1}{2}, q \pm \tfrac{1}{2}, r \mp \tfrac{1}{2}) ,
      \\ &\phantom{\to\ \Big\{}
      (p - 1, q \pm 1 , r) ,
      (p - 1, q, r \pm 1) ,
      (p - 1, q, r)^{\oplus 2} ,
      \\ &\phantom{\to\ \Big\{}
      (p - \tfrac{3}{2}, q \pm \tfrac{1}{2}, r \pm \tfrac{1}{2}) ,
      (p - \tfrac{3}{2}, q \pm \tfrac{1}{2}, r \mp \tfrac{1}{2}) ,
      (p - 2, q, r)
    \Big\} \,.
  \end{split}
\end{equation}
This decomposition is only valid for a generic module and sometimes has to be
modified in two important ways~\cite{VanDerJeugt:1985hq}. Firstly, for a
heighest-weight module $(p ; q ; r)$, if $q<1$ or $r<1$ the right-hand-side
of equation~\eqref{eq:gend21adecomp} will contain at most terms $(p',q',r')$
for which $q'\,,\,r'\ge 0$.\footnote{%
  For special values of $q$ and $r$ it may also not contain some $(p',q',r')$
  for which $q'\,,\,r'\ge 0$. This may happen if a particular $(p',q',r')$ for
  which $q'\,,\,r'\ge 0$ is a descendant of a sub-module $(p'',q'',r'')$ for
  which $q''\,,\,r''< 0$.%
} %
Secondly, as in all superalgebras, there are heighest weight $\algD{\alpha}$
modules for which the heighest weight state is anihilated not only by the
raising operators, but also by a subset of the fermionic lowering operators;
such modules are called atypical, short or BPS. Atypical modules occur only when
$p,q$ and $r$ satisfy so-called shortening conditions. In the present case these
are~\cite{Kac:1977em,VanDerJeugt:1985hq}
\begin{align}
\label{eq:d21ashort}
  0 &= p + \alpha q + (1 - \alpha) r \,, \\
  0 &= p - \alpha (q + 1) + (1 - \alpha) r \,, \\
  0 &= p + \alpha q - (1 - \alpha) (r + 1) \,, \\
  0 &= p - \alpha (q + 1) - (1 - \alpha) (r + 1) \,.
\end{align}

Let us now consider the $\algD{\alpha}$ modules that will feature in our
spin-chain. The discussion at the start of the present section strongly suggests
that the spin-chain we are after is alternating. Comparing the weak-coupling
BA~\eqref{eq:BE-d21a-one-loop} with the general BA\@ for any weights given
in~\eqref{eq:BE-any-group}, we may then read-off the weights of the
$\algD{\alpha}$ representations at the even and odd sites to be
\begin{equation}
  \left(-\tfrac{\alpha}{2} ; \tfrac{1}{2} ; 0\right)
  \qquad\qquad\mbox{and}\qquad\qquad
  \left(-\tfrac{1-\alpha}{2} ; 0 ; \tfrac{1}{2}\right)\,.
\end{equation}
These representations satisfy shortening conditions in
equation~\ref{eq:d21ashort} and are particularly simple since $q$ and $r$ are
also very small for them. In terms of representations of the bosonic part of the
algebra they decompose as
\begin{align}
  \left(-\tfrac{\alpha}{2} ; \tfrac{1}{2} ; 0\right) &= \left\{(-\tfrac{\alpha}{2},\tfrac{1}{2},0),(-\tfrac{\alpha+1}{2},0,\tfrac{1}{2})\right\}\,,\\
  \left(-\tfrac{1-\alpha}{2} ; 0 ; \tfrac{1}{2}\right) &= \left\{(-\tfrac{1-\alpha}{2} , 0 , \tfrac{1}{2}) \,,\, (-\tfrac{2-\alpha}{2} , 0 , \tfrac{1}{2})\right\}\,.
\end{align}
A representation for these representations is given in
appendix~\ref{sec:d21a-algebra}. The heighest-weight state of the
modules $(-\tfrac{\alpha}{2},\tfrac{1}{2},0)$ and
$(-\tfrac{\alpha+1}{2},0,\tfrac{1}{2})$ of the bosonic sub-algebra
will be denoted by $\ket{\phi_{\alpha=+}^{(0)}}$ and
$\ket{\psi_{\dot\alpha=+}^{(0)}}$, respectively. The subscripts
${}_{\alpha}$ and ${}_{\dot\alpha}$ indicate that these states are
part of a doublet representation under $\gen{L}_m$, and
$\gen{R}_{{\dot m}}$, respectively. When acted on by $S_-$ both these
states generate a discrete infinite dimensional representation of
$\algSL(2)$, with descendants denoted by the super-script ${}^{(n)}$
for $n\in\Naturals$. In figure~\ref{fig:short-d21a} we denote
pictorially this representation.
\begin{figure}
  \centering
  \begin{tikzpicture}
    \node (phi0) at (-\hordist,0) {$\left< \ket{\phi_\alpha^{(0)}} \right>$};
    \node (phi1) at (-\hordist,-2\vertdist) {$\left< \ket{\phi_\alpha^{(1)}} \right>$};
    \node (phi2) at (-\hordist,-4\vertdist) {$\left< \ket{\phi_\alpha^{(2)}} \right>$};

    \node (psi0) at (\hordist,-\vertdist) {$\left< \ket{\psi_{\dot{\alpha}}^{(0)}} \right>$};
    \node (psi1) at (\hordist,-3\vertdist) {$\left< \ket{\psi_{\dot{\alpha}}^{(1)}} \right>$};

    \coordinate (q1) at (0,-0.5\vertdist);
    \coordinate (q2) at (0,-1.5\vertdist);
    \coordinate (q3) at (0,-2.5\vertdist);
    \coordinate (q4) at (0,-3.5\vertdist);
    \coordinate (q5) at ($(0,-4.25\vertdist)+(0,-0.4ex)$);

    \coordinate (s1) at ($(-\hordist,-\vertdist)+(0,-0.4ex)$);
    \coordinate (s2) at ($( \hordist,-2\vertdist)+(0,-0.4ex)$);
    \coordinate (s3) at ($(-\hordist,-3\vertdist)+(0,-0.4ex)$);
    \coordinate (s4) at ($( \hordist,-3.75\vertdist)+(0,-0.4ex)$);
    \coordinate (s5) at ($(-\hordist,-4.75\vertdist)+(0,-0.4ex)$);

    \coordinate (L1) at (-1.6\hordist,0);
    \coordinate (L2) at (-1.6\hordist,-2\vertdist);
    \coordinate (L3) at (-1.6\hordist,-4\vertdist);

    \coordinate (R1) at (1.6\hordist,-\vertdist);
    \coordinate (R2) at (1.6\hordist,-3\vertdist);

    \draw [->] ($(phi0.east)-(0,0.30cm)$) -- ($(psi0.west)+(0,0.15cm)$);
    \node at ($(q1)+(-0.4cm,0.1cm)$) [anchor=south west] {$\scriptstyle \epsilon_{\dot{\alpha}\dot{\beta}}Q_{+\alpha\dot{\beta}}$};
    
    \draw [<-] ($(phi0.east)-(0,0.15cm)$) -- ($(psi0.west)+(0,0.30cm)$);
    \node at ($(q1)+(0.2cm,-0cm)$) [anchor=north east] {$\scriptstyle \epsilon_{\alpha\beta}Q_{-\beta\dot{\alpha}}$};

    \draw [->] ($(phi1.east)+(0,0.30cm)$) -- ($(psi0.west)-(0,0.15cm)$);
    \node at ($(q2)+(0.2cm,0.1cm)$) [anchor=south east] {$\scriptstyle \epsilon_{\beta\alpha}Q_{+\beta\dot{\alpha}}$};

    \draw [<-] ($(phi1.east)+(0,0.15cm)$) -- ($(psi0.west)-(0,0.30cm)$);
    \node at ($(q2)+(-0.4cm,-0.2cm)$) [anchor=north west] {$\scriptstyle \epsilon_{\dot{\alpha}\dot{\beta}}Q_{-\alpha\dot{\beta}}$};

    \draw [->] ($(phi1.east)-(0,0.30cm)$) -- ($(psi1.west)+(0,0.15cm)$);
    \node at ($(q3)+(-0.4cm,0.1cm)$) [anchor=south west] {$\scriptstyle \epsilon_{\dot{\alpha}\dot{\beta}}Q_{+\alpha\dot{\beta}}$};

    \draw [<-] ($(phi1.east)-(0,0.15cm)$) -- ($(psi1.west)+(0,0.30cm)$);
    \node at ($(q3)+(0.2cm,-0cm)$) [anchor=north east] {$\scriptstyle \epsilon_{\alpha\beta}Q_{-\beta\dot{\alpha}}$};

    \draw [->] ($(phi2.east)+(0,0.30cm)$) -- ($(psi1.west)-(0,0.15cm)$);
    \node at ($(q4)+(0.2cm,0.1cm)$) [anchor=south east] {$\scriptstyle \epsilon_{\beta\alpha}Q_{+\beta\dot{\alpha}}$};

    \draw [<-] ($(phi2.east)+(0,0.15cm)$) -- ($(psi1.west)-(0,0.30cm)$);
    \node at ($(q4)+(-0.4cm,-0.2cm)$) [anchor=north west] {$\scriptstyle \epsilon_{\dot{\alpha}\dot{\beta}}Q_{-\alpha\dot{\beta}}$};

    \draw [->] ($(phi0.south)+(-0.08cm,0)$) -- ($(phi1.north)+(-0.08cm,0)$);
    \draw [<-] ($(phi0.south)+(+0.08cm,0)$) -- ($(phi1.north)+(+0.08cm,0)$);
    \node at (s1) [anchor=west] {$\scriptstyle S_+$};
    \node at (s1) [anchor=east] {$\scriptstyle S_-$};

    \draw [->] ($(psi0.south)+(-0.08cm,0)$) -- ($(psi1.north)+(-0.08cm,0)$);
    \draw [<-] ($(psi0.south)+(+0.08cm,0)$) -- ($(psi1.north)+(+0.08cm,0)$);
    \node at (s2) [anchor=west] {$\scriptstyle S_+$};
    \node at (s2) [anchor=east] {$\scriptstyle S_-$};

    \draw [->] ($(phi1.south)+(-0.08cm,0)$) -- ($(phi2.north)+(-0.08cm,0)$);
    \draw [<-] ($(phi1.south)+(+0.08cm,0)$) -- ($(phi2.north)+(+0.08cm,0)$);
    \node at (s3) [anchor=west] {$\scriptstyle S_+$};
    \node at (s3) [anchor=east] {$\scriptstyle S_-$};



    \draw [->,out=170,in=90] ($(phi0.west)+(0,0.2cm)$) to (L1) to [out=270,in=190] ($(phi0.west)-(0,0.2cm)$);
    \node at (L1) [anchor=east] {$\scriptstyle L_m$};

    \draw [->,out=170,in=90] ($(phi1.west)+(0,0.2cm)$) to (L2) to [out=270,in=190] ($(phi1.west)-(0,0.2cm)$);
    \node at (L2) [anchor=east] {$\scriptstyle L_m$};

    \draw [->,out=170,in=90] ($(phi2.west)+(0,0.2cm)$) to (L3) to [out=270,in=190] ($(phi2.west)-(0,0.2cm)$);
    \node at (L3) [anchor=east] {$\scriptstyle L_m$};

    \draw [->,out=10,in=90] ($(psi0.east)+(0,0.2cm)$) to (R1) to [out=270,in=350] ($(psi0.east)-(0,0.2cm)$);
    \node at (R1) [anchor=west] {$\scriptstyle R_{\dot{m}}$};

    \draw [->,out=10,in=90] ($(psi1.east)+(0,0.2cm)$) to (R2) to [out=270,in=350] ($(psi1.east)-(0,0.2cm)$);
    \node at (R2) [anchor=west] {$\scriptstyle R_{\dot{m}}$};

    \node at (s4) {$\vdots$};
    \node at (s5) {$\vdots$};
    \node at ($(q5)+(-0.1cm,0)$) {$\vdots$};

  \end{tikzpicture}

  \caption{An illustration of the short
    $\left(-\tfrac{\alpha}{2};\tfrac{1}{2},0\right)$ $\algD{\alpha}$
    module. For clarity, in the figure we have neglected to include
    the exact coefficients of the linear maps for the
    representation. These can be read-off from
    equations~\eqref{eq:d21a-chiral-rep-LR}-\eqref{eq:d21a-chiral-rep-Q}
    in Appendix~\ref{sec:d21a-algebra}. To emphasise this we use the
    notation $\Bigl<\dots\Bigr>$ to denote the span of $\dots$.  The
    states in the left-hand side column have
    $(L_5,R_8)=(\pm\tfrac{1}{2},0)$ and those in the right-hand side
    column have $(L_5,R_8)=(0,\pm\tfrac{1}{2})$. Starting at the top,
    the rows in the diagram have $S_0$ eigenvalues equal to
    $-\tfrac{\alpha}{2}\,,\,-\tfrac{\alpha+1}{2} ; -\tfrac{\alpha+2}{2}\,,\dots$}
  \label{fig:short-d21a}
\end{figure}

The $\left(-\tfrac{1-\alpha}{2};0;\tfrac{1}{2}\right)$ module can be
easily obtained from the
$\left(-\tfrac{\alpha}{2};\tfrac{1}{2};0\right)$ module by exchanging
the two sets of $\algSU(2)$ generators and replacing $\alpha$ with
$1-\alpha$.

\subsection{The integrable $\algD{\alpha}$ hamiltonian}
\label{sec:integhamiltonian}

In the previous sub-section we presented the free alternating
spin-chain relevant to our problem. In this section we will construct
an integrable Hamiltonian for this spin-chain. We do this using the
R-matrix approach. In section~\ref{sec:sl21subsector} we first
construct the Hamiltonian in an $\algSL(2|1)$ subsector of
$\algD{\alpha}$ using the universal R-matrix found
in~\cite{Derkachov:2000ne}. In section~\ref{sec:lift-to-d21a}, we show
that there is a unique lift of this R-matrix to the full
$\algD{\alpha}$; this lift is similar in spirit to those
in~\cite{Beisert:2003jj} and~\cite{Zwiebel:2009vb}. As an example, in
section~\ref{sec:su2-su2-sector} we write down explicitly the
Hamiltonian for the $\algSU(2)^2$ subsector of the alternating
$\algD{\alpha}$ spin-chain. Finally, in
section~\ref{sec:full-spin-chain} we construct the spin-chain and
Hamiltonian of the full $\algD{\alpha}^2$ symmetry relevant to the
$\AdS/\CFT$ duality.

\subsubsection{The $\algSL(2|1)$ subsector}
\label{sec:sl21subsector}

We begin this sub-section by reviewing some facts about the $\algSL(2|1)$ algebra, its embedding into $\algD{\alpha}$ and its representations. Recall that $\algSL(2|1)$ is a maximal regular sub-algebra of $\algD{\alpha}$~\cite{Frappat:1996pb}; an explicit embedding is given by
\begin{equation}
\label{eq:sl21embedding}
  \begin{gathered}
    \begin{aligned}
      \gen{J}_0 &= -\gen{S}_0 \,, &
      \gen{J}_\pm &= \mp\gen{S}_1 + i \, \gen{S}_2 \,, &
      \gen{B} &= -\alpha \, \gen{L}_5 + (1 - \alpha) \, \gen{R}_8 \,,
    \end{aligned} \\
    \begin{aligned}
      \gen{Q}^+ &= \gen{Q}_{--+} \,, &
      \gen{Q}^- &= \gen{Q}_{-+-} \,, &
      \gen{S}^+ &= \gen{Q}_{+-+} \,, &
      \gen{S}^- &= \gen{Q}_{++-} \,.
    \end{aligned}
  \end{gathered}
\end{equation}
It is easy to check that these generators satisfy the $\algSL(2|1)$ algebra
\begin{equation}
  \begin{gathered}
    \begin{aligned}
      \comm{\gen{J}_0}{\gen{J}_{\pm}} &= \pm \gen{J}_{\pm} \,, &
      \comm{\gen{J}_+}{\gen{J}_-} &= 2\gen{J}_0 \,,
    \end{aligned} \\
    \begin{aligned}
      \comm{\gen{B}}{\gen{Q}^{\pm}} &= \pm \tfrac{1}{2} \gen{Q}^{\pm} \,, &
      \comm{\gen{J}_0}{\gen{Q}^{\pm}} &= +\tfrac{1}{2} \gen{Q}^{\pm} \,, &
      \comm{\gen{J}_-}{\gen{Q}^{\pm}} &= \gen{S}^{\pm} \,, \\
      \comm{\gen{B}}{\gen{S}^{\pm}} &= \pm \tfrac{1}{2} \gen{S}^{\pm} \,, &
      \comm{\gen{J}_0}{\gen{S}^{\pm}} &= -\tfrac{1}{2} \gen{S}^{\pm} \,, &
      \comm{\gen{J}_+}{\gen{S}^{\pm}} &= \gen{Q}^{\pm} \,,
    \end{aligned} \\
    \begin{aligned}
      \acomm{\gen{Q}^+}{\gen{Q}^-} &= +\gen{J}_+ \,, &
      \acomm{\gen{Q}^+}{\gen{S}^-} &= -\gen{J}_0 + \gen{B} \,, \\
      \acomm{\gen{S}^+}{\gen{S}^-} &= -\gen{J}_- \,, &
      \acomm{\gen{S}^+}{\gen{Q}^-} &= -\gen{J}_0 - \gen{B} \,.
    \end{aligned}
  \end{gathered}
\end{equation}
The irreducible representations of $\algSL(2|1)$ are labeled by the highest
weights $(j,b)$ under $\gen{J}_0$ and $\gen{B}$. Atypical representations have
$b = \mp j$, and are called chiral and anti-chiral. The chiral representation
$(\frac{\alpha}{2},-\frac{\alpha}{2})$ and the anti-chiral representation
$(\frac{1-\alpha}{2},\frac{1-\alpha}{2})$ will be useful in the construction of
the integrable Hamiltonian below. The chiral representation
$(\frac{\alpha}{2},-\frac{\alpha}{2})$ can be obtained from the $\algD{\alpha}$
representation $\left(-\tfrac{\alpha}{2} ; \tfrac{1}{2} ; 0\right)$
constructed in Appendix~\ref{sec:d21a-algebra}. In particular, the
highest-weight state, $\ket{\phi_{\alpha=+}^{(0)}}$, of the
$\left(-\tfrac{\alpha}{2} ; \tfrac{1}{2} ; 0\right)$ module is taken as the
highest-weight state of the $(\frac{\alpha}{2},-\frac{\alpha}{2})$ module. The
only fermionic generator that does not annihilate this state is $\gen{Q}^-$
which acts as
\begin{equation}
\gen{Q}^-\ket{\phi_{\alpha=+}^{(0)}}=-\sqrt{\alpha}\ket{\psi_{\dot\alpha=+}^{(0)}}\,.
\end{equation}
Acting with the bosonic lowering operator $\gen{J}_+$ we obtain the
(countably infinite) descendents $\ket{\phi_{\alpha=+}^{(n)}}$ and
$\ket{\psi_{\dot\alpha=+}^{(n)}}$, where $n\ge 0$, which form the
basis for the $(\frac{\alpha}{2},-\frac{\alpha}{2})$ module. In
summary then, starting from the $\algD{\alpha}$ module
$\left(-\tfrac{\alpha}{2} ; \tfrac{1}{2} ; 0\right)$ whose basis are
the states $\ket{\phi_{\alpha}^{(n)}}$ and
$\ket{\psi_{\dot\alpha}^{(n)}}$ we can obtain the $\algSL(2|1)$ module
$(\frac{\alpha}{2},-\frac{\alpha}{2})$ by restricting to
${}_{\alpha=+}$ and ${}_{\dot\alpha=+}$ states. In
figure~\ref{fig:short-sl21} we provide a pictorial representation of
this module. Anti-chiral modules
$(\frac{1-\alpha}{2},\frac{1-\alpha}{2})$ can be obtained in an
analogous fashion from the $\algD{\alpha}$ modules
$\left(-\tfrac{1-\alpha}{2} ; 0 ; \tfrac{1}{2}\right)$.
\begin{figure}
  \centering
  \begin{tikzpicture}
    \node (phi0) at (-\hordist,0) {$\left< \ket{\phi_+^{(0)}} \right>$};
    \node (phi1) at (-\hordist,-2\vertdist) {$\left< \ket{\phi_+^{(1)}} \right>$};
    \node (phi2) at (-\hordist,-4\vertdist) {$\left< \ket{\phi_+^{(2)}} \right>$};

    \node (psi0) at (\hordist,-\vertdist) {$\left< \ket{\psi_+^{(0)}} \right>$};
    \node (psi1) at (\hordist,-3\vertdist) {$\left< \ket{\psi_+^{(1)}} \right>$};

    \coordinate (q1) at (0,-0.5\vertdist);
    \coordinate (q2) at (0,-1.5\vertdist);
    \coordinate (q3) at (0,-2.5\vertdist);
    \coordinate (q4) at (0,-3.5\vertdist);
    \coordinate (q5) at ($(0,-4.25\vertdist)+(0,-0.4ex)$);

    \coordinate (s1) at ($(-\hordist,-\vertdist)+(0,-0.4ex)$);
    \coordinate (s2) at ($( \hordist,-2\vertdist)+(0,-0.4ex)$);
    \coordinate (s3) at ($(-\hordist,-3\vertdist)+(0,-0.4ex)$);
    \coordinate (s4) at ($( \hordist,-3.75\vertdist)+(0,-0.4ex)$);
    \coordinate (s5) at ($(-\hordist,-4.75\vertdist)+(0,-0.4ex)$);

    \draw [->] ($(phi0.east)-(0,0.30cm)$) -- ($(psi0.west)+(0,0.15cm)$);
    \node at (q1) [anchor=south,outer sep=0.15cm] {$\scriptstyle S^-$};
    
    \draw [<-] ($(phi0.east)-(0,0.15cm)$) -- ($(psi0.west)+(0,0.30cm)$);
    \node at (q1) [anchor=north,outer sep=0.1cm] {$\scriptstyle Q^+$};

    \draw [->] ($(phi1.east)+(0,0.30cm)$) -- ($(psi0.west)-(0,0.15cm)$);
    \node at (q2) [anchor=south,outer sep=0.1cm] {$\scriptstyle S^+$};

    \draw [<-] ($(phi1.east)+(0,0.15cm)$) -- ($(psi0.west)-(0,0.30cm)$);
    \node at (q2) [anchor=north,outer sep=0.1cm] {$\scriptstyle Q^-$};

    \draw [->] ($(phi1.east)-(0,0.30cm)$) -- ($(psi1.west)+(0,0.15cm)$);
    \node at (q3) [anchor=south,outer sep=0.15cm] {$\scriptstyle S^-$};

    \draw [<-] ($(phi1.east)-(0,0.15cm)$) -- ($(psi1.west)+(0,0.30cm)$);
    \node at (q3) [anchor=north,outer sep=0.1cm] {$\scriptstyle Q^+$};

    \draw [->] ($(phi2.east)+(0,0.30cm)$) -- ($(psi1.west)-(0,0.15cm)$);
    \node at (q4) [anchor=south,outer sep=0.1cm] {$\scriptstyle S^+$};

    \draw [<-] ($(phi2.east)+(0,0.15cm)$) -- ($(psi1.west)-(0,0.30cm)$);
    \node at (q4) [anchor=north,outer sep=0.1cm] {$\scriptstyle Q^-$};

    \draw [->] ($(phi0.south)+(-0.08cm,0)$) -- ($(phi1.north)+(-0.08cm,0)$);
    \draw [<-] ($(phi0.south)+(+0.08cm,0)$) -- ($(phi1.north)+(+0.08cm,0)$);
    \node at (s1) [anchor=west] {$\scriptstyle J_-$};
    \node at (s1) [anchor=east] {$\scriptstyle J_+$};

    \draw [->] ($(psi0.south)+(-0.08cm,0)$) -- ($(psi1.north)+(-0.08cm,0)$);
    \draw [<-] ($(psi0.south)+(+0.08cm,0)$) -- ($(psi1.north)+(+0.08cm,0)$);
    \node at (s2) [anchor=west] {$\scriptstyle J_-$};
    \node at (s2) [anchor=east,baseline=-2cm] {$\scriptstyle J_+$};

    \draw [->] ($(phi1.south)+(-0.08cm,0)$) -- ($(phi2.north)+(-0.08cm,0)$);
    \draw [<-] ($(phi1.south)+(+0.08cm,0)$) -- ($(phi2.north)+(+0.08cm,0)$);
    \node at (s3) [anchor=west] {$\scriptstyle J_-$};
    \node at (s3) [anchor=east] {$\scriptstyle J_+$};

    \node at (s4) {$\vdots$};
    \node at (s5) {$\vdots$};
    \node at ($(q5)+(-0.1cm,0)$) {$\vdots$};

  \end{tikzpicture}
  \caption{An illustration of the short
    $\left(\tfrac{\alpha}{2},-\tfrac{\alpha}{2}\right)$ $\algSL(2|1)$
    module. For clarity, in the figure we have neglected to include the exact
    coefficients of the linear maps for the representation. These can be
    read-off from
    equations~\eqref{eq:d21a-chiral-rep-LR}-\eqref{eq:d21a-chiral-rep-Q} in
    Appendix~\ref{sec:d21a-algebra} using the
    embedding~\eqref{eq:sl21embedding}. To emphasise this we use the notation
    $\Bigl<\dots\Bigr>$ to denote the span of $\dots$.  The states in the
    left-hand side column have $B=-\tfrac{\alpha}{2}$ and those in the
    right-hand side column have $B=\tfrac{1-\alpha}{2}$. Starting at the top,
    the rows in the diagram have $J_0$ eigenvalues equal to
    $\tfrac{\alpha}{2}\,,\,\tfrac{\alpha+1}{2}\,,\,\tfrac{\alpha+2}{2}\,,\dots$}
  \label{fig:short-sl21}
\end{figure}

In constructing R-matrices it is often useful to know the tensor
product decompositions of the constituent representations. The tensor
products of two atypical representations are given by
\begin{align}
  (j_1,\pm j_1) \directprod (j_2,\pm j_2) &= (j,\pm j) \oplus\bigoplus_{n=0}^\infty \left(j + \tfrac{1}{2} + n, \pm(j - \tfrac{1}{2})\right) \,, \\
  (j_1,\pm j_1) \directprod (j_2,\mp j_2) &= \bigoplus_{n=0}^\infty (j + n, \pm \bar{\jmath}) \,,
\end{align}
where $ j = j_1 + j_2$ and $\bar{\jmath} = j_1 - j_2$. For the
representations we are interested in this implies,
\begin{align}
\label{eq:sl21-tensor-ac-ac}
  \left(\tfrac{\alpha}{2},-\tfrac{\alpha}{2}\right) \directprod \left(\tfrac{\alpha}{2},-\tfrac{\alpha}{2}\right) &= 
  (\alpha,-\alpha) \oplus\bigoplus_{n=1}^{\infty} \left( \alpha - \tfrac{1}{2} + n, \tfrac{1}{2} - \alpha \right) \,, \\
\label{eq:sl21-tensor-c-c}
  \left(\tfrac{1-\alpha}{2},\tfrac{1-\alpha}{2}\right) \directprod \left(\tfrac{1-\alpha}{2},\tfrac{1-\alpha}{2}\right) &= 
  (1 - \alpha, 1 - \alpha) \oplus \bigoplus_{n=1}^{\infty} \left( \tfrac{1}{2} - \alpha + n, \tfrac{1}{2} - \alpha \right) \,, \\
\label{eq:sl21-tensor-ac-c}
  \left(\tfrac{\alpha}{2},-\tfrac{\alpha}{2}\right) \directprod \left(\tfrac{1 - \alpha}{2},\tfrac{1 - \alpha}{2}\right) &= 
  \bigoplus_{n=0}^{\infty} \left( \tfrac{1}{2} + n, \tfrac{1}{2} - \alpha\right) \,.
\end{align}
The $R$-matrix $R_{ab}(u)$ acts on the tensor product $\mathcal{V}_a \directprod \mathcal{V}_b$ and can be decomposed as
\begin{equation}\label{eq:rmatrixasproj}
  R_{ab}(u) = \sum_c R_{ab}^c (u) \mathcal{P}_c \,,
\end{equation}
where the sum is over all irreducible representations in the
decomposition of the product, and $\mathcal{P}_c$ is a projector.  To
simplify the notation we will use the labels $a, b = \pm$ for the
matrix acting on chiral respectively anti-chiral representations, and
write the projectors as
\begin{align}
  \mathcal{P}_{(\tfrac{1}{2} - \alpha + n,\tfrac{1}{2}-\alpha)} &\equiv \mathcal{P}_{n^+} \,, &
  \mathcal{P}_{(\alpha - \tfrac{1}{2} + n,\tfrac{1}{2}-\alpha)} &\equiv \mathcal{P}_{n^-} \,, &
  \mathcal{P}_{(\tfrac{1}{2}  + n,\tfrac{1}{2}-\alpha)} &\equiv \mathcal{P}_{n} \,,
\end{align}
with corresponding notation for the label $c$ of the R-matrix.

In~\cite{Derkachov:2000ne} the $R$-matrix acting on an arbitrary
representation of $\algSL(2|1)$ was derived. For two chiral or
anti-chiral states we get ($n \ge 1$)
\begin{align}
  R_{--}^{n^-}(u) &= (-1)^n N_{--}(\tilde{u}) \frac{\Gamma(+\tilde{u} + \alpha + n)}{\Gamma(-\tilde{u} + \alpha + n)} \,, \\
  R_{++}^{n^+}(u) &= (-1)^n N_{++}(\tilde{u}) \frac{\Gamma(+\tilde{u} + 1 - \alpha + n)}{\Gamma(-\tilde{u} + 1 - \alpha + n)} \,,
\intertext{while a chiral with an anti-chiral state gives ($n \ge 0$)}
  R_{+-}^{n}(u) = R_{-+}^{n}(u) &= (-1)^n N_{-+}(\tilde{u}) \frac{\Gamma(+\tilde{u} + 1 + n)}{\Gamma(-\tilde{u} + 1 + n)} \,.
\end{align}
Here $\tilde{u} = u/c$ and $c$ and $N_{\pm\pm}(\tilde{u})$ are arbitrary and correspond to trivial symmetries of the Yang-Baxter equation. We find it convenient to choose
\begin{align}
  N_{--}(\tilde{u}) &= \frac{\Gamma(-\tilde{u} + \alpha)}{\Gamma(+\tilde{u} + \alpha)} \,, &
  N_{++}(\tilde{u}) &= \frac{\Gamma(-\tilde{u} + 1 - \alpha)}{\Gamma(+\tilde{u} + 1 - \alpha)} \,, &
  N_{\mp\pm}(\tilde{u}) &= \frac{\Gamma(-\tilde{u} + 1)}{\Gamma(+\tilde{u} + 1)} \,,
\end{align}
and set $c = 2$. Then the above expressions can be written as
\begin{align}
  R_{--}^{n^-}(u) &= \prod_{k=0}^{n-1} \frac{u + 2\alpha + 2k}{u - 2\alpha - 2k} \,, \label{eq:SL21-R-matrix-mm} \\
  R_{++}^{n^+}(u) &= \prod_{k=0}^{n-1} \frac{u + 2(1 - \alpha) + 2k}{u - 2(1 - \alpha) - 2k} \,, \label{eq:SL21-R-matrix-pp} \\
  R_{\mp\pm}^{n}(u) &=\prod_{k=0}^n \frac{u + 2k}{u - 2k} \,. \label{eq:SL21-R-matrix-pm}
\end{align}
With the above normalization we also have
\begin{align}
  \label{eq:SL21-R-matrix-0}
  R_{--}^{(\alpha,-\alpha)} &= 1 \,, &
  R_{++}^{(1-\alpha,1-\alpha)} &= 1 \,.
\end{align}

We will now use these $R$-matrices to derive a Hamiltonian for the alternating
spin-chain. Our notation closely follows the notation
of~\cite{Minahan:2009te}. The spin-chain has chiral representations sitting at
odd sites and anti-chiral representations at even sites, which we will label by
$a_i$ and $\bar{a}_i$, respectively. For a chain with $L$ sites (of each kind)
we can write down two transfer matrices, with the auxiliary space in either the
chiral representation (with an $a$ index), or the anti-chiral representation
(indicate by an $\bar{a}$ index),
\begin{align}
  T_a(u) &= R_{a a_1}\!(u) \, R_{a \bar{a}_1}\!(u) \, R_{a a_2}\!(u) R_{a \bar{a}_2}\!(u) \dotsm R_{a a_L}\!(u) R_{a \bar{a}_L}\!(u) \,, \\
  T_{\bar{a}}(u) &= R_{\bar{a} a_1}\!(u) \, R_{\bar{a} \bar{a}_1}\!(u) \, R_{\bar{a} a_2}\!(u) R_{\bar{a} \bar{a}_2}\!(u) \dotsm R_{\bar{a} a_L}\!(u) R_{\bar{a} \bar{a}_L}\!(u) \,.
\end{align}
Taking the traces over the auxiliary spaces we define
\begin{align}
  \tau(u) &= \tr_a T_a(u) \,, &
  \bar{\tau}(u) &= \tr\nolimits_{\bar{a}} T_{\bar{a}}(u) \,.
\end{align}
The Yang-Baxter equation now ensures the commutation relations
\begin{equation}
  \comm{\tau(u)}{\tau(v)} = \comm{\tau(u)}{\bar{\tau}(v)} = \comm{\bar{\tau}(u)}{\bar{\tau}(v)} = 0.
\end{equation}
The Hamiltonian is given by
\begin{equation}
  H = C (\tau(0) \bar{\tau}(0) )^{-1}  \left. \frac{d}{du} (\tau(u) \bar{\tau}(u)) \right|_{u=0} \,,
\end{equation}
where $C$ is a normalization constant. Since $\tau(u)$ and $\bar{\tau}(v)$ commute, we can write this as
\begin{align}
  H 
  &= C (\tau(0) \bar{\tau}(0) )^{-1} \left[ \tau'(0) \bar{\tau}(0) + \tau(0) \bar{\tau}'(0) \right] \\
  &= C (\tau(0) \bar{\tau}(0) )^{-1} \left[ \tau'(0) \bar{\tau}(0) + \bar{\tau}'(0) \tau(0) \right] \,.
\end{align}
We note that
\begin{align}
  R_{ab}(0) &= \mathcal{P}_{(\alpha,-\alpha)} + \sum_{n=1}^{\infty} (-1)^n \mathcal{P}_{n^-} \,, \\
  R_{\bar{a}\bar{b}}(0) &= \mathcal{P}_{(1-\alpha,1-\alpha)} + \sum_{n=1}^{\infty} (-1)^n \mathcal{P}_{n^+} \,.
\end{align}
The representations appearing in the sums are symmetric for even $n$ and anti-symmetric for odd $n$. Hence the above operators act as two-site exchange operators
\begin{align}
  R_{ab}(0) &= P_{ab} \,, &
  R_{\bar{a}\bar{b}}(0) &= P_{\bar{a}\bar{b}} \,.
\end{align}
The $R$-matrix acting on a chiral and an anti-chiral representation is, at $u=0$,
\begin{equation}
  R_{a\bar{b}}(0) = \sum_{n=0}^{\infty} (-1)^n \mathcal{P}_{n} \,.
\end{equation}
Introducing explicit indices $I_a$ and $\bar{I}_a$ for the states of the chiral and anti-chiral representations, we can write the above operators as
\begin{align}
  R_{ab}(0) &= \delta_{I_a}^{J_b} \delta_{I_b}^{J_a} \,, &
  R_{\bar{a}\bar{b}}(0) &= \delta_{\bar{I}_a}^{\bar{J}_b} \delta_{\bar{I}_b}^{\bar{J}_a} \,, &
  R_{a\bar{b}}(0) &= \mathcal{M}_{I_a \bar{I}_b}^{J_a \bar{J}_b} \,,
\end{align}
where the last equality defines  $\mathcal{M}$. We also note that
\begin{equation}
  \mathcal{M}_{I_a \bar{I}_b}^{K_a \bar{K}_b}  \mathcal{M}_{K_a \bar{K}_b}^{J_a \bar{J}_b} = \delta_{I_a}^{J_a} \delta_{\bar{I}_b}^{\bar{J}_b} \,.
\end{equation}
It then follows that
\begin{equation}
  (\tau(0) \bar{\tau}(0))_{I_1 \bar{I}_1 I_2 \bar{I}_2 \dotsm I_L \bar{I}_L}^{J_1 \bar{J}_1 J_2 \bar{J}_2 \dotsm J_L \bar{J}_L} =
  \delta_{I_1}^{J_2} \delta_{\bar{I}_1}^{\bar{J}_2} \delta_{I_2}^{J_3} \delta_{\bar{I}_2}^{\bar{J}_3} \dotsm \delta_{I_L}^{J_1} \delta_{\bar{I}_L}^{\bar{J}_1} \,,
\end{equation}
which acts as a two site shift operator.

We also need the derivatives of the $R$-matrix at $u=0$,
\begin{align}
  \mathcal{A} \equiv R'_{ab}(0) &= \sum_{n=1}^{\infty} (-1)^n \left[ \psi(\alpha + n) - \psi(\alpha) \right] \mathcal{P}_{n^-} \,, \\
  \bar{\mathcal{A}} \equiv R'_{\bar{a}\bar{b}}(0) &= \sum_{n=1}^{\infty} (-1)^n \left[\psi(1 - \alpha + n) - \psi(1 - \alpha) \right] \mathcal{P}_{n^+} \,, \\
  \mathcal{B} \equiv R'_{a\bar{b}}(0) &= \sum_{j=1}^{\infty} (-1)^n \left[ \psi(n + 1) + \gamma_{\mathrm{E}} \right] \mathcal{P}_{n} \,,
\end{align}
where $\psi(z)$ is the digamma function and $\gamma_{\mathrm{E}}$ is the Euler-Mascheroni constant.\footnote{%
  We can also express these coefficients in terms of analytically continued harmonic numbers using the relation $h(z) = \psi(z + 1) + \gamma_{\mathrm{E}}$.
}
Putting everything together, the Hamiltonian can now be written as
\begin{equation}
  \begin{split}
    H &= 
    C \sum_{l=1}^L \left( 
      \mathcal{M}_{I_l \bar{I}_l}^{K \bar{K}} \mathcal{A}_{K I_{l+1}}^{J_{l+1} L} \mathcal{M}_{L \bar{K}}^{J_{l} \bar{J}_{l}} + 
      \mathcal{B}_{I_l \bar{I}_l}^{K \bar{K}} \mathcal{M}_{K \bar{K}}^{J_l \bar{J}_l}
    \right) \\
    & +
    C \sum_{l=1}^L \left( 
      \mathcal{M}_{\bar{I}_l I_{l+1}}^{\bar{K} K} \bar{\mathcal{A}}_{\bar{K} \bar{I}_{l+1}}^{\bar{J}_{l+1} \bar{L}} \mathcal{M}_{\bar{L} K}^{\bar{J}_{l} J_{l+1}} + 
      \mathcal{B}_{\bar{I}_l I_{l+1}}^{\bar{K} K} \mathcal{M}_{\bar{K} K}^{\bar{J}_l J_{l+1}}
    \right) \,.
  \end{split}
\end{equation}

We observe in particular that for $\alpha = 1/2$,
\begin{align}
  \mathcal{A} = \bar{\mathcal{A}} &= \sum_{n=1}^{\infty} (-1)^n \left( 2h(2n-1) - h(n-1) \right) \mathcal{P}_{(n,0)} \,, \\
  \mathcal{B} &= \sum_{n=1}^{\infty} (-1)^n h(n) \mathcal{P}_{(n + \tfrac{1}{2},0)} \,.
\end{align}
For this value of $\alpha$ the Hamiltonian exactly coincides with the Hamiltonian for the $\algSL(2|1)$ sector of ABJM~\cite{Minahan:2009te}.

\subsection{The lift to $\algD{\alpha}$}
\label{sec:lift-to-d21a}

The lift of the $\algSL(2|1)$ R-matrix in
\eqref{eq:SL21-R-matrix-mm}--\eqref{eq:SL21-R-matrix-0} to the full
$\algD{\alpha}$ is now straightforward. Comparing the tensor product
decompositions \eqref{eq:sl21-tensor-ac-ac}--\eqref{eq:sl21-tensor-ac-c} with
the corresponding products in the larger group,
\eqref{eq:d21a-tensor-ac-c}--\eqref{eq:d21a-tensor-c-c}, it is easy to see that
there is a direct map between the individual states. The $\algD{\alpha}$
R-matrix is then given by replacing the projectors in the R-matrix of the previous
sector by the projector of the full group. It would be interesting to verify the
validity of this uplifting procedure by a direct check of the YBE equation for
the $\algD{\alpha}$ R-matrix.\footnote{We would like to thanks Kostya Zarembo
  for a discussion of this point.}

\subsection{The $\algSU(2) \times \algSU(2)$ sector}
\label{sec:su2-su2-sector}

The largest compact subalgebra of $\algD{\alpha}$ is $\algSU(2) \times
\algSU(2)$. In this sector, the tensor product between odd and even
sites is truncated to
\begin{align}
  (-\tfrac{\alpha}{2};\tfrac{1}{2};0) \directprod (-\tfrac{1-\alpha}{2};0;\tfrac{1}{2}) &\to  (-\tfrac{1}{2};\tfrac{1}{2};\tfrac{1}{2}) \,, \\
  (-\tfrac{\alpha}{2};\tfrac{1}{2};0) \directprod (-\tfrac{\alpha}{2};\tfrac{1}{2};0) &\to (-\alpha;1;0) \directsum (-\alpha;0;0) \,, \\
  (-\tfrac{1-\alpha}{2};0;\tfrac{1}{2}) \directprod (-\tfrac{1-\alpha}{2};0;\tfrac{1}{2}) &\to (-(1-\alpha);0;1) \directsum (-(1-\alpha);0;0) \,. 
\end{align}
The corresponding products in $\algSL(2|1)$ read
\begin{align}
  \left(\tfrac{\alpha}{2},-\tfrac{\alpha}{2}\right) \directprod \left(\tfrac{\alpha}{2},-\tfrac{\alpha}{2}\right) &\to 
  (\alpha,-\alpha) \directsum \left( \alpha + \tfrac{1}{2}, \tfrac{1}{2} - \alpha \right) \,, \\
  \left(\tfrac{1-\alpha}{2},\tfrac{1-\alpha}{2}\right) \directprod \left(\tfrac{1-\alpha}{2},\tfrac{1-\alpha}{2}\right) &\to
  (1 - \alpha, 1 - \alpha) \directsum \left( \tfrac{3}{2} - \alpha, \tfrac{1}{2} - \alpha \right) \,, \\
  \left(\tfrac{\alpha}{2},-\tfrac{\alpha}{2}\right) \directprod \left(\tfrac{1 - \alpha}{2},\tfrac{1 - \alpha}{2}\right) &\to 
  \left( \tfrac{1}{2}, \tfrac{1}{2} - \alpha\right) \,.
\end{align}
As we previously noted, the representations $\left( \alpha + \tfrac{1}{2}, \tfrac{1}{2} - \alpha \right)$ and $\left( \tfrac{3}{2} - \alpha, \tfrac{1}{2} - \alpha \right)$ are anti-symmetric. Hence
\begin{align}
  \mathcal{P}_{(\tfrac{1}{2} + \alpha, \tfrac{1}{2} - \alpha)} &= \tfrac{1}{2}( 1 - P_{I_{l} I_{l+1}} ) \,, &
  \mathcal{P}_{(\tfrac{3}{2} - \alpha, \tfrac{1}{2} - \alpha)} &= \tfrac{1}{2}( 1 - P_{\bar{I}_{l} \bar{I}_{l+1}} ) \,,
\end{align}
This gives the Hamiltonian
\begin{equation}
  H = \frac{C}{2} \sum_l \left[ \frac{1}{\alpha} (1 - P_{I_l I_{l+1}}) + \frac{1}{1-\alpha} (1 - P_{\bar{I}_l \bar{I}_{l+1}}) \right] \,.
\end{equation}
As expected this is the sum of two Heisenberg spin-chain Hamiltonians.

\subsection{The full $\algD{\alpha}^2$ spin-chain}
\label{sec:full-spin-chain}

The symmetry of the weak coupling Bethe equations constructed above is
$\grD{\alpha}$. However, the symmetry group of superstrings on $AdS_3 \times S^3
\times S^3$ is $\grD{\alpha}^2$. The two factors of the full group act
independently on the left- respectively right-moving sectors of the theory. As
seen in the full Bethe equations in section~\ref{sec:BA-d21a-all-loop}, the two
sets of equations are coupled via the dressing phase and via fermionic inversion
symmetry links. To leading order at weak coupling these interactions are
trivial, and we can treat the left- and right-movers separately.\footnote{%
  We assume that the dressing phase is trivial at weak coupling, as is the case
  in $AdS_5 \times S^5$ and $AdS_4 \times \CP^3$.%
} %
Hence, the situation is similar to the $\algSU(2) \times \algSU(2)$ sector of
ABJM, which also consists of two independent subsectors which at weak coupling
only couple through the momentum constraint~\cite{Minahan:2008hf}.

Hence, the weak coupling Bethe equations in this section only describe the
left-moving sector of the theory. To describe the full spectrum, we need an
additional set of equations describing the right-movers. As argued above, at
weak coupling the left- and right-mover equations are independent. However,
there is still a coupling between them from the momentum constraint which
requires that the total momentum for any physical operator vanishes,\footnote{%
  On the string theory side of the $\AdS_3/\CFT_2$ duality the momentum
  constraint arises from the level matching condition fo closed strings. Since
  the $\CFT_2$ is unknown it is not clear what this corresponds to at weak coupling.%
}%
\begin{equation}
  e^{i P_{\text{tot}}} = e^{i( P_{\text{L}} - P_{\text{R}})} =
  \frac{
    \prod_{k=1}^{K_1} \frac{u_{1,k} + i\alpha}{u_{1,k} - i\alpha}
  }{
    \prod_{k=1}^{\bar{K}_1} \frac{u_{\bar{1},k} + i\alpha}{u_{\bar{1},k} - i\alpha} 
  }
  \frac{
    \prod_{k=1}^{K_3} \frac{u_{3,k} + i(1-\alpha)}{u_{3,k} - i(1-\alpha)}
  }{ 
    \prod_{k=1}^{\bar{K}_3} \frac{u_{\bar{3},k} + i(1-\alpha)}{u_{\bar{3},k} - i(1-\alpha)}
  } = 1 \,,
\label{eq:momconstr}
\end{equation}
where the momentum of excitations in the left- and right-moving
sectors are counted with different signs.

Since we have two sets of Bethe equations coupled only via the
momentum constraint, there are many more physical operators than there
would have been if we required each sector to have vanishing momentum
by itself. In fact, \emph{any} solution to the Bethe equations for the
left-movers can be turned into a physical solution with vanishing
momentum, provided we can find a solution to the equations for the
right-movers with equal momentum. This is easily accomplished by
just setting $u_{\bar{l},i} = u_{l,i}$.

The string theory background also contains an $S^1$ factor,
which couples to the rest of the geometry only via the Virasoro
constraints. At weak coupling this means that also solutions to the
Bethe equations which do not have zero total momentum should be
considered, since any additional momentum can be attributed to one or
more massless excitation on this circle.

Even though a physical operator only needs to have zero total momentum, it is
interesting to consider solutions to the left-moving Bethe equations that
satisfy the momentum constraint by themselves. As we will see below, such
solutions for example display extra degeneracies not directly explained by the
manifest symmetries of the model. In addition the similarities of the Bethe
equations proposed here and those describing $\superN=4$ SYM and the ABJM model
are made more apparent when we concentrate on such a chiral sector.

\section{Fermionic duality}
\label{sec:fermionic-duality}

\begin{figure}
  \centering

  \subfloat[\label{fig:dynkin-d21a-orig}]{
    \begin{tikzpicture}
      [
      thick,
      node/.style={shape=circle,draw,thick,inner sep=0pt,minimum size=5mm}
      ]
      \useasboundingbox (-1.1cm,-1.1cm) rectangle (1.6cm,1.1cm);

      \node (v1) at (-0.38cm,  0.65cm) [node] {};
      \node (v2) at ( 0.75cm,  0.00cm) [node] {};
      \node (v3) at (-0.38cm, -0.65cm) [node] {};

      \draw (v2.south west) -- (v2.north east);
      \draw (v2.north west) -- (v2.south east);

      \draw (v1) -- (v2);
      \draw (v2) -- (v3);

      \node at (v1.west) [anchor=east] {$1$};
      \node at (v3.west) [anchor=east] {$1$};
    \end{tikzpicture}
  }
  \hspace{2cm}
  \subfloat[\label{fig:dynkin-d21a-dual}]{
    \begin{tikzpicture}
      [
      thick,
      node/.style={shape=circle,draw,thick,inner sep=0pt,minimum size=5mm}
      ]
      \useasboundingbox (-1.1cm,-1.1cm) rectangle (1.6cm,1.1cm);

      \node (v1) at (-0.38cm,  0.65cm) [node] {};
      \node (v2) at ( 0.75cm,  0.00cm) [node] {};
      \node (v3) at (-0.38cm, -0.65cm) [node] {};

      \draw (v1.south west) -- (v1.north east);
      \draw (v1.north west) -- (v1.south east);

      \draw (v2.south west) -- (v2.north east);
      \draw (v2.north west) -- (v2.south east);

      \draw (v3.south west) -- (v3.north east);
      \draw (v3.north west) -- (v3.south east);

      \draw (v1) -- (v2);
      \draw (v2) -- (v3);
      \draw [double,double distance=3pt] (v3) -- (v1);

      \node at (v1.west) [anchor=east] {$1$};
      \node at (v3.west) [anchor=east] {$1$};
    \end{tikzpicture}
  }
  
  \caption{Two of the Dynkin diagrams for $\algD{\alpha}$. The crossed
    notes are fermionic and the labels indicate the momentum carrying
    roots in the Bethe equations. The original equations
    \eqref{eq:BE-d21a-one-loop-orig} corresponds to the diagram
    \subref{fig:dynkin-d21a-orig}, while the dualized equations
    \eqref{eq:BE-d21a-one-loop-dual} corresponds to
    \subref{fig:dynkin-d21a-dual}.}
  \label{fig:dynkin-d21a}
\end{figure}
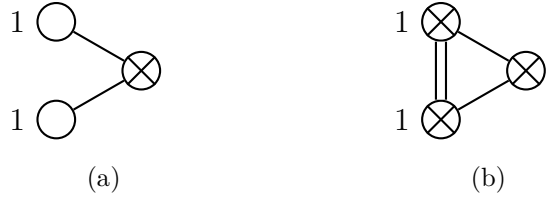

The Dynkin diagram of a superalgebra is not unique, since there are
multiple inequivalent choices of the simple roots. For example, in
figure~\ref{fig:dynkin-d21a} we show two examples of Dynkin diagrams
for $\algD{\alpha}$. The structure of the Bethe equations is
intimately related to a chosen Dynkin diagram and one can transform
the Bethe equations between different such choices using a fermionic
duality. The Bethe equations we have been dealing with so far are
related to the Dynkin diagram in figure~\ref{fig:dynkin-d21a-orig}. In
this section we will perform a fermionic duality on our Bethe
equations in order to obtain a form of them related to the Dynkin
diagram in figure~\ref{fig:dynkin-d21a-dual}. The new form of Bethe
equations will allow us, in the next section, to take the
$\alpha\rightarrow 1$ limit more easily.

The procedure for fermionic duality on Bethe equations used in this section
closely follows the general prescription discussed
in~\cite{Tsuboi:1998ne,Tsuboi:1997iq,Tsuboi:1998sc,Beisert:2005di}. The case at
hand is very similar to the $\algOSp(6|4)$ spin-chain studied
in~\cite{Minahan:2008hf,Gromov:2008qe}. The original Bethe equations read
\begin{equation}\label{eq:BE-d21a-one-loop-orig}
  \begin{aligned}
    \left(\frac{u_{1,i} + i\alpha}{u_{1,i} - i\alpha}\right)^L &= 
    \prod_{\substack{k = 1\\k \neq i}}^{K_1} \frac{u_{1,i} - u_{1,k} + 2i\alpha}{u_{1,i} - u_{1,k} - 2i\alpha}
    \prod_{k=1}^{K_2} \frac{u_{1,i} - u_{2,k} - i\alpha}{u_{1,i} - u_{2,k} + i\alpha} \,, \\
    1 &= 
    \prod_{k=1}^{K_1} \frac{u_{2,i} - u_{1,k} - i\alpha}{u_{2,i} - u_{1,k} + i\alpha}
    \prod_{k=1}^{K_3} \frac{u_{2,i} - u_{3,k} - i(1-\alpha)}{u_{2,i} - u_{3,k} + i(1-\alpha)} \,, \\
    \left(\frac{u_{3,i} + i(1-\alpha)}{u_{3,i} - i(1-\alpha)}\right)^L &= 
    \prod_{\substack{k = 1\\k \neq i}}^{K_3} \frac{u_{3,i} - u_{3,k} + 2i(1-\alpha)}{u_{3,i} - u_{3,k} - 2i(1-\alpha)}
    \prod_{k=1}^{K_2} \frac{u_{3,i} - u_{2,k} - i(1-\alpha)}{u_{3,i} - u_{2,k} + i(1-\alpha)} \,.
  \end{aligned}
\end{equation}
The middle equation can be expressed as
\begin{equation}
  P(u_{2,i}) = 0 \,,
\end{equation}
where $P(u)$ is the polynomial
\begin{equation}
  \begin{split}
    P(u) = 
    &\prod_{k=1}^{K_1} (u - u_{1,k} + i\alpha) \prod_{k=1}^{K_3} (u - u_{3,k} + i(1-\alpha))  \\
    &- \prod_{k=1}^{K_1} (u - u_{1,k} - i\alpha) \prod_{k=1}^{K_3} (u - u_{3,k} - i(1-\alpha)) \,.
  \end{split}
\end{equation}
$P(u)$ has in total $K_1 + K_3 - 1$ zeros. Of these, $K_2$ are the original roots $u_{2,k}$. Hence we can write $P(u)$ as
\begin{equation}
  P(u) = 2i(\alpha K_1 + (1-\alpha) K_3) \prod_{k=1}^{K_2} (u - u_{2,k}) \prod_{k=1}^{\tilde{K}_2} (u - \tilde{u}_{2,k}) \,,
\end{equation}
where we have introduced $\tilde{K}_2 = K_1 + K_3 - K_2 - 1$ dual roots
$\tilde{u}_{2,k}$. An important observation is that these dual roots satisfy the
same Bethe equation as the original roots. By evaluating $P(u)$ in $u_{1,i} \pm
i\alpha$ and $u_{3,i} \pm i(1-\alpha)$ we can now transform our Bethe equations to the dual
form
\begin{equation}\label{eq:BE-d21a-one-loop-dual}
  \begin{aligned}
    \left(\frac{u_{1,i} + i\alpha}{u_{1,i} - i\alpha}\right)^L &= 
    \prod_{k=1}^{K_3} \frac{u_{1,i} - u_{3,k} - i}{u_{1,i} - u_{3,k} + i}
    \prod_{k=1}^{\tilde{K}_2} \frac{u_{1,i} - \tilde{u}_{2,k} + i\alpha}{u_{1,i} - \tilde{u}_{2,k} - i\alpha} \,, \\
    1 &= 
    \prod_{k=1}^{K_1} \frac{\tilde{u}_{2,i} - u_{1,k} + i\alpha}{\tilde{u}_{2,i} - u_{1,k} - i\alpha}
    \prod_{k=1}^{K_3} \frac{\tilde{u}_{2,i} - u_{3,k} + i(1-\alpha)}{\tilde{u}_{2,i} - u_{3,k} - i(1-\alpha)} \,, \\
    \left(\frac{u_{3,i} + i(1-\alpha)}{u_{3,i} - i(1-\alpha)}\right)^L &= 
    \prod_{k=1}^{K_1} \frac{u_{3,i} - u_{1,k} - i}{u_{3,i} - u_{1,k} + i}
    \prod_{k=1}^{\tilde{K}_2} \frac{u_{3,i} - \tilde{u}_{2,k} + i(1-\alpha)}{u_{3,i} - \tilde{u}_{2,k} - i(1-\alpha)} \,.
  \end{aligned}
\end{equation}
Compared to the equations we started with we note a few changes:
\begin{itemize}
\item There are no self-interactions in any equation. This means that this corresponds to a Dynkin diagram where all nodes are fermionic.
\item The interaction of the $u_{2,k}$ roots with $u_{1,k}$ and $u_{3,k}$ has switched sign.
\item There is a new interaction between the $u_{1,k}$ and $u_{3,k}$ roots.
\end{itemize}

\subsection{Dualization of the full Bethe equations?}
\label{sec:sec41}

The dualization of the full Bethe equations is more tricky, due to the links
between, \eg, the $x_{1,k}$ and $x_{\bar{2},k}$ nodes. Here we will work in
the sector where one of the $\algD{\alpha}$ factors carry no excitations,
\ie, $K_{\bar{1}} = K_{\bar{2}} = K_{\bar{3}} = 0$. The procedure is then very
similar to the one-loop case and we get the dual equations
\begin{equation}
  \begin{split}
    \left(\frac{x_{1,i}^+}{x_{1,i}^-}\right)^L &=
    \prod_{\substack{k = 1\\k \neq i}}^{K_1} \frac{1 - \frac{1}{x_{1,i}^+ x_{1,k}^-}}{1 - \frac{1}{x_{1,i}^- x_{1,k}^+}} \sigma_1^2(x_{1,i},x_{1,k})
    \prod_{k=1}^{\tilde{K}_2} \frac{x_{1,i}^+ - \tilde{x}_{2,k}}{x_{1,i}^- - \tilde{x}_{2,k}}
    \prod_{k=1}^{K_3} \frac{x_{1,i}^- - x_{3,k}^+}{x_{1,i}^+ - x_{3,k}^-} \,, \\
    1 &=
    \prod_{k=1}^{K_1} \frac{\tilde{x}_{2,i} - x_{1,k}^+}{\tilde{x}_{2,i} - x_{1,k}^-}
    \prod_{k=1}^{K_3} \frac{\tilde{x}_{2,i} - x_{3,k}^+}{\tilde{x}_{2,i} - x_{3,k}^-} \,, \\
    \left(\frac{x_{3,i}^+}{x_{3,i}^-}\right)^L &=
    \prod_{\substack{k = 1\\k \neq i}}^{K_3} \frac{1 - \frac{1}{x_{3,i}^+ x_{3,k}^-}}{1 - \frac{1}{x_{3,i}^- x_{3,k}^+}} \sigma_3^2(x_{3,i},x_{3,k})
    \prod_{k=1}^{\tilde{K}_2} \frac{x_{3,i}^+ - \tilde{x}_{2,k}}{x_{3,i}^- - \tilde{x}_{2,k}}
    \prod_{k=1}^{K_1} \frac{x_{3,i}^- - x_{1,k}^+}{x_{3,i}^+ - x_{1,k}^-} \,.
  \end{split}
\end{equation}
Note that these equations are valid for any value of $\alpha$.

Let us now consider the $\alpha = 1/2$ case, and put $K_1 = K_3 = K$,
$\tilde{K}_2 = 0$ and $x_{1,k}^{\pm} = x_{3,k}^{\pm} = x_k^{\pm}$.  The above
equations then reduce to
\begin{equation}
  \label{eq:scaling-func-eq}
  \left(\frac{x_i^+}{x_i^-}\right)^L =
  - \prod_{\substack{k = 1\\k \neq i}}^{K} \frac{x_i^- - x_k^+}{x_i^+ - x_k^-} \frac{1 - \frac{1}{x_i^+ x_k^-}}{1 - \frac{1}{x_i^- x_k^+}} \sigma^2(x_i,x_k) \,. 
\end{equation}
An equation of exactly this form appears in ABJM~\cite{Gromov:2008qe}, and,
apart from the minus sign on the right hand side, it is the starting point for
deriving the Eden-Staudacher~\cite{Eden:2006rx},
Beisert-Eden-Staudacher~\cite{Beisert:2006ez} and
Freyhult-Rej-Staudacher~\cite{Freyhult:2007pz} equations. However, these
equations heavily rely on the exact form of the BES/BHL dressing
phase~\cite{Beisert:2006ez,Beisert:2006ib}, and there is no particular reason
that the dressing phase in~\eqref{eq:scaling-func-eq} should take the same form
as the corresponding phases in $\superN=4$ SYM and ABJM.\footnote{%
  In order to reproduce the classical Bethe equations
  in~\eqref{eq:BE-d21a-integral} the leading strong coupling behavior of
  $\sigma$ should take the AFS form~\cite{Arutyunov:2004vx}.%
} %
Hence, the solutions of~\eqref{eq:scaling-func-eq} can in general be very
different from the previously known cases.


\section{The $\alpha\rightarrow 1$ limit and $AdS_3 \times S^3 \times T^4$}
\label{sec:t4-limit}

In the $\alpha \to 1$ limit, the $\algD{\alpha}$ algebra turns into
$\algPSU(1,1|2)$.\footnote{The $\alpha\rightarrow 0$ limit also results in such
  a reduction; the only difference is which $\algSU(2)$ sub-algebra one
  looses. Since the $\algD{\alpha}$ is isomorphic to the $\algD{1-\alpha}$
  algebra the two limits are related and correspond to chosing which of the two
  $S^3$ decompactifies.} This allowed for a unified treatment of the
Green-Schwarz superstring action on both the $AdS_3 \times S^3 \times
S^3 \times S^1$ and $AdS_3 \times S^3 \times T^4$
backgrounds~\cite{Babichenko:2009dk}. In this section we will investigate the
$\alpha \to 1$ limit for the weakly coupled Bethe Ansatz and corresponding
spin-chain described in sections~\ref{sec:BA-d21a-one-loop}
and~\ref{sec:alternating-spin-chain} above. Firstly, in
sub-section~\ref{sec:weakcouplais1eqs}, we will show that taking the
weak-coupling limit of the all-loop Bethe equations for $AdS_3 \times S^3
\times T^4$~\cite{Babichenko:2009dk} gives the same equations as one obtains by
taking the $\alpha\rightarrow 1$ limit of the weak-coupling $\alpha\neq 1$ Bethe
equations discussed in section~\ref{sec:fermionic-duality}. Then, in
sub-section~\ref{sec:psu112spinchain} we construct a homogenous spin-chain which
leads to the $\alpha=1$ weak-coupling Bethe equations of the previous
sub-section. In sub-section~\ref{sec:psu112spinchainasasubchain} we discuss how
this homogenous spin-chain arrises from the $\alpha\rightarrow 1$ limit of
the alternating chain constructed in
section~\ref{sec:alternating-spin-chain}. We show that the integrable structure
underlying the alternating spin-chain constructed in
section~\ref{sec:alternating-spin-chain} remains finite in the
$\alpha\rightarrow 1$ limit, and argue that the ``missing massless states''
puzzle mentioned in the introduction can be resolved by a careful analysis of
this limit.

\subsection{The weakly-coupled $\alpha=1$ Bethe equations}
\label{sec:weakcouplais1eqs}

To find the Bethe equations for $\alpha=1$, we start with the dualized form of the
equations in \eqref{eq:BE-d21a-one-loop-dual}. To get the conventional
notation for the Dynkin labels of $\algPSU(1,1|2)$ we exchange the
labels 1 and 2. Setting $\alpha = 1$ and rescaling the Bethe roots by
a factor $2$, the equations read
\begin{subequations}\label{eq:BE-su22-fff}
  \begin{align}
    1 &= 
    \prod_{k=1}^{K_2} \frac{\tilde{u}_{1,i} - u_{2,k} + \tfrac{i}{2}}{\tilde{u}_{1,i} - u_{2,k} - \tfrac{i}{2}} \,, \\
    \left(\frac{u_{2,i} + \tfrac{i}{2}}{u_{2,i} - \tfrac{i}{2}}\right)^L &= 
    \prod_{k=1}^{\tilde{K}_1} \frac{u_{2,i} - \tilde{u}_{1,k} + \tfrac{i}{2}}{u_{2,i} - \tilde{u}_{1,k} - \tfrac{i}{2}} 
    \prod_{k=1}^{K_3} \frac{u_{2,i} - u_{3,k} - \tfrac{i}{2}}{u_{2,i} - u_{2,k} + \tfrac{i}{2}} \,, \\
    1 &=
    \prod_{k=1}^{K_2} \frac{u_{3,i} - u_{2,k} - \tfrac{i}{2}}{u_{3,i} - u_{2,k} + \tfrac{i}{2}} \,.
  \end{align}
\end{subequations}
This is the standard Bethe equations for a nearest-neighbor spin-chain
in the $\algPSU(1,1|2)$ representation with Dynkin labels $[0;1;0]$,
and the Dynkin diagram in figure~\ref{fig:dynkin-su22-fff}. By
performing a fermionic duality transformation on either of the outer
nodes of the diagram we can make either the compact $\algSU(2)$ sector
or the non-compact $\algSL(2)$ sector manifest. In the first case,
corresponding to the diagram figure~\ref{fig:dynkin-su22-su} the Bethe
equations read
\begin{subequations}\label{eq:BE-su22-su2}
  \begin{align}
    1 &= \prod_{k=1}^{K_2} \frac{u_{1,i} - u_{2,k} - \tfrac{i}{2}}{u_{1,i} - u_{2,k} + \tfrac{i}{2}} \,, \\
    \left( \frac{u_{2,i} + \tfrac{i}{2}}{u_{2,i} - \tfrac{i}{2}} \right)^L &=
    \prod_{\substack{k = 1\\k \neq i}}^{K_2} \frac{u_{2,i} - u_{2,k} + i}{u_{2,i} - u_{2,k} - i} \prod_{k=1}^{K_1} \frac{u_{2,i} - u_{1,k} - \tfrac{i}{2}}{u_{2,i} - u_{1,k} + \tfrac{i}{2}} \prod_{k=1}^{K_3} \frac{u_{2,i} - u_{3,k} - \tfrac{i}{2}}{u_{2,i} - u_{3,k} + \tfrac{i}{2}} \,, \\
    1 &= \prod_{k=1}^{K_2} \frac{u_{3,i} - u_{2,k} - \tfrac{i}{2}}{u_{3,i} - u_{2,k} + \tfrac{i}{2}} \,,
  \end{align}
\end{subequations}
while the equations corresponding to figure~\ref{fig:dynkin-su22-sl} are
\begin{subequations}\label{eq:BE-su22-sl2}
  \begin{align}
    1 &= \prod_{k=1}^{K_2} \frac{\tilde{u}_{1i} - u_{2,k} + \tfrac{i}{2}}{\tilde{u}_{1i} - u_{2,k} - \tfrac{i}{2}} \,, \\
    \left( \frac{u_{2,i} + \tfrac{i}{2}}{u_{2,i} - \tfrac{i}{2}} \right)^L &=
    \prod_{\substack{k = 1\\k \neq i}}^{K_2} \frac{u_{2,i} - u_{2,k} - i}{u_{2,i} - u_{2,k} + i} \prod_{k=1}^{\tilde{K}_1} \frac{u_{2,i} - \tilde{u}_{1k} + \tfrac{i}{2}}{u_{2,i} - \tilde{u}_{1k} - \tfrac{i}{2}} \prod_{k=1}^{\tilde{K}_3} \frac{u_{2,i} - \tilde{u}_{3k} + \tfrac{i}{2}}{u_{2,i} - \tilde{u}_{3k} - \tfrac{i}{2}} \,, \\
    1 &= \prod_{k=1}^{K_2} \frac{\tilde{u}_{3i} - u_{2,k} + \tfrac{i}{2}}{\tilde{u}_{3i} - u_{2,k} - \tfrac{i}{2}} \,.
  \end{align}
\end{subequations}
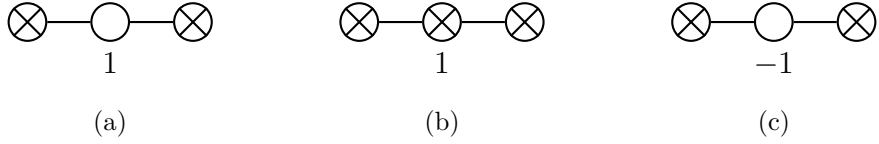
\begin{figure}
  \centering

  \subfloat[\label{fig:dynkin-su22-su}]{
    \begin{tikzpicture}
      [
      thick,
      node/.style={shape=circle,draw,thick,inner sep=0pt,minimum size=5mm}
      ]

      \useasboundingbox (-1.5cm,-1cm) rectangle (1.5cm,1cm);

      \node (v1) at (-1.1cm, 0cm) [node] {};
      \node (v2) at (  0.0cm, 0cm) [node] {};
      \node (v3) at (  1.1cm, 0cm) [node] {};

      \draw (v1.south west) -- (v1.north east);
      \draw (v1.north west) -- (v1.south east);

      \draw (v3.south west) -- (v3.north east);
      \draw (v3.north west) -- (v3.south east);

      \draw (v1) -- (v2);
      \draw (v2) -- (v3);

      \node at (v2.south) [anchor=north] {$1$};
    \end{tikzpicture}
  }
  \hspace{1cm}
  \subfloat[\label{fig:dynkin-su22-fff}]{
    \begin{tikzpicture}
      [
      thick,
      node/.style={shape=circle,draw,thick,inner sep=0pt,minimum size=5mm}
      ]

      \useasboundingbox (-1.5cm,-1cm) rectangle (1.5cm,1cm);

      \node (v1) at (-1.1cm, 0cm) [node] {};
      \node (v2) at (  0.0cm, 0cm) [node] {};
      \node (v3) at (  1.1cm, 0cm) [node] {};

      \draw (v1.south west) -- (v1.north east);
      \draw (v1.north west) -- (v1.south east);

      \draw (v2.south west) -- (v2.north east);
      \draw (v2.north west) -- (v2.south east);

      \draw (v3.south west) -- (v3.north east);
      \draw (v3.north west) -- (v3.south east);

      \draw (v1) -- (v2);
      \draw (v2) -- (v3);

      \node at (v2.south) [anchor=north] {$1$};
    \end{tikzpicture}
  }
  \hspace{1cm}
  \subfloat[\label{fig:dynkin-su22-sl}]{
    \begin{tikzpicture}
      [
      thick,
      node/.style={shape=circle,draw,thick,inner sep=0pt,minimum size=5mm}
      ]

      \useasboundingbox (-1.5cm,-1cm) rectangle (1.5cm,1cm);

      \node (v1) at (-1.1cm, 0cm) [node] {};
      \node (v2) at (  0.0cm, 0cm) [node] {};
      \node (v3) at (  1.1cm, 0cm) [node] {};

      \draw (v1.south west) -- (v1.north east);
      \draw (v1.north west) -- (v1.south east);

      \draw (v3.south west) -- (v3.north east);
      \draw (v3.north west) -- (v3.south east);

      \draw (v1) -- (v2);
      \draw (v2) -- (v3);

      \node at (v2.south) [anchor=north] {$-1$};
    \end{tikzpicture}
  }
  
  \caption{Three Dynkin diagrams for $\algPSU(1,1|2)$.}
\end{figure}

In~\cite{Babichenko:2009dk} a set of all-loop Bethe equations for $AdS_3 \times
S^3 \times T^4$ were proposed. From these we readily obtain the equations
presented above by taking a weak coupling limit of the left-moving sector of the
full equations in a very similar way to the procedure discussed in
section~\ref{sec:BA-d21a-one-loop}. As discussed in
section~\ref{sec:full-spin-chain}, the complete model has another set of
identical equations describing the right-moving sector, with the two sectors
being coupled only through the momentum constraint given in
equation~\eqref{eq:momconstr}. From this analysis, we are lead to conclude that
taking the $\alpha\rightarrow 1$ limit commutes with taking the weak-coupling
limit in the full Bethe ansatz. This was not guaranteed \emph{a priori}, and
suggests that investigating the $\alpha\rightarrow 1$ limit at weak-coupling is
a physically meaningful procedure.

The Bethe equations derived above have appeared
in~\cite{Beisert:2005fw}, in the context of the $\algPSU(1,1|2)$
sector of $\superN=4$ SYM. We are then lead to the natural result
that the integrable spin-chain which gives rise to the Bethe
equations~\eqref{eq:BE-su22-sl2} or~\eqref{eq:BE-su22-su2} is of the
homogenous type. A confirmation of this observation can be found by
noting that, unlike the Bethe equations for $\alpha\neq
1$~\eqref{eq:BE-d21a-one-loop}, these Bethe equations do not have any
$L=1$ solutions. In the next sub-section, we will discuss the
spin-chain which leads to the Bethe equations~\eqref{eq:BE-su22-sl2}
and~\eqref{eq:BE-su22-su2}.

\subsection{The integrable $\algPSU(1,1|2)$ spin-chain}
\label{sec:psu112spinchain}

In this sub-section we construct the R-matrix and hamiltonian for a
spin-chain which can be solved by the $\algPSU(1,1|2)$ BA
equations~\eqref{eq:BE-su22-sl2} and~\eqref{eq:BE-su22-su2}. As
discussed at the end of the previous sub-section these equations
correspond to a \emph{non-alternating}, homogenous, spin-chain with
$\algPSU(1,1|2)$ symmetry. From the Bethe ansatz equations of the
previous sub-section we find that the weights of the $\algPSU(1,1|2)$
representations at each site are
$(-\tfrac{1}{2};\tfrac{1}{2})$.\footnote{%
  We use the notation $(p;q)$ to label a $\algPSU(1,1|2)$ module, with
  $p$ and $q$ being the highest weights of the bosonic sub-algebras
  $\algSU(1,1)$ and $\algSU(2)$.%
} %
This is a short representation, and we refer the reader to
appendix~\ref{sec:su112-algebra}, where we present an explicit
realisation for it.

To construct the R-matrix for this spin-chain we make two observations. Firstly, note the decomposition into irreducible representations of the tensor product of two $(-\tfrac{1}{2};\tfrac{1}{2})$ $\algPSU(1,1|2)$ modules
\begin{equation}
  \label{eq:su22-tensor-decomp}
  (-\tfrac{1}{2};\tfrac{1}{2}) \directprod (-\tfrac{1}{2};\tfrac{1}{2})
  =  (-1;1)\oplus \bigoplus_{j \ge 0} (-1-j;0) \,.
\end{equation}
Secondly, in the construction we will also use the embedding of the maximal sub-algebra $\algSL(2|1)$ into $\algPSU(1,1|2)$. The embedding in terms of generators is obtained by setting $\alpha=1$ in equation~\eqref{eq:sl21embedding}. Under this embedding the $\algSL(2|1)$ module $(\tfrac{1}{2},-\tfrac{1}{2})$ is a submodule of the $\algPSU(1,1|2)$ module $(-\tfrac{1}{2};\tfrac{1}{2})$. The decomposition of the tensor product of these sub-modules can be obtained by setting $\alpha=1$ in equation~\eqref{eq:sl21-tensor-ac-ac} to get
\begin{equation}
\label{eq:sl21-tensor-su22}
  \left(\tfrac{1}{2},-\tfrac{1}{2}\right) \directprod \left(\tfrac{1}{2},-\tfrac{1}{2}\right) = 
  (1,-1) \oplus\bigoplus_{n=1}^{\infty} \left( \tfrac{1}{2} + n, -\tfrac{1}{2} \right) \,.
\end{equation}
The R-matrix can be constructed using the universal
R-matrix~\cite{Derkachov:2000ne}. It has the general form given in
equation~\eqref{eq:rmatrixasproj}. This R-matrix is just the $R_{--}(u)$
R-matrix of section~\ref{sec:sl21subsector} evaluated with $\alpha=1$: the
coefficients in front of the projection operators are given in
equation~\eqref{eq:SL21-R-matrix-mm} evaluated at $\alpha=1$. The transfer
matrix is then the conventional one for a homogenous spin-chain
\begin{align}
  T_a(u) &= R_{a a_1}\!(u) \, R_{a a_2}\!(u) \dotsm R_{a a_L}\!(u) \,.
\end{align}
Taking the traces over the auxiliary space, which is also taken in the $(\tfrac{1}{2}\,,\,-\tfrac{1}{2})$ representation of $\algSL(2|1)$, we define
\begin{equation}
  \tau(u) = \tr_a T_a(u) \,.
\end{equation}
The Hamiltonian is given by
\begin{equation}
  H = C (\tau(0) )^{-1}  \left. \frac{d}{du} \tau(u)  \right|_{u=0} \,,
\end{equation}
where $C$ is a normalization constant. Comparing the tensor product
decompositions of the $\algSL(2|1)$ modules given in
equation~\eqref{eq:sl21-tensor-su22} with those of the
$\algPSU(1,1|2)$ modules given in
equation~\eqref{eq:su22-tensor-decomp} we see that there is an
isomorphism between the irreducible representations of the two
decompositions. As a result, just as in
section~\ref{sec:lift-to-d21a}, we can uniquely lift the $\algSL(2|1)$
R-matrix and Hamiltonian to a $\algPSU(1,1|2)$ R-matrix and
Hamiltonian. As expected, the resulting $\algPSU(1,1|2)$ Hamiltonian
is precisely the one studied in~\cite{Beisert:2007sk}.

\subsection{The $\algPSU(1,1|2)$ spin-chain from the $\algD{\alpha}$ spin-chain}
\label{sec:psu112spinchainasasubchain}

In this subsection we collect some observations about the
$\alpha\rightarrow 1$ limit.\footnote{Similar comments apply to the
  equivalent $\alpha\rightarrow 0$ limit.}  In this limit the algebra
changes from $\algD{\alpha}$ to $\algPSU(1,1|2)$. One key feature of
this limit is that the number of massless BMN states
changes~\cite{Babichenko:2009dk}. Recall that for generic $\alpha$ the
BMN states consist of $2+2+2=6$ massive bosons and two massless bosons
as well as their fermionic superpartners. The massive bosons have
masses squared proportional to $1$, $\alpha$ and $1-\alpha$. As
$\alpha\rightarrow 1$, two bosons and two fermions become massless. We
will argue below that understanding the $\alpha\rightarrow 1$ limit is
intimately related to understanding the way the massless modes should
enter the $\AdS_3/\CFT_2$ correspondence.

\subsubsection{The $\alpha\rightarrow 1$ limit in the BA equations}

Consider first taking the $\alpha\rightarrow 1$ limit in the BA
equations. As we saw in section~\ref{sec:weakcouplais1eqs} above, it
is possible to write down BA equations for which this limit can be
taken smoothly. From this we can conclude that any solution of the
$\alpha=1$ BA equations~\eqref{eq:BE-su22-fff} can be uplifted to a
solution of the $\alpha\neq 1$ BA
equations~\eqref{eq:BE-d21a-one-loop}.
However, this does not mean that all solutions of the $\alpha\neq 1$
BA equations~\eqref{eq:BE-d21a-one-loop} map smoothly to solutions of
the $\alpha=1$ BA equations~\eqref{eq:BE-su22-fff}. In
section~\ref{sec:BAalpha1limit} below, we illustrate this by
considering some simple solutions which lie in a compact $\algSU(2)$
subsector of the $\algD{\alpha}$ BA~\eqref{eq:BE-d21a-one-loop} and
investigate their $\alpha\rightarrow 1$ limit.  We observe that the
energies of some solutions in this sector diverge, while other
solutions' energies remain finite. The presence of solutions to the
BA~\eqref{eq:BE-d21a-one-loop-dual} whose energies diverge in this
limit shows that the $\alpha=1$ BA~\eqref{eq:BE-su22-fff} ``loses'' some
of the states from the $\alpha\neq 1$ BA.

This divergent behaviour can be expected on general grounds. To see this we note that the magnon dispersion relation of the $u_{3,i}$ Bethe roots is
given by
\begin{equation}
  \epsilon_3(p) = \sqrt{1-\alpha + 4h^2 \sin^2\frac{p}{2}} \,,
\end{equation}
In the weak coupling limit, we expand this as
\begin{equation}
  \epsilon_3(p) = \sqrt{1-\alpha} + \frac{2h^2}{\sqrt{1-\alpha}} \sin^2 \frac{p}{2} + \order(h^4) \,.
\end{equation}
When $\alpha \to 1$, the $u_{3,i}$ excitations become massless, and this
expansion is not valid anymore. Instead
\begin{equation}
  \epsilon_3(p) \to 2 h \left\lvert \sin \frac{p}{2} \right\rvert \,.
\end{equation}
However, the weak coupling Bethe equations, and the corresponding spin-chain
Hamiltonian derived in section~\ref{sec:alternating-spin-chain} gives the
leading correction to the energy for generic $\alpha$, and is therefore
proportional to $h^2$. Hence, the divergences appearing in the energies in the
$\alpha \to 1$ limit is an indication of the presence of extra massless modes.

What is more, as we see in some examples in section~\ref{sec:BAalpha1limit}
below, the BA solutions with divergent terms will generically have, in the
$\alpha\rightarrow 1$ limit, coincident roots. Such behaviour is not allowed in
the conventional BA~\cite{Reshetikhin:1983vw, Reshetikhin:1986vd}. We take it as
a strong hint that one should be able to construct, at $\alpha=1$, a
``generalised'' BA which allows for such coincident roots, and in this way
captures the ``lost'' states described above. We expect this generalised ansatz to
contain useful information about the massless states which constitute some of
the ``lost'' states mentioned above.

Naively, one might think that the massless modes in the $AdS_3 \times S^3 \times
T^4$ theory would be decoupled from the other degrees of freedom, interacting
perhaps at most via the level-matching condition. While it is true that the
bosons on $T^4$ are free, in the GS formulation the fermions are 10d spinors. As
such, reduced on $AdS_3 \times S^3 \times T^4$ they will transform as
tri-spinors of the three components of this spacetime. As a result, once we fix
kappa gauge, there will not be any fermions that decouple from the $AdS_3 \times
S^3$ directions. World-sheet superconformal invariance then implies that the
bosons of $T^4$ will also have to have a non-trivial coupling to the $AdS_3
\times S^3$ directions.

\subsubsection{The $\alpha\rightarrow 1$ limit in the integrable spin-chain}

From the above discussion it should be clear that the missing
description of massless states is related to the subtleties of the
$\alpha\rightarrow 1$ limit. As we have argued above, in general the
spin-chain Hamiltonian will diverge in this limit. The simplest way to
see this is to consider the $\algSU(2)\times \algSU(2)$ subsector
discussed in section~\ref{sec:su2-su2-sector} -- the Hamiltonian is
clearly divergent. One may expect that this divergence comes from an
order of limits problem between taking the weak-coupling limit and the
$\alpha\rightarrow 1$ limit. In general, this would prevent us from
extracting useful information about massless modes in the weak
coupling limit. However, the system we are considering is far from
generic -- it is in fact integrable. In this sub-section we point out
that the underlying integrable structure of the spin-chain remains
\emph{finite} in the $\alpha\rightarrow 1$ limit. This indicates that
a more detailed analysis of the spin-chain in this limit should yield
exact information about the missing massless states of the
$\AdS_3/\CFT_2$ correspondence.

Let us begin by relating the alternating $\alpha\neq 1$ spin-chain constructed
in section~\ref{sec:alternating-spin-chain} to the homogenous $\alpha=1$
spin-chain constructed in section~\ref{sec:psu112spinchain}.  In the alternating
$\algD{\alpha}$ spin-chain, the odd and even sites transform in the
$(-\frac{\alpha}{2} ; \frac{1}{2} ; 0)$ and $(-\frac{1-\alpha}{2} ; 0 ;
\frac{1}{2})$ representations, respectively. In
appendix~\ref{sec:su112-algebra}, we study the $\alpha \to 1$ limit of these
representations. The first representation, describing the odd sites of the
spin-chain, turns into the $(-\frac{1}{2} ; \frac{1}{2})$ spin representation of
$\algPSU(1,1|2)$, while the second representation, which sits at even site,
becomes \emph{reducible}, decomposing into a pair of singlet states and a
$(-\frac{1}{2} ; \frac{1}{2})$ module whose heighest weight state is fermionic
in the original grading. The $\algPSU(1,1|2)$ Bethe equations and corresponding
spin-chain only describe operators in which all even sites contain the singlet
state $\bar{\phi}_+$. Indeed one can always replace a homogenous spin-chain
by an alternating spin-chain where the extra sites are just singlets of the
underlying global symmetry. The energies of the two spin-chains and the BA
equations will be indistinguishable.

Having identified the homogenous spin-chain constructed in
section~\ref{sec:psu112spinchain} as a subsector of the $\alpha\rightarrow 1$
limit of the alternating spin-chain constructed in
section~\ref{sec:alternating-spin-chain}, we now turn to the integrable
structure's behaviour in the $\alpha\rightarrow 1$ limit. The full
$\algD{\alpha}$ spin-chain actually contains more information about the $\alpha
\to 1$ spectrum than the equations~\eqref{eq:BE-su22-fff}. To see this, let us
consider the $\alpha \to 1$ limit of the $\algSL(2|1)$ R-matrix:
\begin{align}
  \tilde{R}_{--}^{n^-}(u) &= \prod_{k=0}^{n-1} \frac{u + 2(k+1)}{u - 2(k+1)} \,, &
  \tilde{R}_{++}^{n^+}(u) &= \prod_{k=0}^{n-1} \frac{u + 2k}{u - 2k} \,, &
  \tilde{R}_{\mp\pm}^{n}(u) &=\prod_{k=0}^n \frac{u + 2k}{u - 2k} \,.
\end{align}
We use the same notation for the states appearing in the tensor product as in
section~\ref{sec:sl21subsector}, though some of the projectors now project onto
reducible representations. 

The key observation is that while the Hamiltonian of the alternating spin-chain
in the $\alpha\rightarrow 1$ limit is divergent, the R-matrix remains
\emph{finite} and \emph{non-trivial}. This indicates that in order to construct
the complete Hamiltonian for the $\alpha\rightarrow 1$ limit of the alternating
spin-chain, one needs to start with the above R-matrix, which is well defined in
this limit, and construct the transfer matrix and Hamiltonian from it. This
Hamiltonian will ``know'' about the massless states appearing in the
$\alpha\rightarrow 1$ limit. From the finite R-matrix above, one should also be
able to construct a set of ``generalised'' BA equations for the $\alpha=1$
theory which also ``know'' about the massless modes. We are pursuing these
directions presently and hope to report more fully on these developments in the
near future~\cite{toappear}.\footnote{%
  The above procedure will tell us how to incorporate two massless bosonic modes
  (and their superpartners) into the integrability machinery. The $AdS_3 \times
  S^3 \times T^4$ model has four identical massless bosonic modes (and their
  superpartners). We strongly suspect that once we know how to incorporate two
  massless modes into the integrability machinery, we will also be able to
  incorporate the other two.%
}

\section{Solutions to the Bethe equations}
\label{sec:BA-solutions}

In this section we collect some solutions to the Bethe equations
presented in the paper.

\subsection{The $\alpha \to 1$ limit}
\label{sec:BAalpha1limit}

We will now consider some solutions in the $\alpha \to 1$ limit. The
simplest case is to consider the $\grSU(2)$ that remains
compact. Setting $L = 4$, $K_1 = 2$ and $K_2 = K_3 = 0$, and applying
the momentum constraint, we have a single solution
\begin{align}
  u_{1,1} &= \frac{\alpha}{\sqrt{3}} \,, &
  u_{1,2} &= -\frac{\alpha}{\sqrt{3}} \,.
\end{align}
This state is in the $(-2;0;2)$ representation of $\algD{\alpha}$ and
the energy is $E = \frac{3}{\alpha} h^2$.

To find more interesting solutions we set $L=4$, $K_1 = K_3 = 1$ and
$K_2 = 0$, corresponding to the $(-2;1;1)$ representation. Then there
are three solutions with zero total momentum, an unpaired solution
\begin{align}
  u_{1,1}^0 &= 0 \,, &
  u_{3,1}^0 &= 0 \,,
  \intertext{and the parity pair}
  u_{1,1}^{\pm} &= \pm \alpha \,, &
  u_{3,1}^{\pm} &= \mp (1 - \alpha) \,.
\end{align}
The energies are
\begin{align}
  E^0 &= \frac{2}{\alpha} h^2 + \frac{2}{1 - \alpha} h^2 \,, &
  E^{\pm} &= \frac{1}{\alpha} h^2 + \frac{1}{1 - \alpha} h^2 \,.
\end{align}
As $\alpha \to 1$ these energies diverge as expected from the discussion in section~\ref{sec:t4-limit}.

\subsection{Twist-one operators}
\label{sec:twist-one}

We will now consider the $\algSL(2)$ sector and set $L=1$. We will
work with the dualized Bethe equations
\eqref{eq:BE-d21a-one-loop-dual}. For the state $(-K ; 0 ; 0)$ the
excitation numbers are $K_1 = K_3 = K$, $\tilde{K }_2 = 0$. The
equations we want to solve are
\begin{subequations}
  \label{eq:twist-1-BE}
  \begin{align}
    \frac{u_{1,i} + i\alpha}{u_{1,i} - i\alpha} &= 
    \prod_{j \neq i}^{K} \frac{u_{1,i} - u_{3,j} - i}{u_{1,i} - u_{3,j} + i} \,, \\
    \frac{u_{3,i} + i(1-\alpha)}{u_{3,i} - i(1-\alpha)} &= 
    \prod_{j \neq i}^{K} \frac{u_{3,i} - u_{1,j} - i}{u_{3,i} - u_{1,j} + i} \,.
  \end{align}
\end{subequations}
To find solutions to these equations we introduce the Baxter
polynomials~\cite{Korchemsky:1995be,Eden:2006rx,Zwiebel:2009vb}
\begin{align}
  Q_{1K}(u) &= c_{1K} \prod_{k=1} (u - u_{1,k} ) \,, &
  Q_{3K}(u) &= c_{3K} \prod_{k=1} (u - u_{3,k} ) \,,
\end{align}
where $c_{1K}$ and $c_{3K}$ are irrelevant normalization
constants. The Bethe equations \eqref{eq:twist-1-BE} can then be
rewritten as two coupled difference equations for the polynomials
$Q_{1K}$ and $Q_{3K}$,
\begin{subequations} \label{eq:twist-1-diff-eqs}
  \begin{align}
    \label{eq:twist-1-diff-eq-1}
    T_{1K} Q_{1K}(u) &= \left(u + i\alpha\right) Q_{3K}(u + i) - \left(u - i\alpha\right) Q_{3K}(u-i) \,, \\
    \label{eq:twist-1-diff-eq-3}
    T_{3K} Q_{3K}(u) &= \left(u + i(1-\alpha)\right) Q_{1K}(u + i) - \left(u - i(1-\alpha)\right) Q_{1K}(u - i) \,.
  \end{align}
\end{subequations}
In general $T_{1K}$ and $T_{3K}$ would be functions of $u$, but
comparing powers of $u$ on the two sides of the equations we find that
here $T_{1K}$ and $T_{3K}$ are independent of $u$.

The energy of a state can be calculated from the polynomials $Q_{1K}$ and $Q_{3K}$ as
\begin{equation}
  E_K = 2i h^2 \left. \frac{d}{du} \log Q_{1K}(u) \right|_{u = i\alpha} + 2ih^2 \left. \frac{d}{du} \log Q_{3K}(u) \right|_{u = i(1-\alpha)} \,.
\end{equation}

For $\alpha=1/2$ we can set $Q_{1K}(u) = Q_{3K}(u) = Q_K(u)$. The resulting
equation has a solution in terms of the Meixner
polynomials~\cite{Zwiebel:2009vb}
\begin{align} \label{eq:twist-1-diff-eq-sol-a-1}
  Q_{K}(u) &= {}_2F_1(-K, iu+\tfrac{1}{2}; 1; 2) \,, &
  T_K &= (2K+1)i \,,
\end{align}
and gives the energy
\begin{equation}
  E_K^{(\alpha=1/2)} = 8 h^2 ( S_1(K) - S_{-1}(K)) \,,
\end{equation}
where $S_a(M)$ is the harmonic sum
\begin{equation}
  S_a(M) = \sum_{j=1}^M \frac{(\operatorname{sgn} a)^j}{j^{|a|}} \,.
\end{equation}

For general $\alpha$ we need to solve for both polynomials $Q_{1K}$
and $Q_{2K}$. For a solution to exist, the constants $T_{lK}$ have to
satisfy
\begin{align}
  T_{1K} &= 2i(K + \alpha) \,, &
  T_{3K} &= 2i(K + 1 - \alpha) \,.
\end{align}
Using \eqref{eq:twist-1-diff-eq-1} to eliminate the $Q_{1K}$ dependence in
\eqref{eq:twist-1-diff-eq-3} and introducing
\begin{align}
  v &= -\frac{i u}{2} \,, &
  R_K(v) &= Q_{3K}(2 i v) \,.
\end{align}
we get the equation
\begin{equation}
  \begin{split}
    \left( \tfrac{\alpha^2-1}{4} - v^2 \right) &\Big( R_K(v+1) - 2R_K(v) + R_K(v-1) \Big) \\ - v&\Big( R_K(v+1) - R_K(v-1) \Big) + K(K+1) R_K(v) = 0\,.
  \end{split}
\end{equation}
This is a discrete form of the Legendre differential equation
\begin{equation}
  \left( \tfrac{\alpha^2-1}{4} - v^2\right) R_n''(v) - 2v R_n'(v) + n(n+1) R_n(v) = 0 \,.
\end{equation}
From this equation we can derive a recurrence relation for $R_K$~\cite{Lorente:2001}
\begin{equation}
  R_{K+1}(v) - \frac{K^2}{16} \frac{K^2 - \alpha^2}{K^2 - \frac{1}{4}} R_{K-1}(v) - v R_K(v) = 0 \,.
\end{equation}
We have not been able to find a closed form solution of this equation,
but it is straight forward to find solutions for any $K$.  The first
few solutions for $Q_{3K}$ are
\begin{align}
  Q_{31}(u) &= -\frac{i}{2} u \,, \\
  Q_{32}(u) &= -\frac{1}{4} u^2 + \frac{1 - \alpha}{12} \,, \\
  Q_{33}(u) &= \frac{i}{8} u^3 + \frac{i(3\alpha^2 - 7)}{40} u \,, \\
  Q_{34}(u) &= \frac{1}{16} u^4 + \frac{3\alpha^2 - 13}{56} u^2 + \frac{3(1 - \alpha^2)(9 - \alpha^2)}{560} \,.
\end{align}
The $Q_{1K}$ polynomials are obtained from $Q_{3K}$ by substituting $\alpha \to
1-\alpha$.  The corresponding energies are
\begin{align}
  E_1 &= 2\left(\frac{1}{\alpha} + \frac{1}{1-\alpha}\right) \,, \\
  E_2 &= 6\left(\frac{1}{1+\alpha} + \frac{1}{2-\alpha}\right) \,, \\
  E_3 &= 2\left(\frac{1}{\alpha} + \frac{1}{1-\alpha}\right) - 10\left(\frac{1}{1+\alpha} + \frac{1}{2-\alpha}\right) 
    + 20\left(\frac{1}{2+\alpha} + \frac{1}{3-\alpha}\right) \,, \\
  E_4 &= 20\left(\frac{1}{1+\alpha} + \frac{1}{2-\alpha}\right) - 70\left(\frac{1}{2+\alpha} + \frac{1}{3-\alpha}\right)
    + 70\left(\frac{1}{3+\alpha} + \frac{1}{4-\alpha}\right) \,.
\end{align}
We note that these energies are of the form
\begin{equation}
  E_K = \sum_{m=1}^K c_m \left( \frac{1}{(m - 1) + \alpha} + \frac{1}{m - \alpha} \right) \,, \qquad c_m \in \Integers \,,
\end{equation}
with $c_1 = 0$ for even $K$. Hence the energies of the odd states diverge as $\alpha$ approaches $0$ or $1$.

\subsection{Degeneracies in the spectrum}

Let us a ssume that we have a solution to the Bethe equations \eqref{eq:BE-d21a-one-loop} where one of the $u_{2,k}$ roots sits at zero:
\begin{equation}
  u_{2,K_2} = 0 \,,
\end{equation}
and with total momentum
\begin{equation}
  \prod_{i=1}^{K_1} \frac{u_{2,i} + i\alpha}{u_{2,i} - i\alpha}
  \prod_{i=1}^{K_3} \frac{u_{2,i} + i(1-\alpha)}{u_{2,i} - i(1-\alpha)} = 1 \,.
\end{equation}
The equations for the roots at the first and third nodes then read
\begin{subequations}
\begin{align}
  \left( \frac{u_{1,i} + i\alpha}{u_{1,i} - i\alpha} \right)^{L+1} &= 
  \prod_{\substack{k=1 \\ k \neq i}}^{K_1} \frac{u_{1,i} - u_{1,k} + 2i\alpha}{u_{1,i} - u_{1,k} - 2i\alpha}
  \prod_{k=1}^{K_2-1} \frac{u_{1,i} - u_{2,k} - i\alpha}{u_{1,i} - u_{2,k} + i\alpha} \,, \\
   \left( \frac{u_{3,i} + i(1-\alpha)}{u_{3,i} - i(1-\alpha)} \right)^{L+1} &= 
   \prod_{\substack{k=1 \\ k \neq i}}^{K_3} \frac{u_{3,i} - u_{3,k} + 2i(1-\alpha)}{u_{3,i} - u_{3,k} - 2i(1-\alpha)}
   \prod_{k=1}^{K_2-1} \frac{u_{3,i} - u_{2,k} - i(1-\alpha)}{u_{3,i} - u_{2,k} + i(1-\alpha)} \,.\\
\intertext{and the equation for $u_{2,K_2}$ becomes}
  1 &= 
   \prod_{k=1}^{K_1} \frac{u_{2,k} + i\alpha}{u_{2,k} - i\alpha}
   \prod_{k=1}^{K_3} \frac{u_{2,k} + i(1-\alpha)}{u_{2,k} - i(1-\alpha)} \,.
 \end{align}
\end{subequations}
The last equation is automatically fulfilled since we assumed the solution to
satisfy the momentum condition. Hence the resulting equations are the same as
what we would get for a state with the same $\algD{\alpha}$ charges but with the
length increased by one and the excitation number of the middle node lowered
by one.

The same degeneracy is seen in the $\algOSp(4|2)$ sector of
ABJM~\cite{Zwiebel:2009vb}. However, in that case the two degenerate states are
actually part of the same representation of the full $\algOSp(6|4)$ algebra. It
would be interesting to undertand the origin of the degeneracy in the spectrum
of the $\algD{\alpha}$ spin-chain.

\section{Conclusions}
\label{sec:conclusions}

In this paper we have investagated the $\AdS_3/\CFT_2$ correspondences with 16
supersymmetries using integrable techniques. There are two classes of such dual
pairs: those with $\algPSU(1,1|2)^2$ symmetry and those with $\algD{\alpha}^2$
symmetry, corresponding to $AdS_3\times S^3\times T^4$ and $AdS_3\times
S^3\times S^3\times S^1$ spacetimes, respectively. We have presented a set of
all-loop BA equations valid for all values of the parameter $\alpha$,
generalising the $\alpha=1/2$ BA given in~\cite{Babichenko:2009dk}, resolving
one of the puzzles left open in that paper.

In the remainder of the paper we have considered the weakly coupled limit of the
$\algPSU(1,1|2)^2$ and $\algD{\alpha}^2$ BA equations. This is analogous to the
perturbative gauge theory limit in the $\AdS_5/\CFT_4$ correspondence. We have
constructed integrable spin-chains with local interactions whose energies
reproduce both sets of weakly coupled BA equations. In
section~\ref{sec:t4-limit} we addressed the second unresolved puzzle mentioned
in the introduction: incorporating massless modes into the integrability
description of the gauge/string correspondence. Recall that the $AdS_3\times
S^3\times T^4$ BA is missing four massless bosonic modes (and four fermionic
superpartners). On the other hand, the $AdS_3\times S^3\times S^3\times S^1$ BA
is missing only two massless bosons (and their superpartners). In the
$\alpha\rightarrow 1$ limit the Hamiltonian of the alternating spin-chain and
certain solutions of the weakly-coupled $\algD{\alpha}^2$ BA diverge. We argue
that these divergences signal the appearance of massless modes in the
$\alpha\rightarrow 1$ limit. These massless states are precisely the two
\emph{extra} massless bosons (and their superpartners) that the $AdS_3\times
S^3\times T^4$ BA is missing compared to the $AdS_3\times S^3\times S^3\times
S^1$ BA.

In general, it might be very difficult to retain control over such divergences
coming from extra massless states appearing as one varies a parameter in the
theory. If that were the case, there would be little hope for understanding the
massless modes by investigating the $\alpha\rightarrow 1$ limit. However, the
theory we are considering is integrable; and while the Hamiltonian and certain
solutions of the BA diverge in this limit, we show that the integrable structure
underlying the alternating $\algD{\alpha}^2$ spin-chain remains
\emph{non-singular} in the $\alpha\rightarrow 1$ limit. This leads us to posit
that the $\alpha\rightarrow 1$ limit of the alternating $\algD{\alpha}^2$
spin-chain R-matrix that we have constructed in this paper, describes not just
the massive modes coming from the homogenous spin-chain coming from the weakly
coupled $AdS_3\times S^3\times T^4$ BA, but also two of the four missing zero
modes (and their superpartners). We will return to a more detailed study of this
later~\cite{toappear}.

\section*{Acknowledgments}

We would like to thank Joe Chuang, Andreas Fring, Joe Minahan, Alessandro
Torrielli, Arkady Tseytlin and Kostya Zarembo for interesting discussions
throughout this project and comments on the manuscript. We would like to thanks
Charlotte Kristjansen, the organisers of the 27th Nordic Network Meeting on
``Strings, Fields and Branes'' and the Niels Bohr Insititute for hospitality
during a part of this project. O.O.S. was partially supported by a Pump Prime
grant from City University. B.S. is supported by an ARF grant from the EPSRC.

\appendix

\section{Representations of the $\algD{\alpha}$ superalgebra}
\label{sec:d21a-algebra}

We are mostly interested in the short representations
$(-\tfrac{\alpha}{2};\tfrac{1}{2};0)$ and
$(-\tfrac{1-\alpha}{2};0;\tfrac{1}{2})$. The tensor product between these state
decomposes as
\begin{equation}
  \label{eq:d21a-tensor-ac-c}
  (-\tfrac{\alpha}{2};\tfrac{1}{2};0) \directprod (-\tfrac{1-\alpha}{2};0;\tfrac{1}{2})
  =
  (-\tfrac{1}{2};\tfrac{1}{2};\tfrac{1}{2})
  \directsum_{j \ge 0}
  (-1-j;0;0) \,.
\end{equation}
We also need the decompositions
\begin{align}
  \label{eq:d21a-tensor-ac-ac}
  (-\tfrac{\alpha}{2};\tfrac{1}{2};0) \directprod (-\tfrac{\alpha}{2};\tfrac{1}{2};0)
  &=
  (-\alpha;1;0) \directsum_{j \ge 0} (-\alpha-j;0;0) \,, \\
  \label{eq:d21a-tensor-c-c}
  (-\tfrac{1-\alpha}{2};0;\tfrac{1}{2}) \directprod (-\tfrac{1-\alpha}{2};0;\tfrac{1}{2})
  &=
  (-(1-\alpha);0;1) \directsum_{j \ge 0} (-(1-\alpha)-j;0;0) \,.
\end{align}

\subsection{Representations}

The $(-\frac{\alpha}{2} ; \frac{1}{2} ; 0)$ representation consists of the
bosonic states $\ket{\phi_{\alpha}^{(n)}}$, which transform as a doublet under
$\gen{L}_m$, and the fermions $\ket{\psi_{\dot\alpha}^{(n)}}$, transforming as a
doublet under $\gen{R}_{\dot m}$. The non-vanishing action of the
$\algD{\alpha}$ generators on these states are given by
\begin{equation}\label{eq:d21a-chiral-rep-LR}
  \begin{aligned}
    L_5 \ket{\phi_{\pm}^{(n)}} &= \pm \frac{1}{2} \ket{\phi_{\pm}^{(n)}} \,, &
    L_+ \ket{\phi_{-}^{(n)}} &= \ket{\phi_+^{(n)}} \,, &
    L_- \ket{\phi_{+}^{(n)}} &= \ket{\phi_-^{(n)}} \,, \\
    R_8 \ket{\psi_{\pm}^{(n)}} &= \pm \frac{1}{2} \ket{\psi_{\pm}^{(n)}} \,, &
    R_+ \ket{\psi_{-}^{(n)}} &= \ket{\psi_+^{(n)}} \,, &
    R_- \ket{\psi_{+}^{(n)}} &= \ket{\psi_-^{(n)}} \,,
  \end{aligned}
\end{equation}
\begin{equation}\label{eq:d21a-chiral-rep-S}
  \begin{aligned}
    S_0 \ket{\phi_{\alpha}^{(n)}} &= - \left( \frac{\alpha}{2} + n \right) \ket{\phi_{\alpha}^{(n)}} \,, \\
    S_- \ket{\phi_{\alpha}^{(n)}} &= -\sqrt{(n + \alpha)(n + 1)} \ket{\phi_{\alpha}^{(n+1)}} \,, \\
    S_+ \ket{\phi_{\alpha}^{(n)}} &= \sqrt{(n - 1 + \alpha) n} \ket{\phi_{\alpha}^{(n-1)}} \,, \\
    S_0 \ket{\psi_{\dot\alpha}^{(n)}} &= - \left( \frac{\alpha + 1}{2} + n \right) \ket{\psi_{\dot\alpha}^{(n)}} \,, \\
    S_- \ket{\psi_{\dot\alpha}^{(n)}} &= -\sqrt{(n+1)(n + 1+ \alpha)} \ket{\psi_{\dot\alpha}^{(n+1)}} \,, \\
    S_+ \ket{\psi_{\dot\alpha}^{(n)}} &= \sqrt{n (n + \alpha)} \ket{\psi_{\dot\alpha}^{(n-1)}} \,,
  \end{aligned}
\end{equation}
\begin{equation}\label{eq:d21a-chiral-rep-Q}
  \begin{aligned}
    Q_{-\pm\dot\alpha} \ket{\phi_{\mp}^{(n)}} &= \pm \sqrt{n+\alpha} \ket{\psi_{\dot\alpha}^{(n)}} \,, &
    Q_{+\pm\dot\alpha} \ket{\phi_{\mp}^{(n)}} &= \pm \sqrt{n} \ket{\psi_{\dot\alpha}^{(n-1)}} \,, \\
    Q_{-\alpha\pm} \ket{\psi_{\mp}^{(n)}} &= \mp \sqrt{n+1} \ket{\phi_{\alpha}^{(n+1)}} \,, &
    Q_{+\alpha\pm} \ket{\psi_{\mp}^{(n)}} &= \mp \sqrt{n+\alpha} \ket{\phi_{\alpha}^{(n)}} \,.
  \end{aligned}
\end{equation}
It is straightforward to check that the above expressions satisfy the $\algD{\alpha}$ algebra.

To get the $(-\frac{1-\alpha}{2} ; 0 ; \frac{1}{2})$ representation, we
from the above expressions, exchange the $\gen{L}_m$ and
$\gen{R}_{\dot{m}}$ generators and replace $\alpha \to 1 -
\alpha$. The corresponding state are
$\ket{\bar{\phi}_{\dot\alpha}^{(n)}}$ and
$\ket{\bar{\psi}_{\alpha}^{(n)}}$, and the actions of the generators
are
\begin{equation}\label{eq:d21a-anti-chiral-rep-LR}
  \begin{aligned}
    L_5 \ket{\bar{\psi}_{\pm}^{(n)}} &= \pm \frac{1}{2} \ket{\bar{\psi}_{\pm}^{(n)}} \,, &
    L_+ \ket{\bar{\psi}_{-}^{(n)}} &= \ket{\bar{\psi}_+^{(n)}} \,, &
    L_- \ket{\bar{\psi}_{+}^{(n)}} &= \ket{\bar{\psi}_-^{(n)}} \,, \\
    R_8 \ket{\bar{\phi}_{\pm}^{(n)}} &= \pm \frac{1}{2} \ket{\bar{\phi}_{\pm}^{(n)}} \,, &
    R_+ \ket{\bar{\phi}_{-}^{(n)}} &= \ket{\bar{\phi}_+^{(n)}} \,, &
    R_- \ket{\bar{\phi}_{+}^{(n)}} &= \ket{\bar{\phi}_-^{(n)}} \,,
  \end{aligned}
\end{equation}
\begin{equation}\label{eq:d21a-anti-chiral-rep-S}
  \begin{aligned}
    S_0 \ket{\bar{\phi}_{\dot\alpha}^{(n)}} &= - \left( \frac{1-\alpha}{2} + n \right) \ket{\bar{\phi}_{\dot\alpha}^{(n)}} \,, \\
    S_- \ket{\bar{\phi}_{\dot\alpha}^{(n)}} &= -\sqrt{(n + 1 - \alpha)(n + 1)} \ket{\bar{\phi}_{\dot\alpha}^{(n+1)}} \,, \\
    S_+ \ket{\bar{\phi}_{\dot\alpha}^{(n)}} &= \sqrt{(n - \alpha) n} \ket{\bar{\phi}_{\dot\alpha}^{(n-1)}} \,, \\
    S_0 \ket{\bar{\psi}_{\alpha}^{(n)}} &= - \left( \frac{2 - \alpha}{2} + n \right) \ket{\bar{\psi}_{\alpha}^{(n)}} \,, \\
    S_- \ket{\bar{\psi}_{\alpha}^{(n)}} &= -\sqrt{(n+1)(n + 2 - \alpha)} \ket{\bar{\psi}_{\alpha}^{(n+1)}} \,, \\
    S_+ \ket{\bar{\psi}_{\alpha}^{(n)}} &= \sqrt{n (n + 1 - \alpha)} \ket{\bar{\psi}_{\alpha}^{(n-1)}} \,,
  \end{aligned}
\end{equation}
\begin{equation}\label{eq:d21a-anti-chiral-rep-Q}
  \begin{aligned}
    Q_{-\alpha\pm} \ket{\bar{\phi}_{\mp}^{(n)}} &= \pm \sqrt{n+1-\alpha} \ket{\bar{\psi}_{\alpha}^{(n)}} \,, &
    Q_{+\alpha\pm} \ket{\bar{\phi}_{\mp}^{(n)}} &= \pm \sqrt{n} \ket{\bar{\psi}_{\alpha}^{(n-1)}} \,, \\
    Q_{-\pm\dot\alpha} \ket{\bar{\psi}_{\mp}^{(n)}} &= \mp \sqrt{n+1} \ket{\bar{\phi}_{\dot\alpha}^{(n+1)}} \,, &
    Q_{+\pm\dot\alpha} \ket{\bar{\psi}_{\mp}^{(n)}} &= \mp \sqrt{n+1-\alpha} \ket{\bar{\phi}_{\dot\alpha}^{(n)}} \,.
  \end{aligned}
\end{equation}

\section{The $\algPSU(1,1|2)$ superalgebra}
\label{sec:su112-algebra}

We obtain the $\algPSU(1,1|2)$ algebra by taking the $\alpha \to 1$ (or,
equivalently, $\alpha \to 0$) limit of the $\algD{\alpha}$ algebra. However,
depending on how the limit is approach, slightly different versions of the
resulting algebra can be obtained. By comparing with the corresponding coset
sigma model in~\cite{Babichenko:2009dk} we see that to get the relevant limit we
need to rescale the generators $\gen{R}_{\dot{m}} \to \gen{R}_{\dot{m}} /
\sqrt{1-\alpha}$ before letting $\alpha \to 1$. The resulting algebra then
reads\footnote{%
  If we take the limit without rescaling the $\gen{R}_{\dot{m}}$ generators,
  these generators would appear as an $\algSU(2)$ outer automorphism of the
  resulting algebra. Another alternative would be to instead rescale
  $\gen{R}_{\dot{m}} \to \gen{R}_{\dot{m}} / (1-\alpha)$. Then
  $\gen{R}_{\dot{m}}$ would appear as three commuting central charges in the
  anti-commutators of the supercharges
  (see~\cite{Beisert:2005tm,Beisert:2006qh}).}
\begin{equation}
  \label{eq:su112-comm-rel}
  \begin{gathered}
    \begin{aligned}
      \comm{\gen{S}_\mu}{\gen{S}_\nu} &= i \, \epsilon_{\mu\nu\rho} \, \gen{S}^\rho \,, & \qquad
      \comm{\gen{S}_\mu}{\gen{Q}_{a\alpha\dot{\alpha}}} &= \frac{1}{2} \gen{Q}_{b\alpha\dot{\alpha}} \gamma^b_{\mu\,a} \,, \\
      \comm{\gen{L}_m}{\gen{L}_n} &= i \, \epsilon_{mnp} \, \gen{L}^p \,, & \qquad
      \comm{\gen{L}_m}{\gen{Q}_{a\alpha\dot{\alpha}}} &= \frac{1}{2} \gen{Q}_{a\beta\dot{\alpha}} \gamma^\beta_{m\,\alpha} \,, \\
    \end{aligned} \\
    \acomm{\gen{Q}_{a\alpha\dot{\alpha}}}{\gen{Q}_{b\beta\dot{\beta}}} = 
    - \left(
      \gen{S}_\mu \, (\epsilon \gamma^\mu)_{ab} \, \epsilon_{\alpha\beta}
      + \gen{L}_m \, \epsilon_{ab} \, (\epsilon \gamma^m)_{\alpha\beta}
    \right) \epsilon_{\dot{\alpha}\dot{\beta}} \,.
  \end{gathered}
\end{equation}

The $\algPSU(1,1|2)$ algebra is equipped with an outer $\algSU(2)$
automorphism, which commutes with the bosonic generators and under
which the supercharges transform as a doublet, as indicated by the
third, dotted, index. Hence the action of these $\algSU(2)$ generators
on the algebra is the same as the action of the $\gen{R}_{\dot{m}}$
generators of the $\algD{\alpha}$ algebra in
\eqref{eq:d21a-comm-rel}. Note, however, that these additional
generators are \emph{not} identical with $\gen{R}_{\dot{m}}$, since
the latter, in the limit $\alpha \to 1$, commute both among themselves
and with $\algPSU(1,1|2)$.

Let us now consider what happens to the representations in the last
section as $\alpha \to 1$. With the above scaling, the $\gen{R}_{\dot
  m}$ generators annihilate all states. The rest of the states in the
$(-\frac{\alpha}{2} ; \frac{1}{2} ; 0)$ representation in
\eqref{eq:d21a-chiral-rep-LR}--\eqref{eq:d21a-chiral-rep-Q} transform as
\begin{equation}\label{eq:su112-chiral-rep-L}
  \begin{aligned}
    L_5 \ket{\phi_{\pm}^{(n)}} &= \pm \frac{1}{2} \ket{\phi_{\pm}^{(n)}} \,, &
    L_+ \ket{\phi_{-}^{(n)}} &= \ket{\phi_+^{(n)}} \,, &
    L_- \ket{\phi_{+}^{(n)}} &= \ket{\phi_-^{(n)}} \,,
  \end{aligned}
\end{equation}
\begin{equation}\label{eq:su112-chiral-rep-S}
  \begin{aligned}
    S_0 \ket{\phi_{\alpha}^{(n)}} &= - \left( \frac{1}{2} + n \right) \ket{\phi_{\alpha}^{(n)}} \,, &
    S_0 \ket{\psi_{\dot\alpha}^{(n)}} &= - \left( 1 + n \right) \ket{\psi_{\dot\alpha}^{(n)}} \,, \\
    S_- \ket{\phi_{\alpha}^{(n)}} &= -(n+1) \ket{\phi_{\alpha}^{(n+1)}} \,, &
    S_- \ket{\psi_{\dot\alpha}^{(n)}} &= -\sqrt{(n+1)(n + 2)} \ket{\psi_{\dot\alpha}^{(n+1)}} \,, \\
    S_+ \ket{\phi_{\alpha}^{(n)}} &= n \ket{\phi_{\alpha}^{(n-1)}} \,, &
    S_+ \ket{\psi_{\dot\alpha}^{(n)}} &= \sqrt{n (n + 1)} \ket{\psi_{\dot\alpha}^{(n-1)}} \,,
  \end{aligned}
\end{equation}
\begin{equation}\label{eq:su112-chiral-rep-Q}
  \begin{aligned}
    Q_{+\pm\dot\alpha} \ket{\phi_{\mp}^{(n)}} &= \pm \sqrt{n} \ket{\psi_{\dot\alpha}^{(n-1)}} \,, &
    Q_{-\pm\dot\alpha} \ket{\phi_{\mp}^{(n)}} &= \pm \sqrt{n+1} \ket{\psi_{\dot\alpha}^{(n)}} \,, \\
    Q_{+\alpha\pm} \ket{\psi_{\mp}^{(n)}} &= \mp \sqrt{n+1} \ket{\phi_{\alpha}^{(n)}} \,, &
    Q_{-\alpha\pm} \ket{\psi_{\mp}^{(n)}} &= \mp \sqrt{n+1} \ket{\phi_{\alpha}^{(n+1)}} \,.
  \end{aligned}
\end{equation}
This is the same spin representation that appears in the
$\algPSU(1,1|2)$ sector of $\superN=4$ SYM~\cite{Beisert:2007sk}.

The limit of the $(-\frac{1-\alpha}{2} ; 0 ; \frac{1}{2})$
representation is more interesting. Setting $\alpha=1$ in
\eqref{eq:d21a-anti-chiral-rep-LR}--\eqref{eq:d21a-anti-chiral-rep-Q} we
obtain
\begin{equation}\label{eq:su11-anti-chiral-rep-L}
  \begin{aligned}
    L_5 \ket{\bar{\psi}_{\pm}^{(n)}} &= \pm \frac{1}{2} \ket{\bar{\psi}_{\pm}^{(n)}} \,, &
    L_+ \ket{\bar{\psi}_{-}^{(n)}} &= \ket{\bar{\psi}_+^{(n)}} \,, &
    L_- \ket{\bar{\psi}_{+}^{(n)}} &= \ket{\bar{\psi}_-^{(n)}} \,, \\
  \end{aligned}
\end{equation}
\begin{equation}\label{eq:su112-anti-chiral-rep-S}
  \begin{aligned}
    S_0 \ket{\bar{\phi}_{\dot\alpha}^{(n)}} &= - n \ket{\bar{\phi}_{\dot\alpha}^{(n)}} \,, &
    S_0 \ket{\bar{\psi}_{\alpha}^{(n)}} &= - \left( \frac{1}{2} + n \right) \ket{\bar{\psi}_{\alpha}^{(n)}} \,, \\
    S_- \ket{\bar{\phi}_{\dot\alpha}^{(n)}} &= -\sqrt{n(n+1)} \ket{\bar{\phi}_{\dot\alpha}^{(n+1)}} \,, &
    S_- \ket{\bar{\psi}_{\alpha}^{(n)}} &= -(n+1) \ket{\bar{\psi}_{\alpha}^{(n+1)}} \,, \\
    S_+ \ket{\bar{\phi}_{\dot\alpha}^{(n)}} &= \sqrt{n (n - 1)} \ket{\bar{\phi}_{\dot\alpha}^{(n-1)}} \,, &
    S_+ \ket{\bar{\psi}_{\alpha}^{(n)}} &= n \ket{\bar{\psi}_{\alpha}^{(n-1)}} \,,
  \end{aligned}
\end{equation}
\begin{equation}\label{eq:su112-anti-chiral-rep-Q}
  \begin{aligned}
    Q_{+\pm\dot\alpha} \ket{\bar{\phi}_{\mp}^{(n)}} &= \pm \sqrt{n} \ket{\bar{\psi}_{\alpha}^{(n-1)}} \,, &
    Q_{-\alpha\pm} \ket{\bar{\phi}_{\mp}^{(n)}} &= \pm \sqrt{n} \ket{\bar{\psi}_{\alpha}^{(n)}} \,, \\
    Q_{+\pm\dot\alpha} \ket{\bar{\psi}_{\mp}^{(n)}} &= \mp \sqrt{n} \ket{\bar{\phi}_{\dot\alpha}^{(n)}} \,, &
    Q_{-\pm\dot\alpha} \ket{\bar{\psi}_{\mp}^{(n)}} &= \mp \sqrt{n+1} \ket{\bar{\phi}_{\dot\alpha}^{(n+1)}} \,.
  \end{aligned}
\end{equation}
Note, that the states $\ket{\bar{\phi}_{\pm}^{(0)}}$ are annihilated
by all generators. Hence, this representation is \emph{reducible}. The
highest weight state of the general $\alpha$ representation becomes a
pair of singlets.  To identify the rest of the
representation we define a new set of bosonic states
$\ket{\bar{\varphi}_{\pm}^{(n)}} =
-\ket{\bar{\phi}_{\pm}^{(n+1)}}$. The algebra then reads
\begin{equation}\label{eq:su11-anti-chiral-rep2-L}
  \begin{aligned}
    L_5 \ket{\bar{\psi}_{\pm}^{(n)}} &= \pm \frac{1}{2} \ket{\bar{\psi}_{\pm}^{(n)}} \,, &
    L_+ \ket{\bar{\psi}_{-}^{(n)}} &= \ket{\bar{\psi}_+^{(n)}} \,, &
    L_- \ket{\bar{\psi}_{+}^{(n)}} &= \ket{\bar{\psi}_-^{(n)}} \,, \\
  \end{aligned}
\end{equation}
\begin{equation}\label{eq:su112-anti-chiral-rep2-S}
  \begin{aligned}
    S_0 \ket{\bar{\psi}_{\alpha}^{(n)}} &= - \left( \frac{1}{2} + n \right) \ket{\bar{\psi}_{\alpha}^{(n)}} \,, &
    S_0 \ket{\bar{\varphi}_{\dot\alpha}^{(n)}} &= - (n+1) \ket{\bar{\varphi}_{\dot\alpha}^{(n)}} \,, \\
    S_- \ket{\bar{\psi}_{\alpha}^{(n)}} &= -(n+1) \ket{\bar{\psi}_{\alpha}^{(n+1)}} \,, &
    S_- \ket{\bar{\varphi}_{\dot\alpha}^{(n)}} &= -\sqrt{(n+1)(n+2)} \ket{\bar{\varphi}_{\dot\alpha}^{(n+1)}} \,, \\
    S_+ \ket{\bar{\psi}_{\alpha}^{(n)}} &= n \ket{\bar{\psi}_{\alpha}^{(n-1)}} \,, &
    S_+ \ket{\bar{\varphi}_{\dot\alpha}^{(n)}} &= \sqrt{n (n + 1)} \ket{\bar{\varphi}_{\dot\alpha}^{(n-1)}} \,,
  \end{aligned}
\end{equation}
\begin{equation}\label{eq:su112-anti-chiral-rep2-Q}
  \begin{aligned}
    Q_{+\pm\dot\alpha} \ket{\bar{\psi}_{\mp}^{(n)}} &= \pm \sqrt{n} \ket{\bar{\varphi}_{\dot\alpha}^{(n-1)}} \,, &
    Q_{-\pm\dot\alpha} \ket{\bar{\psi}_{\mp}^{(n)}} &= \pm \sqrt{n+1} \ket{\bar{\varphi}_{\dot\alpha}^{(n)}} \,, \\
    Q_{+\pm\dot\alpha} \ket{\bar{\varphi}_{\mp}^{(n)}} &= \mp \sqrt{n+1} \ket{\bar{\psi}_{\alpha}^{(n)}} \,, &
    Q_{-\alpha\pm} \ket{\bar{\varphi}_{\mp}^{(n)}} &= \mp \sqrt{n+1} \ket{\bar{\psi}_{\alpha}^{(n+1)}} \,.
  \end{aligned}
\end{equation}
Comparing these expressions to
\eqref{eq:su112-chiral-rep-L}--\eqref{eq:su112-chiral-rep-Q}, we see
that this is almost the same representation except that the roles of
the bosonic and fermionic states have been interchanged.

\section{Global charges}

The global charges corresponding to a set of Bethe roots can be read off by adding a Bethe root close to $u = \infty$~\cite{Beisert:2005fw}.
From the general equations in \eqref{eq:BE-any-group} we get\footnote{%
  To restrict to highest weight states we assume that no root except $u_{l,i}$ is at $u = \infty$.
}%
\begin{equation}
  \begin{split}
    1 &=
    \left( \frac{u_{l,i} - \frac{i}{2} \vec{\alpha}_l \cdot \vec{w}}{u_{l,i} + \frac{i}{2} \vec{\alpha}_l \cdot \vec{w}} \right)^L
    \prod_{l'} \prod_{\substack{j \\ (j,l') \neq (i,l)}}^{K_{l'}} 
    \frac{u_{l,i} - u_{l',j} + \frac{i}{2} \vec{\alpha}_l \cdot \vec{\alpha}_{l'}}{u_{l,i} - u_{l,j} - \frac{i}{2} \vec{\alpha}_l \cdot \vec{\alpha}_{l'}} \\
    &\approx
    1 - \frac{i}{u_{l,i}} \vec{\alpha}_l \cdot \left( L \vec{w} - \sum_{l'} K_{l'} \vec{\alpha}_{l'} \right) + \dotsb \,.
  \end{split}
\end{equation}
For $\algD{\alpha}$ the charges for a state with excitation numbers $K_1$, $K_2$, $K_3$ are
\begin{equation}
  (p;q;r) = \left( -\tfrac{L + K_2}{2} ; \tfrac{L + K_2}{2} - K_1 ; \tfrac{L + K_2}{2} - K_3 \right) \,.
\end{equation}

\bibliographystyle{nb}
\bibliography{refs}

\makeatletter \@ifundefined{Sphere}{\newcommand{\Sphere}{S}}{}
  \@ifundefined{AdS}{\newcommand{\AdS}{\mathrm{AdS}}}{}
  \@ifundefined{CFT}{\newcommand{\CFT}{\mathrm{CFT}}}{}
  \@ifundefined{CP}{\newcommand{\CP}{\mathrm{CP}}}{}
  \@ifundefined{superN}{\newcommand{\superN}{\mathcal{N}}}{}
  \@ifundefined{grOSp}{\newcommand{\grOSp}{\mathrm{OSp}}}{}
  \@ifundefined{grSU}{\newcommand{\grSU}{\mathrm{SU}}}{}
  \@ifundefined{grU}{\newcommand{\grU}{\mathrm{U}}}{}
  \@ifundefined{grSL}{\newcommand{\grSL}{\mathrm{SL}}}{}
  \@ifundefined{grSp}{\newcommand{\grSp}{\mathrm{Sp}}}{}
  \@ifundefined{grSO}{\newcommand{\grSO}{\mathrm{SO}}}{}
  \@ifundefined{grO}{\newcommand{\grO}{\mathrm{O}}}{}
  \@ifundefined{algOSp}{\newcommand{\algOSp}{\mathrm{osp}}}{}
  \@ifundefined{algSU}{\newcommand{\algSU}{\mathrm{su}}}{}
  \@ifundefined{algSp}{\newcommand{\algSp}{\mathrm{sp}}}{}
  \@ifundefined{algSL}{\newcommand{\algSL}{\mathrm{sl}}}{}
  \@ifundefined{algU}{\newcommand{\algU}{\mathrm{u}}}{}
  \@ifundefined{algSO}{\newcommand{\algSO}{\mathrm{so}}}{}
  \@ifundefined{algO}{\newcommand{\algO}{\mathrm{o}}}{} \makeatother
\begin{thebibliography}{10}
\ifx\href\asklfhas\newcommand{\href}[2]{#2}\fi
\ifx\arxivref\asklfhas\newcommand{\arxivref}[2]{\href{http://arxiv.org/abs/#1}{#2}}\fi
\ifx\doiref\asklfhas\newcommand{\doiref}[2]{\href{http://dx.doi.org/#1}{#2}}\fi
\raggedright
\small
\parskip 0pt

\bibitem{Babichenko:2009dk}
A.~Babichenko, B.~Stefa{\'n}ski,~jr. and K.~Zarembo,
\textit{``Integrability and the {$\AdS_3$/$\CFT_2$} correspondence''},
\textsf{\doiref{10.1007/JHEP03(2010)058}{JHEP~1003,~058~(2010)}},
\texttt{\arxivref{0912.1723}{arxiv:0912.1723}}.

\bibitem{Maldacena:1997re}
J.~M.~Maldacena,
\textit{``The large {N} limit of superconformal field theories and
  supergravity''},
\textsf{Adv.~Theor.~Math.~Phys.~2,~231~(1998)},
\texttt{\arxivref{hep-th/9711200}{hep-th/9711200}}.

\bibitem{Gubser:1998bc}
S.~S.~Gubser, I.~R.~Klebanov and A.~M.~Polyakov,
\textit{``Gauge theory correlators from non-critical string theory''},
\textsf{\doiref{10.1016/S0370-2693(98)00377-3}{Phys.~Lett.~B428,~105~(1998)}},
\texttt{\arxivref{hep-th/9802109}{hep-th/9802109}}.

\bibitem{Witten:1998qj}
E.~Witten,
\textit{``Anti-de {S}itter space and holography''},
\textsf{Adv.~Theor.~Math.~Phys.~2,~253~(1998)},
\texttt{\arxivref{hep-th/9802150}{hep-th/9802150}}.

\bibitem{Aharony:1999ti}
O.~Aharony, S.~S.~Gubser, J.~M.~Maldacena, H.~Ooguri and Y.~Oz,
\textit{``Large {N} field theories, string theory and gravity''},
\textsf{\doiref{10.1016/S0370-1573(99)00083-6}{Phys.~Rept.~323,~183~(2000)}},
\texttt{\arxivref{hep-th/9905111}{hep-th/9905111}}.

\bibitem{'tHooft:1973jz}
G.~'t~Hooft,
\textit{``A Planar Diagram Theory for Strong Interactions''},
\textsf{\doiref{10.1016/0550-3213(74)90154-0}{Nucl.~Phys.~B72,~461~(1974)}}.

\bibitem{Fiamberti:2007rj}
F.~Fiamberti, A.~Santambrogio, C.~Sieg and D.~Zanon,
\textit{``Wrapping at four loops in {$\superN = 4$ SYM}''},
\textsf{\doiref{10.1016/j.physletb.2008.06.061}{Phys.~Lett.~B666,~100~(2007)}},
\texttt{\arxivref{0712.3522}{arxiv:0712.3522}}.

\bibitem{Bajnok:2008bm}
Z.~Bajnok and R.~A.~Janik,
\textit{``Four-loop perturbative {K}onishi from strings and finite size effects
  for multiparticle states''},
\textsf{\doiref{10.1016/j.nuclphysb.2008.08.020}{Nucl.~Phys.~B807,~625~(2009)}},
\texttt{\arxivref{0807.0399}{arxiv:0807.0399}}.

\bibitem{Bajnok:2008qj}
Z.~Bajnok, R.~A.~Janik and T.~{\L}ukowski,
\textit{``Four loop twist two, {BFKL}, wrapping and strings''},
\textsf{\doiref{10.1016/j.nuclphysb.2009.02.005}{Nucl.~Phys.~B816,~376~(2008)}},
\texttt{\arxivref{0811.4448}{arxiv:0811.4448}}.

\bibitem{Fiamberti:2008sh}
F.~Fiamberti, A.~Santambrogio, C.~Sieg and D.~Zanon,
\textit{``Anomalous dimension with wrapping at four loops in {$\superN = 4$}
  {SYM}''},
\textsf{\doiref{10.1016/j.nuclphysb.2008.07.014}{Nucl.~Phys.~B805,~231~(2008)}},
\texttt{\arxivref{0806.2095}{arxiv:0806.2095}}.

\bibitem{Bajnok:2009vm}
Z.~Bajnok, A.~Hegedus, R.~A.~Janik and T.~{\L}ukowski,
\textit{``Five loop {K}onishi from {AdS/CFT}''},
\textsf{\doiref{10.1016/j.nuclphysb.2009.10.015}{Nucl.~Phys.~B827,~426~(2010)}},
\texttt{\arxivref{0906.4062}{arxiv:0906.4062}}.

\bibitem{Lukowski:2009ce}
T.~Lukowski, A.~Rej and V.~N.~Velizhanin,
\textit{``Five-Loop Anomalous Dimension of Twist-Two Operators''},
\textsf{\doiref{10.1016/j.nuclphysb.2010.01.008}{Nucl.~Phys.~B831,~105~(2010)}},
\texttt{\arxivref{0912.1624}{arxiv:0912.1624}}.

\bibitem{Arutyunov:2010gb}
G.~Arutyunov, S.~Frolov and R.~Suzuki,
\textit{``Five-loop {K}onishi from the Mirror {TBA}''},
\textsf{\doiref{10.1007/JHEP04(2010)069}{JHEP~1004,~069~(2010)}},
\texttt{\arxivref{1002.1711}{arxiv:1002.1711}}.

\bibitem{Balog:2010xa}
J.~Balog and A.~Hegedus,
\textit{``{5}-loop {K}onishi from linearized {TBA} and the {XXX} magnet''},
\textsf{\doiref{10.1007/JHEP06(2010)080}{JHEP~1006,~080~(2010)}},
\texttt{\arxivref{1002.4142}{arxiv:1002.4142}}.

\bibitem{Beisert:2010jr}
N.~Beisert et~al.,
\textit{``Review of {AdS/CFT} Integrability: An Overview''},
\texttt{\arxivref{1012.3982}{arxiv:1012.3982}}.

\bibitem{Sorokin:2011rr}
D.~Sorokin, A.~Tseytlin, L.~Wulff and K.~Zarembo,
\textit{``Superstrings in {$\AdS_2 \times \Sphere^2 \times T^6$}''},
\texttt{\arxivref{1104.1793}{arxiv:1104.1793}}.

\bibitem{Giveon:1998ns}
A.~Giveon, D.~Kutasov and N.~Seiberg,
\textit{``Comments on string theory on {$\AdS_3$}''},
\textsf{Adv.~Theor.~Math.~Phys.~2,~733~(1998)},
\texttt{\arxivref{hep-th/9806194}{hep-th/9806194}}.

\bibitem{Elitzur:1998mm}
S.~Elitzur, O.~Feinerman, A.~Giveon and D.~Tsabar,
\textit{``String theory on {$\AdS_3 \times \Sphere^3 \times \Sphere^3 \times
  \Sphere^1$}''},
\textsf{\doiref{10.1016/S0370-2693(99)00101-X}{Phys.~Lett.~B449,~180~(1999)}},
\texttt{\arxivref{hep-th/9811245}{hep-th/9811245}}.

\bibitem{Seiberg:1999xz}
N.~Seiberg and E.~Witten,
\textit{``The {D1/D5} system and singular {CFT}''},
\textsf{\doiref{10.1088/1126-6708/1999/04/017}{JHEP~9904,~017~(1999)}},
\texttt{\arxivref{hep-th/9903224}{hep-th/9903224}}.

\bibitem{Larsen:1999uk}
F.~Larsen and E.~J.~Martinec,
\textit{``{$\grU(1)$} charges and moduli in the {D1}-{D5} system''},
\textsf{\doiref{10.1088/1126-6708/1999/06/019}{JHEP~9906,~019~(1999)}},
\texttt{\arxivref{hep-th/9905064}{hep-th/9905064}}.

\bibitem{Maldacena:2000hw}
J.~M.~Maldacena and H.~Ooguri,
\textit{``Strings in {$\AdS_3$} and {$\grSL(2,R)$} {WZW} model. {I}''},
\textsf{\doiref{10.1063/1.1377273}{J.~Math.~Phys.~42,~2929~(2001)}},
\texttt{\arxivref{hep-th/0001053}{hep-th/0001053}}.

\bibitem{Maldacena:2000kv}
J.~M.~Maldacena, H.~Ooguri and J.~Son,
\textit{``Strings in {$\AdS_3$} and the {$\grSL(2,R)$} {WZW} model. {II}:
  {E}uclidean black hole''},
\textsf{\doiref{10.1063/1.1377039}{J.~Math.~Phys.~42,~2961~(2001)}},
\texttt{\arxivref{hep-th/0005183}{hep-th/0005183}}.

\bibitem{Maldacena:2001km}
J.~M.~Maldacena and H.~Ooguri,
\textit{``Strings in {$\AdS_3$} and the {$\grSL(2,R)$} {WZW} model.~{III}:
  Correlation functions''},
\textsf{\doiref{10.1103/PhysRevD.65.106006}{Phys.~Rev.~D65,~106006~(2002)}},
\texttt{\arxivref{hep-th/0111180}{hep-th/0111180}}.

\bibitem{Adam:2007ws}
I.~Adam, A.~Dekel, L.~Mazzucato and Y.~Oz,
\textit{``Integrability of type {II} superstrings on {R}amond-{R}amond
  backgrounds in various dimensions''},
\textsf{\doiref{10.1088/1126-6708/2007/06/085}{JHEP~0706,~085~(2007)}},
\texttt{\arxivref{hep-th/0702083}{hep-th/0702083}}.

\bibitem{David:2008yk}
J.~R.~David and B.~Sahoo,
\textit{``Giant magnons in the {D1}-D5 system''},
\textsf{\doiref{10.1088/1126-6708/2008/07/033}{JHEP~0807,~033~(2008)}},
\texttt{\arxivref{0804.3267}{arxiv:0804.3267}}.

\bibitem{David:2010yg}
J.~R.~David and B.~Sahoo,
\textit{``{S}-matrix for magnons in the {D1}-{D5} system''},
\textsf{\doiref{10.1007/JHEP10(2010)112}{JHEP~1010,~112~(2010)}},
\texttt{\arxivref{1005.0501}{arxiv:1005.0501}}.

\bibitem{Ashok:2009xx}
S.~K.~Ashok, R.~Benichou and J.~Troost,
\textit{``Conformal Current Algebra in Two Dimensions''},
\textsf{\doiref{10.1088/1126-6708/2009/06/017}{JHEP~0906,~017~(2009)}},
\texttt{\arxivref{0903.4277}{arxiv:0903.4277}}.

\bibitem{Benichou:2010rk}
R.~Benichou and J.~Troost,
\textit{``The conformal current algebra on supergroups with applications to the
  spectrum and integrability''},
\textsf{\doiref{10.1007/JHEP04(2010)121}{JHEP~1004,~121~(2010)}},
\texttt{\arxivref{1002.3712}{arxiv:1002.3712}}.

\bibitem{Benichou:2010ts}
R.~Benichou,
\textit{``Fusion of line operators in conformal sigma-models on supergroups,
  and the {H}irota equation''},
\textsf{\doiref{10.1007/JHEP01(2011)066}{JHEP~1101,~066~(2011)}},
\texttt{\arxivref{1011.3158}{arxiv:1011.3158}}.

\bibitem{David:2011iy}
J.~R.~David and A.~Sadhukhan,
\textit{``Classical integrability in the {BTZ} black hole''},
\texttt{\arxivref{1105.0480}{arxiv:1105.0480}}.

\bibitem{Gauntlett:1998kc}
J.~P.~Gauntlett, R.~C.~Myers and P.~K.~Townsend,
\textit{``Supersymmetry of rotating branes''},
\textsf{\doiref{10.1103/PhysRevD.59.025001}{Phys.~Rev.~D59,~025001~(1999)}},
\texttt{\arxivref{hep-th/9809065}{hep-th/9809065}}.

\bibitem{Cowdall:1998bu}
P.~M.~Cowdall and P.~K.~Townsend,
\textit{``Gauged supergravity vacua from intersecting branes''},
\textsf{\doiref{10.1016/S0370-2693(98)00445-6}{Phys.~Lett.~B429,~281~(1998)}},
\texttt{\arxivref{hep-th/9801165}{hep-th/9801165}}.

\bibitem{Boonstra:1998yu}
H.~J.~Boonstra, B.~Peeters and K.~Skenderis,
\textit{``Brane intersections, anti-de {S}itter spacetimes and dual
  superconformal theories''},
\textsf{\doiref{10.1016/S0550-3213(98)00512-4}{Nucl.~Phys.~B533,~127~(1998)}},
\texttt{\arxivref{hep-th/9803231}{hep-th/9803231}}.

\bibitem{deBoer:1999rh}
J.~de~Boer, A.~Pasquinucci and K.~Skenderis,
\textit{``{AdS/CFT} dualities involving large 2d {$\superN = 4$} superconformal
  symmetry''},
\textsf{Adv.~Theor.~Math.~Phys.~3,~577~(1999)},
\texttt{\arxivref{hep-th/9904073}{hep-th/9904073}}.

\bibitem{Papadopoulos:1999tw}
G.~Papadopoulos, J.~G.~Russo and A.~A.~Tseytlin,
\textit{``Curved branes from string dualities''},
\textsf{\doiref{10.1088/0264-9381/17/7/310}{Class.~Quant.~Grav.~17,~1713~(2000)}},
\texttt{\arxivref{hep-th/9911253}{hep-th/9911253}}.

\bibitem{Giveon:2003ku}
A.~Giveon and A.~Pakman,
\textit{``More on superstrings in {$\AdS_3 \times N$}''},
\textsf{\doiref{10.1088/1126-6708/2003/03/056}{JHEP~0303,~056~(2003)}},
\texttt{\arxivref{hep-th/0302217}{hep-th/0302217}}.

\bibitem{Gukov:2004ym}
S.~Gukov, E.~Martinec, G.~W.~Moore and A.~Strominger,
\textit{``The search for a holographic dual to {$\AdS_3 \times S^3 \times S^3
  \times S^1$}''},
\textsf{Adv.~Theor.~Math.~Phys.~9,~435~(2005)},
\texttt{\arxivref{hep-th/0403090}{hep-th/0403090}}.

\bibitem{Grisaru:1985fv}
M.~T.~Grisaru, P.~S.~Howe, L.~Mezincescu, B.~Nilsson and P.~K.~Townsend,
\textit{``{$\superN=2$} Superstrings in a Supergravity Background''},
\textsf{\doiref{10.1016/0370-2693(85)91071-8}{Phys.~Lett.~B162,~116~(1985)}}.

\bibitem{Duff:1987bx}
M.~J.~Duff, P.~S.~Howe, T.~Inami and K.~S.~Stelle,
\textit{``Superstrings in {$D = 10$} from supermembranes in {$D = 11$}''},
\textsf{\doiref{10.1016/0370-2693(87)91323-2}{Phys.~Lett.~B191,~70~(1987)}}.

\bibitem{Henneaux:1984mh}
M.~Henneaux and L.~Mezincescu,
\textit{``A Sigma Model Interpretation of {G}reen-{S}chwarz Covariant
  Superstring Action''},
\textsf{\doiref{10.1016/0370-2693(85)90507-6}{Phys.~Lett.~B152,~340~(1985)}}.

\bibitem{Metsaev:1998it}
R.~R.~Metsaev and A.~A.~Tseytlin,
\textit{``Type {IIB} superstring action in {$\AdS_5 \times \Sphere^5$}
  background''},
\textsf{\doiref{10.1016/S0550-3213(98)00570-7}{Nucl.~Phys.~B533,~109~(1998)}},
\texttt{\arxivref{hep-th/9805028}{hep-th/9805028}}.

\bibitem{Stefanski:2007dp}
B.~Stefanski,~Jr.,
\textit{``{L}andau-{L}ifshitz sigma-models, fermions and the {A}d{S/CFT}
  correspondence''},
\textsf{\doiref{10.1088/1126-6708/2007/07/009}{JHEP~0707,~009~(2007)}},
\texttt{\arxivref{0704.1460}{arxiv:0704.1460}}.

\bibitem{Arutyunov:2008if}
G.~Arutyunov and S.~Frolov,
\textit{``Superstrings on {$\AdS_4 \times \CP^3$} as a Coset Sigma-model''},
\textsf{\doiref{10.1088/1126-6708/2008/09/129}{JHEP~0809,~129~(2008)}},
\texttt{\arxivref{0806.4940}{arxiv:0806.4940}}.

\bibitem{Stefanski:2008ik}
B.~Stefa{\'n}ski,~jr,
\textit{``{G}reen-{S}chwarz action for Type {IIA} strings on
  {$\AdS_4\times\CP^3$}''},
\textsf{\doiref{10.1016/j.nuclphysb.2008.09.015}{Nucl.~Phys.~B808,~80~(2008)}},
\texttt{\arxivref{0806.4948}{arxiv:0806.4948}}.

\bibitem{Zarembo:2010sg}
K.~Zarembo,
\textit{``Strings on Semisymmetric Superspaces''},
\textsf{\doiref{10.1007/JHEP05(2010)002}{JHEP~1005,~002~(2010)}},
\texttt{\arxivref{1003.0465}{arxiv:1003.0465}}.

\bibitem{Gomis:2008jt}
J.~Gomis, D.~Sorokin and L.~Wulff,
\textit{``The complete {$\AdS_4 \times \CP^3$} superspace for the type {IIA}
  superstring and {D}-branes''},
\textsf{\doiref{10.1088/1126-6708/2009/03/015}{JHEP~0903,~015~(2009)}},
\texttt{\arxivref{0811.1566}{arxiv:0811.1566}}.

\bibitem{Bena:2003wd}
I.~Bena, J.~Polchinski and R.~Roiban,
\textit{``Hidden symmetries of the {$\AdS_5 \times \Sphere^5$} superstring''},
\textsf{\doiref{10.1103/PhysRevD.69.046002}{Phys.~Rev.~D69,~046002~(2004)}},
\texttt{\arxivref{hep-th/0305116}{hep-th/0305116}}.

\bibitem{Berkovits:1999zq}
N.~Berkovits, M.~Bershadsky, T.~Hauer, S.~Zhukov and B.~Zwiebach,
\textit{``Superstring theory on {$\AdS_2 \times \Sphere^2$} as a coset
  supermanifold''},
\textsf{\doiref{10.1016/S0550-3213(99)00683-5}{Nucl.~Phys.~B567,~61~(2000)}},
\texttt{\arxivref{hep-th/9907200}{hep-th/9907200}}.

\bibitem{Gromov:2008qe}
N.~Gromov and P.~Vieira,
\textit{``The all loop {$\AdS_4$/$\CFT_3$} {B}ethe ansatz''},
\textsf{\doiref{10.1088/1126-6708/2009/01/016}{JHEP~0901,~016~(2009)}},
\texttt{\arxivref{0807.0777}{arxiv:0807.0777}}.

\bibitem{Berenstein:2002jq}
D.~E.~Berenstein, J.~M.~Maldacena and H.~S.~Nastase,
\textit{``Strings in flat space and {pp} waves from {$\superN = 4$} super
  {Y}ang {M}ills''},
\textsf{\doiref{10.1088/1126-6708/2002/04/013}{JHEP~0204,~013~(2002)}},
\texttt{\arxivref{hep-th/0202021}{hep-th/0202021}}.

\bibitem{Russo:2002rq}
J.~G.~Russo and A.~A.~Tseytlin,
\textit{``On solvable models of type {IIB} superstring in {NS}-{NS} and {R}-{R}
  plane wave backgrounds''},
\textsf{\doiref{10.1088/1126-6708/2002/04/021}{JHEP~0204,~021~(2002)}},
\texttt{\arxivref{hep-th/0202179}{hep-th/0202179}}.

\bibitem{Lu:2002kw}
H.~Lu and J.~F.~Vazquez-Poritz,
\textit{``Penrose limits of non-standard brane intersections''},
\textsf{\doiref{10.1088/0264-9381/19/15/312}{Class.~Quant.~Grav.~19,~4059~(2002)}},
\texttt{\arxivref{hep-th/0204001}{hep-th/0204001}}.

\bibitem{Gomis:2002qi}
J.~Gomis, L.~Motl and A.~Strominger,
\textit{``pp-wave / {$\CFT_2$} duality''},
\textsf{\doiref{10.1088/1126-6708/2002/11/016}{JHEP~0211,~016~(2002)}},
\texttt{\arxivref{hep-th/0206166}{hep-th/0206166}}.

\bibitem{Gava:2002xb}
E.~Gava and K.~S.~Narain,
\textit{``Proving the pp-wave / {$\CFT_2$} duality''},
\textsf{\doiref{10.1088/1126-6708/2002/12/023}{JHEP~0212,~023~(2002)}},
\texttt{\arxivref{hep-th/0208081}{hep-th/0208081}}.

\bibitem{Sommovigo:2003kd}
L.~Sommovigo,
\textit{``Penrose limit of {$\AdS_3 \times \Sphere^3 \times \Sphere^3 \times
  \Sphere^1$} and its associated sigma-model''},
\textsf{\doiref{10.1088/1126-6708/2003/07/035}{JHEP~0307,~035~(2003)}},
\texttt{\arxivref{hep-th/0305151}{hep-th/0305151}}.

\bibitem{Derkachov:2000ne}
S.~E.~Derkachov, D.~Karakhanian and R.~Kirschner,
\textit{``Heisenberg spin chains based on {$sl(2|1)$} symmetry''},
\textsf{\doiref{10.1016/S0550-3213(00)00346-1}{Nucl.~Phys.~B583,~691~(2000)}},
\texttt{\arxivref{nlin/0003029}{nlin/0003029}}.

\bibitem{Minahan:2008hf}
J.~A.~Minahan and K.~Zarembo,
\textit{``The {B}ethe ansatz for superconformal {C}hern-{S}imons''},
\textsf{\doiref{10.1088/1126-6708/2008/09/040}{JHEP~0809,~040~(2008)}},
\texttt{\arxivref{0806.3951}{arxiv:0806.3951}}.

\bibitem{Pakman:2009mi}
A.~Pakman, L.~Rastelli and S.~S.~Razamat,
\textit{``A Spin Chain for the Symmetric Product {$\CFT_2$}''},
\texttt{\arxivref{0912.0959}{arxiv:0912.0959}}.

\bibitem{Beisert:2007sk}
N.~Beisert and B.~I.~Zwiebel,
\textit{``On Symmetry Enhancement in the {$psu(1,1|2)$} Sector of {$\superN =
  4$} {SYM}''},
\textsf{\doiref{10.1088/1126-6708/2007/10/031}{JHEP~0710,~031~(2007)}},
\texttt{\arxivref{0707.1031}{arxiv:0707.1031}}.

\bibitem{Zwiebel:2009vb}
B.~I.~Zwiebel,
\textit{``Two-loop Integrability of Planar {$\superN=6$} Superconformal
  {C}hern-{S}imons Theory''},
\textsf{\doiref{10.1088/1751-8113/42/49/495402}{J.~Phys.~A42,~495402~(2009)}},
\texttt{\arxivref{0901.0411}{arxiv:0901.0411}}.

\bibitem{Ogievetsky:1986hu}
E.~Ogievetsky and P.~Wiegmann,
\textit{``Factorized {S} matrix and the {B}ethe ansatz for simple {L}ie
  groups''},
\textsf{\doiref{10.1016/0370-2693(86)91644-8}{Phys.~Lett.~B168,~360~(1986)}}.

\bibitem{Beisert:2006ez}
N.~Beisert, B.~Eden and M.~Staudacher,
\textit{``Transcendentality and crossing''},
\textsf{J.~Stat.~Mech.~0701,~P021~(2007)},
\texttt{\arxivref{hep-th/0610251}{hep-th/0610251}}.

\bibitem{Beisert:2006ib}
N.~Beisert, R.~Hern{\'a}ndez and E.~L{\'o}pez,
\textit{``A crossing-symmetric phase for {$\AdS_5 \times \Sphere^5$}''},
\textsf{\doiref{10.1088/1126-6708/2006/11/070}{JHEP~0611,~070~(2006)}},
\texttt{\arxivref{hep-th/0609044}{hep-th/0609044}}.

\bibitem{Arutyunov:2004vx}
G.~Arutyunov, S.~Frolov and M.~Staudacher,
\textit{``{B}ethe ansatz for quantum strings''},
\textsf{\doiref{10.1088/1126-6708/2004/10/016}{JHEP~0410,~016~(2004)}},
\texttt{\arxivref{hep-th/0406256}{hep-th/0406256}}.

\bibitem{Minahan:2002ve}
J.~A.~Minahan and K.~Zarembo,
\textit{``The {B}ethe-ansatz for {$\superN = 4$} super {Y}ang-{M}ills''},
\textsf{\doiref{10.1088/1126-6708/2003/03/013}{JHEP~0303,~013~(2003)}},
\texttt{\arxivref{hep-th/0212208}{hep-th/0212208}}.

\bibitem{VanDerJeugt:1985hq}
J.~Van~der~Jeugt,
\textit{``Irreducible representations of the exceptional {L}ie superalgebras
  {$D(2,1 \alpha)$}''},
\textsf{\doiref{10.1063/1.526547}{J.~Math.~Phys.~26,~913~(1985)}}.

\bibitem{Frappat:1996pb}
L.~Frappat, P.~Sorba and A.~Sciarrino,
\textit{``Dictionary on {L}ie superalgebras''},
\texttt{\arxivref{hep-th/9607161}{hep-th/9607161}}.

\bibitem{Kac:1977em}
V.~G.~Kac,
\textit{``{L}ie Superalgebras''},
\textsf{\doiref{10.1016/0001-8708(77)90017-2}{Adv.~Math.~26,~8~(1977)}}.

\bibitem{Kac:1977qb}
V.~G.~Kac,
\textit{``A Sketch of {L}ie Superalgebra Theory''},
\textsf{\doiref{10.1007/BF01609166}{Commun.~Math.~Phys.~53,~31~(1977)}}.

\bibitem{Beisert:2003jj}
N.~Beisert,
\textit{``The complete one-loop dilatation operator of {$\superN = 4$} super
  {Y}ang-{M}ills theory''},
\textsf{\doiref{10.1016/j.nuclphysb.2003.10.019}{Nucl.~Phys.~B676,~3~(2004)}},
\texttt{\arxivref{hep-th/0307015}{hep-th/0307015}}.

\bibitem{Minahan:2009te}
J.~A.~Minahan, W.~Schulgin and K.~Zarembo,
\textit{``Two loop integrability for {C}hern-{S}imons theories with {$\superN =
  6$} supersymmetry''},
\textsf{\doiref{10.1088/1126-6708/2009/03/057}{JHEP~0903,~057~(2009)}},
\texttt{\arxivref{0901.1142}{arxiv:0901.1142}}.

\bibitem{Tsuboi:1998ne}
Z.~Tsuboi,
\textit{``Analytic {B}ethe Ansatz And Functional Equations Associated With Any
  Simple Root Systems Of The Lie Superalgebra {$\grSL(r+1|s+1)$}''},
\textsf{Physica~A252,~565~(1998)}.

\bibitem{Tsuboi:1997iq}
Z.~Tsuboi,
\textit{``Analytic {B}ethe ansatz and functional equations for {L}ie
  superalgebra {$\algSL(r+1|s+1)$}''},
\textsf{\doiref{10.1088/0305-4470/30/22/031}{J.~Phys.~A30,~7975~(1997)}}.

\bibitem{Tsuboi:1998sc}
Z.~Tsuboi,
\textit{``Analytic {B}ethe ansatz related to a one-parameter family of
  finite-dimensional representations of the {L}ie superalgebra
  {$\algSL(r+1|s+1)$}''},
\textsf{\doiref{10.1088/0305-4470/31/24/010}{J.~Phys.~A31,~5485~(1998)}}.

\bibitem{Beisert:2005di}
N.~Beisert, V.~A.~Kazakov, K.~Sakai and K.~Zarembo,
\textit{``Complete spectrum of long operators in {$\superN=4$} {SYM} at one
  loop''},
\textsf{\doiref{10.1088/1126-6708/2005/07/030}{JHEP~0507,~030~(2005)}},
\texttt{\arxivref{hep-th/0503200}{hep-th/0503200}}.

\bibitem{Eden:2006rx}
B.~Eden and M.~Staudacher,
\textit{``Integrability and transcendentality''},
\textsf{J.~Stat.~Mech.~0611,~P014~(2006)},
\texttt{\arxivref{hep-th/0603157}{hep-th/0603157}}.

\bibitem{Freyhult:2007pz}
L.~Freyhult, A.~Rej and M.~Staudacher,
\textit{``A Generalized Scaling Function for {AdS/CFT}''},
\textsf{\doiref{10.1088/1742-5468/2008/07/P07015}{J.~Stat.~Mech.~0807,~P07015~(2007)}},
\texttt{\arxivref{0712.2743}{arxiv:0712.2743}}.

\bibitem{Beisert:2005fw}
N.~Beisert and M.~Staudacher,
\textit{``Long-range {$PSU(2,2|4)$} {B}ethe ansaetze for gauge theory and
  strings''},
\textsf{\doiref{10.1016/j.nuclphysb.2005.06.038}{Nucl.~Phys.~B727,~1~(2005)}},
\texttt{\arxivref{hep-th/0504190}{hep-th/0504190}}.

\bibitem{Reshetikhin:1983vw}
N.~{\relax Yu}.~Reshetikhin,
\textit{``A method of functional equations in the theory of exactly solvable
  quantum systems''},
\textsf{\doiref{10.1007/BF00400435}{Lett.~Math.~Phys.~7,~205~(1983)}}.

\bibitem{Reshetikhin:1986vd}
N.~{\relax Yu}.~Reshetikhin,
\textit{``Integrable models of quantum one-dimensional magnets with {$\grO(N)$}
  and {$\grSp(2K)$} symmetry''},
\textsf{\doiref{10.1007/BF01017501}{Theor.~Math.~Phys.~63,~555~(1985)}}.

\bibitem{toappear}
O.~Ohlsson~Sax, B.~Stefa\'nski and A.~Torrielli,
To appear.

\bibitem{Korchemsky:1995be}
G.~P.~Korchemsky,
\textit{``Quasiclassical {QCD} pomeron''},
\textsf{\doiref{10.1016/0550-3213(96)00019-3}{Nucl.~Phys.~B462,~333~(1996)}},
\texttt{\arxivref{hep-th/9508025}{hep-th/9508025}}.

\bibitem{Lorente:2001}
M.~Lorente,
\textit{``Raising and lowering operators, factorization and
  differential/difference operators of hypergeometric type''},
\textsf{J.~Phys.~A34,~569~(2001)}.

\bibitem{Beisert:2005tm}
N.~Beisert,
\textit{``The {$su(2|2)$} dynamic {$S$}-matrix''},
\textsf{Adv.~Theor.~Math.~Phys.~12,~945~(2008)},
\texttt{\arxivref{hep-th/0511082}{hep-th/0511082}}.

\bibitem{Beisert:2006qh}
N.~Beisert,
\textit{``The Analytic {B}ethe Ansatz for a Chain with Centrally Extended
  $su(2|2)$ Symmetry''},
\textsf{J.~Stat.~Mech.~0701,~P017~(2007)},
\texttt{\arxivref{nlin/0610017}{nlin/0610017}}.

\end{thebibliography}

\end{document}